\documentclass[fleqn,usenatbib]{mnras}

\usepackage{newtxtext,newtxmath}
\usepackage[T1]{fontenc}

\DeclareRobustCommand{\VAN}[3]{#2}
\let\VANthebibliography\thebibliography
\def\thebibliography{\DeclareRobustCommand{\VAN}[3]{##3}\VANthebibliography}

\usepackage{graphicx}	% Including figure files
\usepackage{amsmath}	% Advanced maths commands

\newlength{\wdth}

\title[Modelling the diffuse continuum emission: the NLS1 Mrk~110]{Modelling the diffuse continuum emission: the NLS1 Mrk~110}

\author[Goad, Korista \& Cackett]{
Michael R.\ Goad,$^{1}$\thanks{E-mail: mg159@le.ac.uk}
Kirk T.\ Korista,$^{2}$
Edward M.\ Cackett$^{3}$
\\
$^{1}$School of Physics and Astronomy, University of Leicester, University Road, Leicester, LE1 7RH, UK\\
$^{2}$Western Michigan University, Department of Physics, 1120 Everett Tower, Kalamazoo, MI 49008-5252, USA\\ 
$^{3}$Wayne State University, Department of Physics and Astronomy,
666 W Hancock St.,
Detroit,
MI 48201, USA\\
}

\date{Accepted 2026, June 18. Received 2026, June 12; in original form 2026, March 30}

\pubyear{2026}

\begin{document}
\label{firstpage}
\pagerange{\pageref{firstpage}--\pageref{lastpage}}
\maketitle

\begin{abstract}

We present detailed model diffuse continuum (DC) plus hydrogen
emission line templates from a summation over a broad range in
hydrogen gas densities. We address the effect of finite gas densities
and the presence of weak higher-order emission-line transitions on the
strength and location of the Balmer and Paschen jumps which act to
shift the jumps red-ward of their respective vacuum wavelength
positions and substantially soften the gradient of the jump. Our
photoionisation model-based DC template favours a lower optical depth
and lower electron temperature than traditionally employed for the
Balmer continuum, directly impacting empirical estimates of the
strength of UV Fe~{\sc ii}. Microturbulent velocities increase the
emission-line–continuum contrast, suppressing the DC contribution and
weakening the jump heights, though the spectral shape of the DC
remains broadly similar.  Even with these additional complexities, a
spectral decomposition of the UV-Optical-IR spectrum of the NLS1 Mrk
110, which includes a substantial DC contribution (30–-50\% of the
total light at 3600\AA), reveals a significant deficit in emission
just longward of the Balmer jump. Interestingly, significant thermal
DC emission acts to flatten the SED through the UV–optical, negating
oft-employed intrinsic reddening. Our best fit 1000\AA--3\micron\,
model requires a black hole mass $\sim$$10^{8}~{\rm M}_{\odot}$,
similar to that inferred when considering features in the H/He broad
emission line profiles that suggest the presence of gravitational
redshift. Finally, we provide diagnostic plots that may assist the
spectral modeler to construct a physically useful DC spectrum and
quantify its contribution to AGN UV-optical spectra.

\end{abstract}


%%%%%%%%%%%%%%%%%%%%%%%%%%%%%%%%%%%%%%%%%%%%%%%%%%

%%%%%%%%%%%%%%%%% BODY OF PAPER %%%%%%%%%%%%%%%%%%

\section{Introduction}
Active Galaxies are among the most luminous persistent sources known
with bolometric luminosities spanning several decades $L_{\rm bol}\sim
10^{41}$--$10^{48}$~erg~s$^{-1}$.  Powered by accretion, this
luminosity is derived within a volume which cannot be spatially
resolved for all but the nearest AGN (e.g., GRAVITY Collaboration
2018, 2020). Thus the study of these sources requires indirect
techniques.

In this respect, reverberation mapping has been particularly
fruitful. By exploiting AGN variability, reverberation mapping enables
us to investigate size scales of $\sim$ few micro-arcseconds, allowing
us to probe, albeit obliquely, the structure of the dusty torus, the
broad emission-line region, and even the accretion flow feeding the
black hole at its centre.

Much recent effort has focused on understanding the structure of the
accreting flow, using correlated continuum variations: disc
reverberation mapping. Pioneered more than 20 years ago (Collier
et~al.\ 1998, 2001), it is only within the last decade that
significant progress has been made, often with surprising results
(Cackett et~al.\ 2007, 2018, 2020; Edelson et al.\ 2015, 2019;
Fausnaugh et~al.\ 2016; Hern\'{a}ndez-Santisteban et~al.\ 2020;
Vincentelli et~al.\ 2021).

In the standard model for accretion (Shakura and Sunyaev 1973,
hereafter SS73), for material to be accreted it must lose significant
angular momentum. Angular momentum is dissipated locally via viscous
torques, which transport material inwards and angular momentum
outwards, and the disc becomes hot. The resulting structure is
geometrically thin and optically thick and produces a thermal spectrum
which over a broad range in frequencies may be approximated by a sum
of blackbodies. The disc, which is hot at the centre and cooler at
larger radii, displays a radial temperature profile that approximates
a power-law in radius $\rm{T(R)\propto R^{-3/4}}$.

Disc reverberation and its interpretation, hinges on the assumption
that the disc is illuminated from above by a compact X-ray source (the
X-ray corona), located a few scale heights above the disc at its
centre. Luminosity variations in the X-ray corona, which irradiates
the disc from above, drive corresponding flux variations within the
disc -- first within the inner, hotter regions of the disc, followed
by the cooler outer disc. This results in wavelength-dependent
spectral continuum variations. These manifest as measurable
wavelength-dependent amplitude flux variations and
wavelength-dependent delays, and the latter may be used to infer the
disc radial temperature profile and thereby test the standard disc
hypothesis (e.g., Collier et~al. 1998; Peterson
et~al.\ 1998). Formally, the wavelength-dependent continuum delays
$\tau(\lambda)$, follow the relation:

\begin{equation}
\tau(\lambda) \propto (M\dot{M})^{1/3}\lambda^{4/3} \; ,
\end{equation}

\noindent where $M$ is the black hole mass, and $\dot{M}$ the mass
accretion rate through the disc. Thus if $M$ is known independently,
e.g., via a virial estimate from: (i) reverberation mapping of the
broad emission line region (BLR), or (ii) a single spectrum and
application of the radius-luminosity relation for nearby AGN (e.g.,
Kaspi et~al. 2000; Vestergaard \& Peterson 2006), which provide the
BLR size and gas line of sight velocity dispersion, then $\dot{M}$ may
be uniquely determined.

Despite much recent success, largely owing to dedicated high cadence
multi-wavelength ground and space-based monitoring campaigns of
individual targets, results to date are rather mixed, generally
raising more questions than they have answered. Chief among these is
that though wavelength-dependent continuum variations appear to follow
the predictions of the standard disc model, with delays increasing
with increasing wavelength, the normalisation appears incorrect,
typically a factor 3 too large, implying disc sizes which are too
large for their luminosities\footnote{Overly large disc sizes have
also been found in microlensing studies of AGN (e.g., Morgan
et~al. 2010; Cornachione and Morgan 2020). However, Zdziarski et
al.\ 2022, suggest that the discrepancy between inferred disc sizes
from microlensing studies and accretion disc models may be resolved by
using more physically realistic discs, e.g., by including the effects
of colour temperature correction (e.g., Chiang 2002; Done
et~al. 2012), inner disc truncation, and the presence of disc winds on
the emergent disc spectrum.}  . Further, in nearly all sources the
U-band shows an excess in delay\footnote{We caution that attempts to
recover the pure disc delay signature that in the fitting process
simply exclude the excess delay in the u-band, are unphysical.} over
and above the standard relation (Vincentelli et~ al. 2021;
Hern\'{a}ndez-Santisteban et~al. 2020; Edelson et al.\ 2019; Cackett
et~al. 2018, though see Kara et~al. 2021 for a possible exception),
and appears strongest in objects with the largest obscuration (e.g.,
Lewin et al. 2025), while extrapolation of the UV--optical continuum
delays to X-ray wavelengths, reveal a stark disconnect, which is hard
to explain (e.g., Edelson et~al. 2017).

To address the normalisation issue, alternate models have been
proposed, e.g., (i) patchy inhomogeneous discs (Hall et al.\ 2018;
Starkey et al.\ 2017), (ii) disc scattering atmospheres (Narayan
1996), or (iii) maximally spinning black holes (Kammoun
et~al. 2021a,b). Alternatively, the larger than expected delays could
be unrelated to the disc emission, and may instead result from
contamination of the pure disc signal by, for example, reprocessed
continuum emission originating in the BLR (the so-called diffuse
continuum, hereafter DC), which for all BLR models must be present at
some and likely significant level (Korista and Goad 2001, 2019;
hereafter KG2001 and KG2019; Lawther et al.\ 2018; Chelouche et
al.\ 2019; Netzer 2022; Netzer et al.\ 2024; Jaiswal et~al. 2025).  A
frequency resolved analysis of the continuum variability data in a
handful of sources (NGC~5548, Mkn~335, and Mkn~817) provide strong
supporting evidence for contributions from a second more distant
reprocessor, and with delays commensurate with an origin in the inner
BLR (e.g., Cackett et~al.\ 2022; Lewin et~al. 2023, 2024).

Solutions to the X-ray--UV/optical disconnect which do not rule out
X-rays as the primary driver of disc variations are also forthcoming:
Gardner and Done (2017) introduced a second reprocessing region, an
extreme ultraviolet torus which acts to increase the delays between
the X-ray--UV/optical continuum bands. Alternatively, Panagiotou et
al. (2022) suggest that the weaker X-ray--UV/optical correlations are
entirely consistent with a dynamic corona, one in which the coronal
height above the disc is time-variable.

While increasing complexity may ultimately be required to explain the
variations that we see, we suggest that first a proper census of the
known contributors to the variable continuum emission should be
carried out (see the appendix in Korista and Goad 2019 for a
description of the main contributors to the broad band continuum
emission). Only once these have been accounted for will we then be in
a position to infer the true nature of the disc continuum variations.

\subsection{Diffuse continuum emission}
Of the known variable continuum emitters in the vicinity of the disc
(e.g., BLR and dusty torus), it is the BLR that is the main
contaminant of the disc-delay signature at UV--optical wavelengths
(KG2001, KG2019).  To date, published models for the BLR diffuse
continuum emission and which importantly are first {\it constrained by
  matching the observed emission-line strengths}, exist for just one
AGN -- the well-studied Sy~1.5 NGC~5548 (KG2001, KG2019; Lawther et
al.\ 2018).  Spectral decompositions of other sources typically use a
scaled version of the KG2001 or KG2019 template (e.g., Cackett
et~al. 2018; Hern\'andez-Santisteban et~al. 2020; Vincentelli
et~al. 2021), under the reasonable assumptions that: (i) the DC is
largely insensitive to the shape of the incident ionizing continuum,
and that (ii) the DC contribution to the UV-optical-IR spectrum and
broad emission-line equivalent widths approximately scale together.

Key to further progress in understanding the disc continuum variations
hinges on our ability to accurately account for, and then remove the
DC contributions. Furthermore, knowledge of the DC can provide
important additional constraints on the BLR gas physics, in particular
on the gas neutral fraction and gas hydrogen density $\rm{n_H}$, and
column density $\rm{N_H}$ (KG2001, KG2019). Thus DC models spanning a
range of BLR gas physical conditions, and for sources which may have
very different SEDs, are key ingredients for constraining the physics
of the BLR gas, and the BLR contribution to the continuum emission
through the UV--optical--IR.

In the following work we first re-visit the widely used model DC
template for NGC~5548 with the aim of incorporating important
additional physical processes that affect the appearance of the DC and
underlying disc spectra. The work is organized as follows: First we
investigate the effect of finite gas densities on the {\em location
  and rate of decline\/} of the DC at the major recombination edges
present in the optical and near infra-red, the hydrogen Balmer and
Paschen jumps (see also Jin et~al. 2012). Next, we include a proper
treatment of the pile-up of broad emission-lines (the higher order
Balmer lines, and He~{\sc i} and He~{\sc ii} lines) long-ward of the
Balmer and Paschen jumps, accounting for both their predicted
intensities and for their mean formation radii, as estimated from
local optimally emitting cloud (LOC) model integrations.  The former
are used to scale the relative emission-line contributions, and the
latter to determine the relative widths of the lines under the
assumption that the broad line region gas is virialised.

Next, we attempt to quantify the DC contribution to the underlying
continuum by applying our improved physically motivated model DC
template to a spectral decomposition of the UV--optical--IR continuum
emission in the Narrow Line Seyfert~1 galaxy Mrk~110, a source with a
significant Balmer jump and broad emission-lines which are
comparatively narrow for a type I object, and therefore easier to
isolate from the underlying continuum. While our spectral
decomposition (a combination of model calculations and empirically
derived model templates) is deliberately focused on Mrk~110, the
adopted procedure may be more widely applicable to the larger AGN
population. Finally, we conclude by constructing general diagnostic
plots to aid observers in quantifying the likely DC contribution to
the UV-optical spectra of AGN.

\section{The role of finite gas densities}

\subsection{A simple model of hydrogenic atoms at finite density}

Given that gas number densities $\rm{n_H}$ likely reach or even exceed
$10^{12}$~cm$^{-3}$ within the broad emission line regions of AGN, it
is reasonable to expect significant alterations of atomic structure
from isolated-atom expectations due to finite external electric field
strengths. Pigarov et~al. (1998, and references therein) describe the
phenomena associated with various electrical interactions for hydrogen
gases of finite density (mainly for electron temperatures near 1~eV),
which affect the discrete Balmer line transition and the continuous
spectra, and their spectral convergence. The simplest effect is the
lowering of the ionisation potential (Inglis \& Teller 1939), and here
we will assume a simple model in which the typical distance between
hydrogen atoms and ions is $d \sim (3/4\pi\/\rm{n_H})^{1/3}$, and we
will assume that no bound states will exist for electron orbital radii
$a_n = a_o n^2 > d/2$ ($n$ is the principle quantum number and $a_o =
0.52946541$~\AA\ is the hydrogen Bohr radius\footnote{In determining
the bound-state energies, we assumed the simple hydrogenic Bohr model
atom using the reduced mass of the electron--proton system, but scaled
from the NIST ionisation potential, 13.5984346 eV for an isolated
hydrogen atom. These are then modified for finite gas density as
described here. This is sufficiently accurate for our purposes.}). The
maximum bound atomic level in hydrogen $n_{max}$ can then be
estimated,

\begin{equation}
n_{max} = [4\pi/3\ (2a_o)^3\rm{n_{ion}}]^{-1/6},
\end{equation}

\noindent where $\rm{n_{ion}}$ is the ion number density, for which we
will simply associate with $\rm{n_H}$. Using the above scheme, at a
hydrogen gas density of $10^{12}$~cm$^{-3}$ the highest bound level is
$n \approx 76$, lowering the ionisation potential from 13.5984~eV to
13.5961~eV. This lowering of the ionisation potential shifts the
wavelength positions of the Balmer and Paschen series limits or
``jumps'' from $\lambda$3647.0~\AA\ to $\lambda$3649.5~\AA\ and from
$\lambda$8205.8~\AA\ to $\lambda$8218.4~\AA\/, respectively. See
Table~\ref{shifts} for values of $n_{max}$ and the shifted positions
of the two hydrogen free-bound continuum jumps at several values in
gas number density which may be found in the broad emission line
regions of AGN. The magnitude of the shift in wavelength of the jumps
grows by a factor of $10^{1/3}$ for each decade increase in gas
density. Given the expected broad range in gas number densities within
the BLR, the superposition of recombination emission edges that are
shifted red-ward in wavelength will act to {\em smooth the observed
  free-bound continuum emission edges in wavelength space, reducing
  the sharpness of the jumps}. This will be in addition to the
relatively smaller amount of smoothing due to the BLR velocity
field\footnote{We note that velocity broadening does not shift the
centroid of the position of the edge, rather it smooths off any sharp
spectral features, altering the shape of the jump. }.

Next, the statistical plasma microfield acts to depopulate a discrete
series' upper levels lying ever nearer to the shifted thresholds,
effectively lowering these levels' statistical weights (Pigarov et
al.\ 1998). This effect depends on the density, and for a hydrogen gas
density of $10^{12}$~cm$^{-3}$ and electron temperatures near 1~eV the
correction factor for the statistical weight falls to $\approx$~0.5
for $n \approx 52$, and decays rapidly to values well below 0.1 for $n
> 63$ (recall that the highest bound level is $n \approx 76$, for the
same gas density). The resulting reduced emission from the higher
lying discrete transitions within a series thus appears instead within
the associated recombination continuum. Thus, at a particular gas
density, instead of a \emph{vertical jump} at
$\lambda$3647.0~\AA\ (vacuum, low-density limit) associated with an
ideal Balmer recombination spectrum, a \emph{gradual ramp} at
wavelengths longward of a \emph{shifted threshold} results. In fact,
the Balmer series emission lines and continuum will overlap one
another to some extent, even before considering Doppler
broadening. Pigarov et al.\ (1998) find that for a density of
$10^{12}$~cm$^{-3}$, the Balmer ``jump'' is, in contrast, a ramp
spanning from a threshold wavelength of approximately 3649.5~\AA\/,
falling to 20$\%$ of the threshold intensity near 3670~\AA\/, and
eventually merging with the underlying Paschen continuum plus
higher-$n$ free-bound plus free-free continua beyond 3700~\AA\/. Here,
we approximate this reduction factor $\delta_n$ in the level's
statistical weight, following the parametrisation in Pigarov et al.:
\begin{equation}
\delta_{n} = {\rm exp}(-f(2a_n/d)^3),
\end{equation}
where $a_n$ and $d$ are defined above, and we have chosen a value of
$f = 7.1$ to bring the behaviour approximately in line with more
detailed predictions in their Figure~3 (for kT = 1 eV). The quantity
($1-\delta_n$) may be thought of as the probability that level $n$ is
coupled to the continuum, rather than to a bound state. In this
approximate manner, in \S2.5 we implement both the reduction in the
discrete series emission and the corresponding augmentation in the
associated continuum at wavelengths {\em longward} of the shifted
series limit.

\begin{table}
    \centering
  \begin{tabular}{rrcccr} \hline
 & & Balmer series & \\
 &  & $\lambda_{\rm vac}=3647.014$ \AA        &          \\ \hline
$\log$ n(H)      &  $n_{\rm max}$       & $\lambda_{\rm finite}$  & $\Delta \lambda$ \\
(cm$^{-3}$) & & (\AA) & (\AA) \\
8  & 355 & 3647.129  & 0.116 \\
9  & 242 & 3647.263  & 0.249 \\
10 & 164 & 3647.550  & 0.537 \\
11 & 112 & 3648.170  & 1.156 \\
12 & 76 & 3649.506  & 2.492 \\
13 & 52 & 3652.387  & 5.373 \\
\\ \hline
& & Paschen series & \\
& & $\lambda_{\rm vac}$=8205.781 \AA     &  \\ \hline
%$\log$ n(H)      &  $n_{\rm max}$       & $\lambda_{\rm finite}$  & $\Delta \lambda$ \\
%(cm$^{-3}$) & & (\AA) & (\AA) \\
8  & 355  & 8206.366 & 0.585\\
9  & 242  & 8207.042 & 1.261\\
10 & 164  & 8208.498 & 2.717\\
11 & 112  & 8211.637 & 5.856 \\
12 & 76 & 8218.407 & 12.626 \\
13 & 52 & 8233.031 & 27.250\\
%14 & 35 & 3658.609  & 11.595 & 8264.715 & 58.934 \\ 
\hline
    \end{tabular}
    \caption{The dependence of the Balmer and Paschen series limits
      with gas hydrogen density ${\rm n_H}$, as predicted by the
      adopted model. Note the factor $\approx$~5.1 larger shift
      ($\Delta\lambda$) in the location of the series limit for finite
      gas densities relative to their ideal vacuum wavelengths for the
      Paschen series as compared to the Balmer series.}
    \label{shifts}

\end{table}

Due to its closer proximity to the lowered ionisation limit, the shift
in the position of the Paschen jump will be greater ($\approx
5.1\times$) than that in the Balmer jump, for a particular hydrogen
density (see Table~\ref{shifts}). Given the wide range of gas
densities expected within the BLRs of AGN, this greater shift in the
Paschen series limit will act to smear out the intrinsically weaker
Paschen emission jump greatly in wavelength space. As we will
demonstrate, this effect, along with the pile-up of higher order
Paschen line transitions, as well as the plasma microfield's
depopulation of levels near the threshold will act to measurably
reduce any spectral emission features at the Paschen and higher-$n$
free-bound continuum edges in the spectra of Type~1 AGN.

Guo et al.\ (2022) discuss the lack of an observable Paschen jump
feature in AGN spectra and demonstrated the effect of the Paschen
series emission line pile-up in reducing the contrast of an expected
Paschen jump. The pile-up of the higher-order Balmer emission lines
(of finite Doppler profile broadening) will also contribute to
smoothing the Balmer continuum emission jump. However, as just
discussed, this is only one of several contributors acting to obscure
these spectral features, particularly the Paschen jump\footnote{Guo
et~al. (2022), as well as Vincentelli et~al. (2021), noted but did not
model the effect of shifted Balmer and Paschen continuum thresholds
due to the presence of dense gas within the BLR.}.

In a multicomponent fit to the Balmer continuum spectral region in
PG1427$+$480, Jin et al.\ (2012) suggested gas number densities in the
range $10^{16}-10^{17}$~cm$^{-3}$ for electron temperatures spanning
$T_{e}\sim 10^{4}-10^{5}$~K, corresponding to $n_{max}\approx$~12, to
place the Balmer jump in this AGN near a wavelength of 3746\AA\/ to
account for the excess smooth-appearing emission centred here. Such
high gas number densities are {\em orders of magnitude in excess of
  those associated with the broad emission line regions of AGN}, but
may be appropriate for the chromosphere of an X-ray illuminated
accretion disc. In their spectral fit near the region of the Balmer
jump, however, Jin et al.\ did not consider the aforementioned effects
of: (1) the pile-up of a truncated Balmer emission line series, (2) a
Paschen plus free-free continuum (both of which must be present), and
(3) the presence of Balmer continuum emission \emph{long-ward} of the
shifted thresholds emitted by gas spanning a range of finite gas
densities (Pigarov et al.\ 1998; their Figures 2 and 5). Here, using
the high-quality \emph{Hubble Space Telescope} spectrum of the AGN
Mrk~110 we will consider the above effects in modelling the Balmer and
Paschen jump regions of the spectrum.

In Section 3 we will show that their inclusion reduces the ``emission
gap'' between the observed and fitted spectrum long-ward of 3647\AA\/,
but that a substantial gap nevertheless remains. In Section 4 we
introduce additional considerations that may help to fill this
emission ``gap'', evident when comparing model predictions with
observations.

Finally, we note that the finite external electric fields will also
break degeneracies in the bound-bound transitions, introduce
additional local line-broadening, and affect the emission line
transfer (the Stark effect\footnote{At BLR gas densities, Stark
broadening (e.g., Balcon et~al. 2007) is expected to be relatively
weak ($\Delta \lambda_{\rm Stark}$ (\AA\/) $= 2\times 10^{-10}
(n_{e})^{2/3}$, where $n_{e}$ is the electron number density in units
of cm$^{-3}$), amounting to $<$ 0.02\AA\ for
$n_{e}=10^{12}$~cm$^{-3}$.}; e.g., Wiese et al.\ 1971). While we do
not explore these effects here, we note that such effects may
contribute to the smoothing of the broad emission-line profiles, since
most velocity bins will have contributions from gas spanning a wide
range in density, and the wings in particular will likely be dominated
by emission from the highest density gases in the BLR (see the
discussion of the conundrum of smooth broad emission-line profiles in,
e.g., Arav et~al.\ 1998).  However, the presence of microturbulent
velocities within the BLR gas, and described in \S4.2, are likely far
more effective at smoothing the emission-line profile.

%/rfs/XROA/mg159/PRESS/OTHO_AUG17/MAGDZIARZ/SOLAR/NH23/raw/plot_h_he
\begin{figure}
    \centering
    \includegraphics[width=\columnwidth]{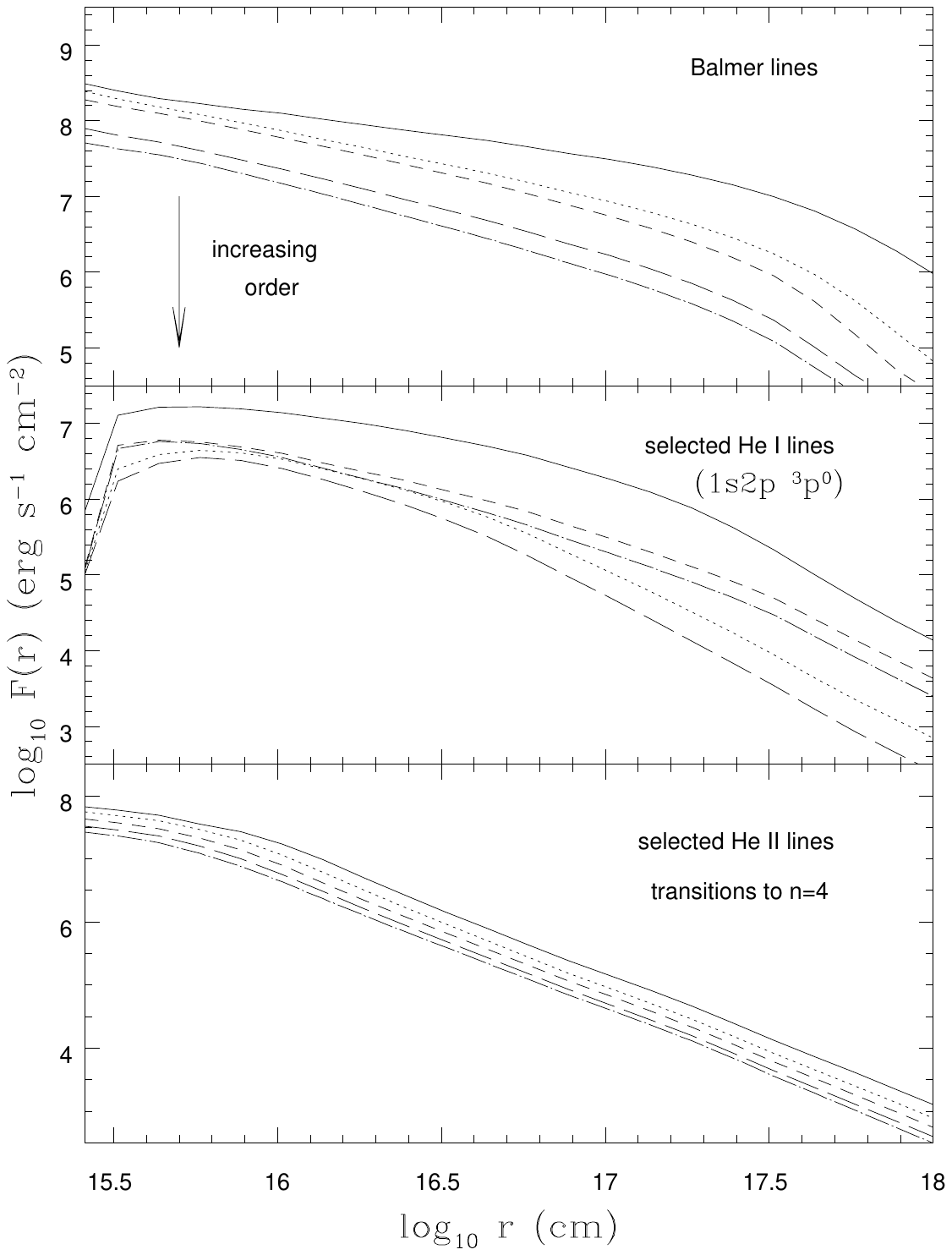}
    \caption{Radial surface emissivity distributions $F(r)$, for
      selected Hydrogen (H$\alpha$ -- H$\epsilon$ upper panel),
      He~{\sc i} (triplet lines 5875.6 (solid), 4713.0 (dotted),
      4471.4 (small dashed), 4120.8 (long dashed), 4026.2 (dot dashed)
      -- middle panel), and He~{\sc ii} (6559.9, 5411.3, 4859.1,
      4541.4, 4338.5 -- lower panel) emission lines. Note that the
      Balmer (H$\alpha$--H$\epsilon$) lines show a decrease in
      strength, and steepening gradient for higher order
      transitions. This equates to smaller mean formation radius for
      the higher order lines. While the He~{\sc ii} lines also
      decrease in strength for higher order transitions, the gradient
      remains unchanged ($F(r) \propto r^{\beta}$), with $\rm{\beta}
      \approx -2$ for all but the innermost radii. The mean formation
      radius for the He~{\sc ii} lines is therefore broadly similar.}
    \label{plot_h_he}
\end{figure}

\subsection{A previous BLR photoionisation model for NGC~5548}

The first DC template was constructed in 2001 for the nearby Seyfert
1.5 galaxy NGC~5548 (KG2001). That model and more recent versions
(e.g., Korista and Goad 2019), are based on LOC integrations (Baldwin
et~al. 1995) over photoionisation model grids calculated using the
photoionisation code {\sc cloudy} version c17.02 (Ferland et
al.\ 2017).

Using an ionizing spectral energy distribution (SED) appropriate for
NGC~5548 (Magdziarz et al. 1998), emergent emission line fluxes are
computed for line emitting entities (here, referred to as clouds),
spanning a range in gas hydrogen density $7 \leq \log \rm{n_H}$
(cm$^{-3}$)~$\leq 14$, and incident hydrogen ionizing photon fluxes
$17 \leq \log \Phi_{\rm H}$~(photon~s$^{-1}$~cm$^{-2}$)$~\leq 24$, for
gas with a fixed total hydrogen column density $\log
\rm{N_H}=23$~cm$^{-2}$.  In KG2019, we also explored models with total
hydrogen column densities spanning $22 \leq \log \rm{N_H}(cm^{-2})
\leq 24$.  Emergent line intensities are represented graphically on
the flux--density ($\log \Phi_{\rm H}$--$\log \rm{n_H}$) plane as
logarithmic contours in equivalent width (EW) referenced to the
incident continuum flux at $\lambda$1215\AA. These, at a glance
indicate the physical conditions for which the production of a
particular line is maximally efficient.  Next we sum over the emergent
emission line intensities using a power-law weighting in gas hydrogen
density with an index $-1$ (equivalent to equal weighting per unit
decade) and consistent with the distribution in gas densities
predicted in magneto-hydrodynamic (MHD) simulations of BLR cloud
stability (Krause et~al. 2012) to produce radial surface emissivity
distributions for lines of interest. We additionally apply cut-offs in
ionisation parameter $\rm{U_H \equiv \Phi_H/n_Hc}$, such that the
product $\rm{U_H c}$ lies in the range $6 \leq \log \rm{U_H c}~{\rm
  (}{\rm cm}~s^{-1}\rm {)} \leq 11.25$, in order to exclude those
regions lying in the bottom right (very low ionisation gas) and upper
left (very high ionisation gas) of the $\log \Phi_{\rm H}$--$\log
\rm{n_H}$ plane (see e.g., Korista and Goad 2004, their
Figure~1). Example radial surface emissivity distributions $F(r)$, for
selected hydrogen and helium lines of interest are illustrated in
Figure~\ref{plot_h_he}. Over much of the BLR the radial surface
emissivity distributions $F(r)$ for the simplest lines (e.g., He~{\sc
  ii}) may be approximated by a simple powerlaw in radius ($F(r)
\propto r^{\beta}$), with $\beta \approx -2$, breaking to a flatter
slope only at small BLR radii. These lines thus have similar mean
formation radii and similar (approximately 1:1) local response
amplitudes. More complex lines are better fit with doubly broken power
laws, with flatter slopes at small BLR radii and steeper slopes at
larger radii. Also clear is a general increasing trend toward steeper
slopes for the higher order lines (e.g., Figure~\ref{plot_h_he},
panels 1 and 2), with obvious implications for their mean formation
radius (decreasing) and response amplitude (increasing).

Finally, we integrate over the radial surface emissivity distributions
from an inner radius $\rm{R_{in}}=1$~lt-day to outer radius
$\rm{R_{out}}=140$ lt-days\footnote{Note that for some slices in
hydrogen gas density $n_{\rm H}$, the surface emissivity will be zero
for a broad range in radii (see e.g., Figure 2).} using a differential
radial covering fraction dependence that is a power-law in radius
$dC(r)\propto r^{\gamma} dr$.  This outer boundary corresponds to an
ionizing photon flux $\log~\Phi_{\rm H} = 17.9$ in the average
luminosity state of NGC~5548, at which many types of grains are
sublimating (e.g., Mor and Netzer 2012). A power-law index of $-1.2$
and covering fraction of $\sim 50$\% was found to give a reasonable
match to the observed emission-line strengths {\it and\/} their
variability behaviour (response amplitude and timescale) for the
strongest UV and optical emission lines in this source (Korista and
Goad 2000, hereafter KG2000).

Having found model parameters that satisfy the constraints imposed by
the emission-line strengths and their variability behaviour, we then
compute from the same model, the diffuse continuum contribution over a
broad range in wavelengths spanning the far ultra-violet ($\sim$
200~\AA) to infra-red ($\sim$ 4~microns). As with the emission-lines,
we are able to determine the total DC contribution at each wavelength,
and its radial dependence, including mean formation radius
(emissivity-weighted and responsivity-weighted -- see also Lawther
et~al. 2018).

The aim of this work is to provide a more physically motivated model
of the DC contribution. As a point of reference we refer the reader to
KG2019, their Figure~3 (upper panel), and shown here as the solid
black line in Figure~\ref{loc_comp}, representing our most recent
model diffuse continuum spectral energy distribution as determined for
the nearby Seyfert~1.5 galaxy NGC~5548. Three prominent features stand
out: a substantial neutral hydrogen (``Rayleigh'') scattering feature
centred near $\sim$1216~\AA\/ (Korista and Ferland 1998), and
significant Balmer and Paschen continuum emission which rise in
strength toward their respective jumps at the series limits for atomic
hydrogen. In KG2019, these jumps are located at their vacuum
wavelengths $\lambda$3647.0\AA\, and $\lambda$8205.8\AA\/,
respectively.  The overall shape of the diffuse continuum from the BLR
is blue, with a characteristic logarithmic slope of $\approx -1$.

\subsection{DC F(r) distributions}
\begin{figure}
    \centering
    \includegraphics[width=\columnwidth]{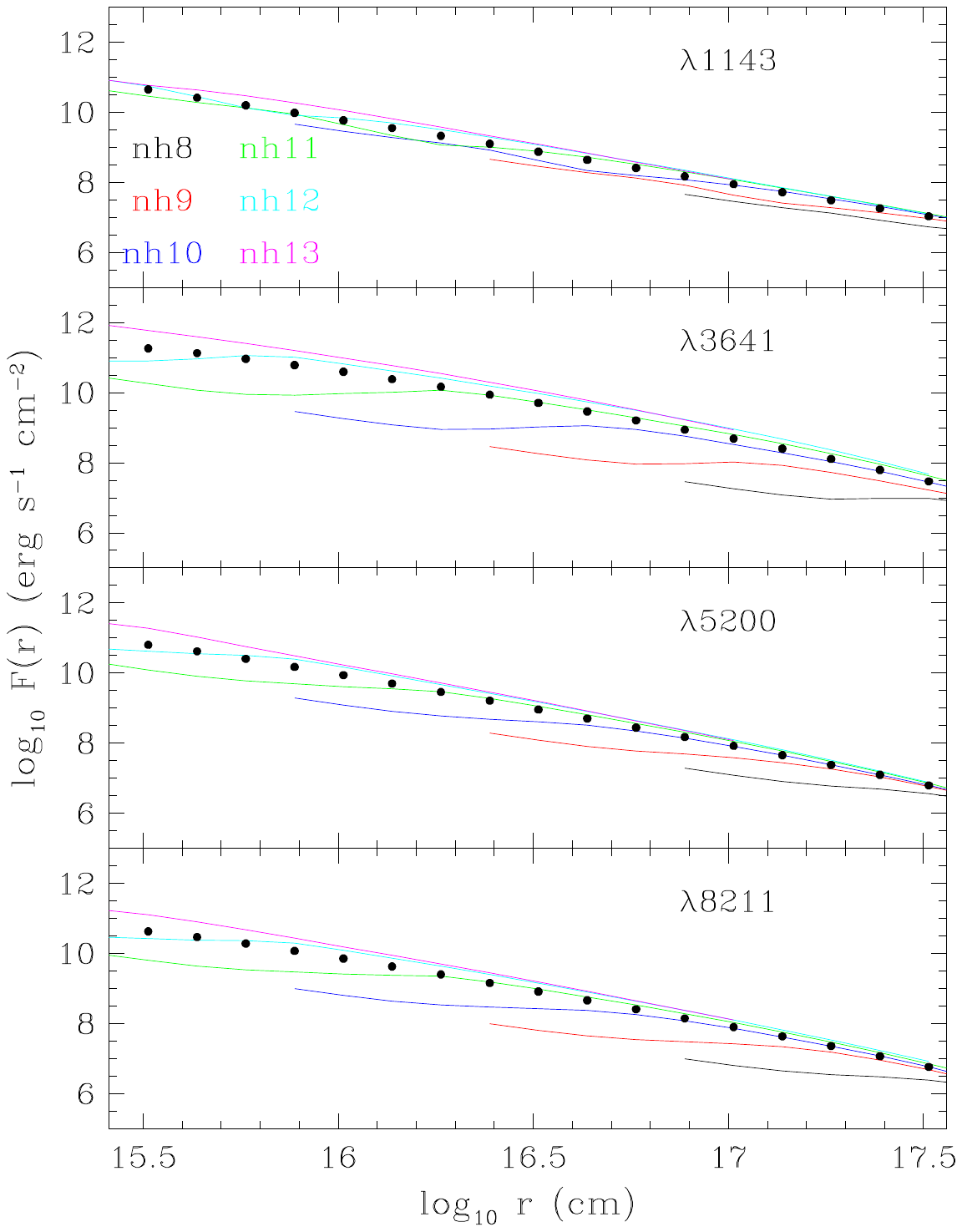}
    \caption{Radial surface emissivity distributions $F(r)$ as a
      function of radial distance $r$ for 4 representative diffuse
      continuum bands. Colours represent constant density slices ($8
      \leq \log n_{\rm H} \rm{(cm}^{-3}\rm{)} \leq 13$) through the
      flux-density ($\log \Phi_{\rm H}$--$\log n_{\rm H}$) plane. For
      comparison, filled circles indicate the radial surface
      emissivity distributions for the same lines for a standard LOC
      integration, with a gas density distribution that follows a
      power-law in densities with index $-1$. Importantly, note that
      not all gas densities contribute to the emergent flux at each
      radius.  }
    \label{dc_fr}
\end{figure}

In Figure~\ref{dc_fr} we compare the radial surface emissivity
distributions, $F(r)$ for representative DC bands for constant density
slices through the flux--density ($\log \Phi_{\rm H}$--$\log n_{\rm
  H}$) plane spanning $7 \leq \log n_{\rm H} ({\rm cm}^{-3})\leq 14$
(coloured lines), together with the radial surface emissivity
distributions for the same 4 continuum bands, for a representative LOC
integration (dotted curves), assuming a gas density distribution that
follows a power-law in densities, with power-law index $-$1. Note that
gas at a given density does not occupy the full range in radii
presumed to span the BLR (Figure~\ref{dc_fr}). This is a consequence
of a combination of local gas physics and our chosen cut-off in
ionization parameter $U_{\rm H}$(max).  A higher cut-off value allows
for the presence of lower-density gas at smaller BLR radii and acts to
increase the surface emissivity in this region.

%/rfs/XROA/mg159/PRESS/OTHO_AUG17/MAGDZIARZ/HALFSOLAR/NH23/raw/plot_dc_nh

Figure~\ref{dc_slices} is a graphical representation of the effect
that the presence of finite gas densities has on the location of the
Balmer and Paschen jumps for constant density slices within the model
of KG2000. In the top panel we show the diffuse continuum spectra
arising from each gas density ($\rm{n_{H}}$) slice within the BLR
radius bounds within the grid.  In blue we show on the same scale the
luminosity of the incident continuum, while in red we show a scaled
version of the incident continuum to guide the eye. For each slice in
hydrogen gas density, the location of the jumps have been shifted to
the wavelength appropriate for that density, and renormalised so that
the total energy in the diffuse continuum is conserved.  For low gas
densities the intensity of the diffuse continuum is weak compared to
that of the incident continuum, surpassing the intensity of the
incident continuum in the vicinity of the jumps for gas densities in
excess of $10^{11}$~cm$^{-3}$. In general, the amplitude of the jump
and overall energy in the spectrum increase with increasing gas
density, though the increase in the overall energy is modest for gas
densities above $\approx10^{12}$~cm$^{-3}$.  Given our fixed outer BLR
boundary, the lower gas density slices contain greater fractions of
(constant density) clouds that are more highly ionized and at higher
electron temperatures, thus containing weaker thermal free-bound
continua and appear more similar to weak electron scattering copies of
the incident continuum. Gas of greater densities at a fixed ionization
parameter is also a greater source of free-bound and free-free
continuum emission. Both of these effects explain the general
behaviour of the diffuse continuum spectra in the top panel.

In the lower panels of Figure~\ref{dc_slices} we show close up views
of the diffuse continuum in the vicinity of the Balmer and Paschen
jumps to emphasise the gas density-dependent shifts in their
positions. Note that the wavelength scale spanning the Paschen jump
region is a factor $\approx5.1$ larger than that shown for the Balmer
jump due to the similar factor larger density-dependent shift in the
Paschen jump compared to that in the Balmer jump.

\subsection{LOC integrations}
%/rfs/XROA/mg159/PRESS/CLOUDY/c17.02/SERIAL/DC_DENSITY/plot_dc_grid_aug
% updated version august 2022 to reflect new location of edges
%\onecolumn
\begin{figure}
    \centering
\includegraphics[width=\columnwidth]{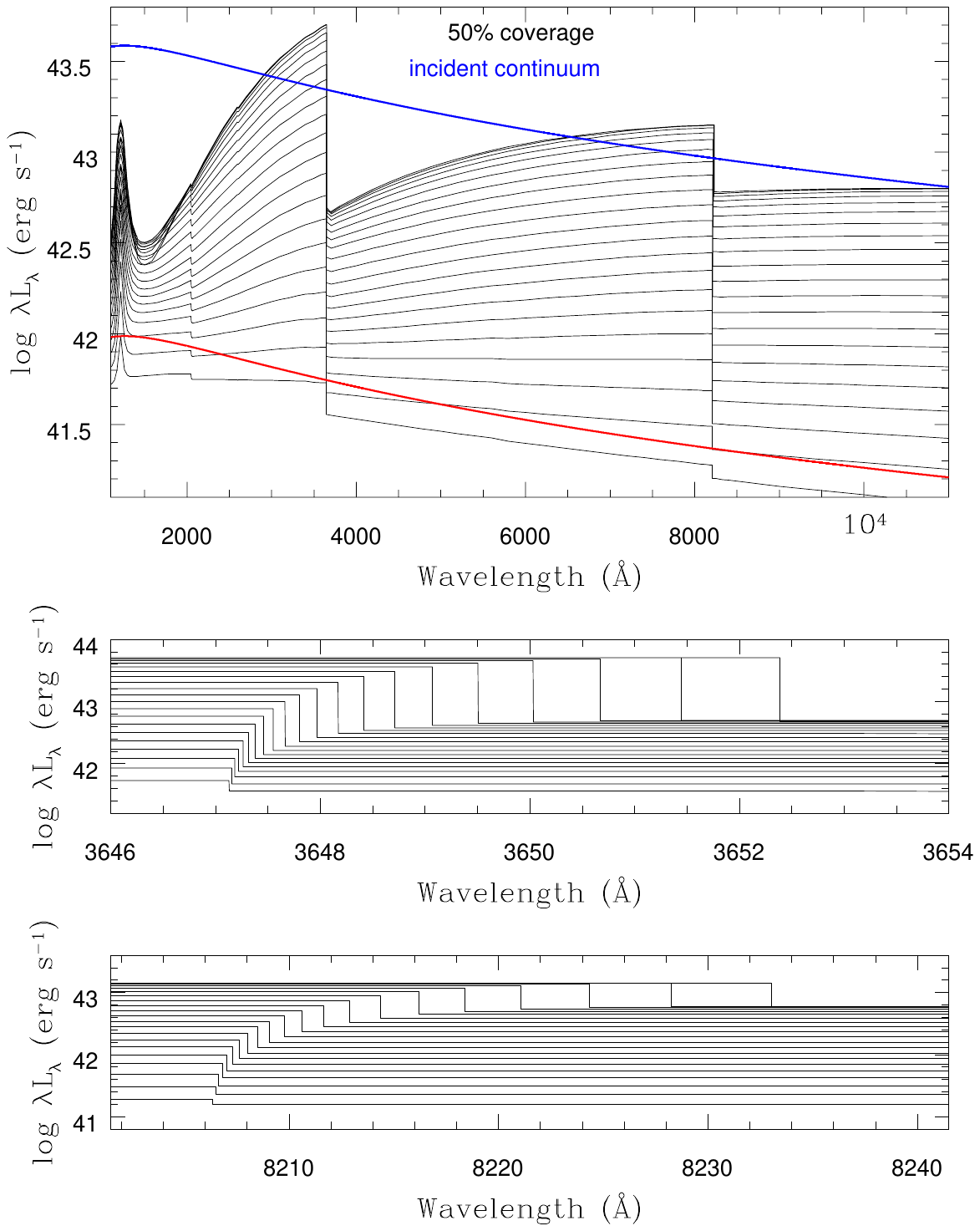}
    \caption{An illustration of the effect of finite gas densities on
      the location of the Balmer and Paschen jumps. Upper panel --
      Spectra of the broad line region wavelength dependent diffuse
      continuum emission for densities in the range $8 \leq \log
      n_{\rm H}$(cm$^{-3}$) $ \leq 13$ in 0.25 dex intervals, and
      assuming a gas covering fraction of 50\%. Each spectrum is
      determined from a vertical slice through the flux density
      ($\Phi_{\rm H}$--${\rm n_H}$) plane (i.e. at constant
      density). The jumps have been shifted to the appropriate
      ionization edge for that gas density (\S2.2). In general, both
      the emitted flux and the location of the jump increase with
      increasing gas density, though the flux increase for hydrogen
      gas densities greater than $\log n_{\rm H}$(cm$^{-3}$) = 12 is
      modest. In blue we show the incident continuum, and in red a
      scaled version of the incident continuum to guide the eye. At
      low gas densities the DC is weak, while at gas densities
      exceeding $\log n_{\rm H}$(cm$^{-3}$) = 11, the DC component is
      comparable to or larger than the incident continuum. Panels 2
      and 3 show enlarged views of the Balmer and Paschen jump
      regions. The horizontal scale for panel 3 spans a factor 5
      larger range in wavelength than for panel 2.}
    \label{dc_slices}
\end{figure}

In the LOC picture, the shape and intensity of the diffuse continuum
is determined by summing over the emergent surface emissivities from a
cloud distribution spanning a broad range in hydrogen gas density at
each radius (\S1.2). Since the output from photoionisation modelling
codes such as {\sc cloudy} do not account for the presence of finite
gas densities, we here mimic this effect, by summing over the constant
density-slices shown in Figure~\ref{dc_slices}, using an appropriate
density-weighting scheme {\em after first modifying the edge for that
  finite density, and assuming that the total flux is conserved\/}.
Here we adopt as a weighting scheme, a simple average of the
wavelength-dependent flux contributions from each of the individual
slices in hydrogen gas density.

%/rfs/XROA/mg159/PRESS/CLOUDY/c17.02/SERIAL/DC_DENSITY/plot_dens.pdf
\begin{figure}
   \centering
\includegraphics[width=\columnwidth]{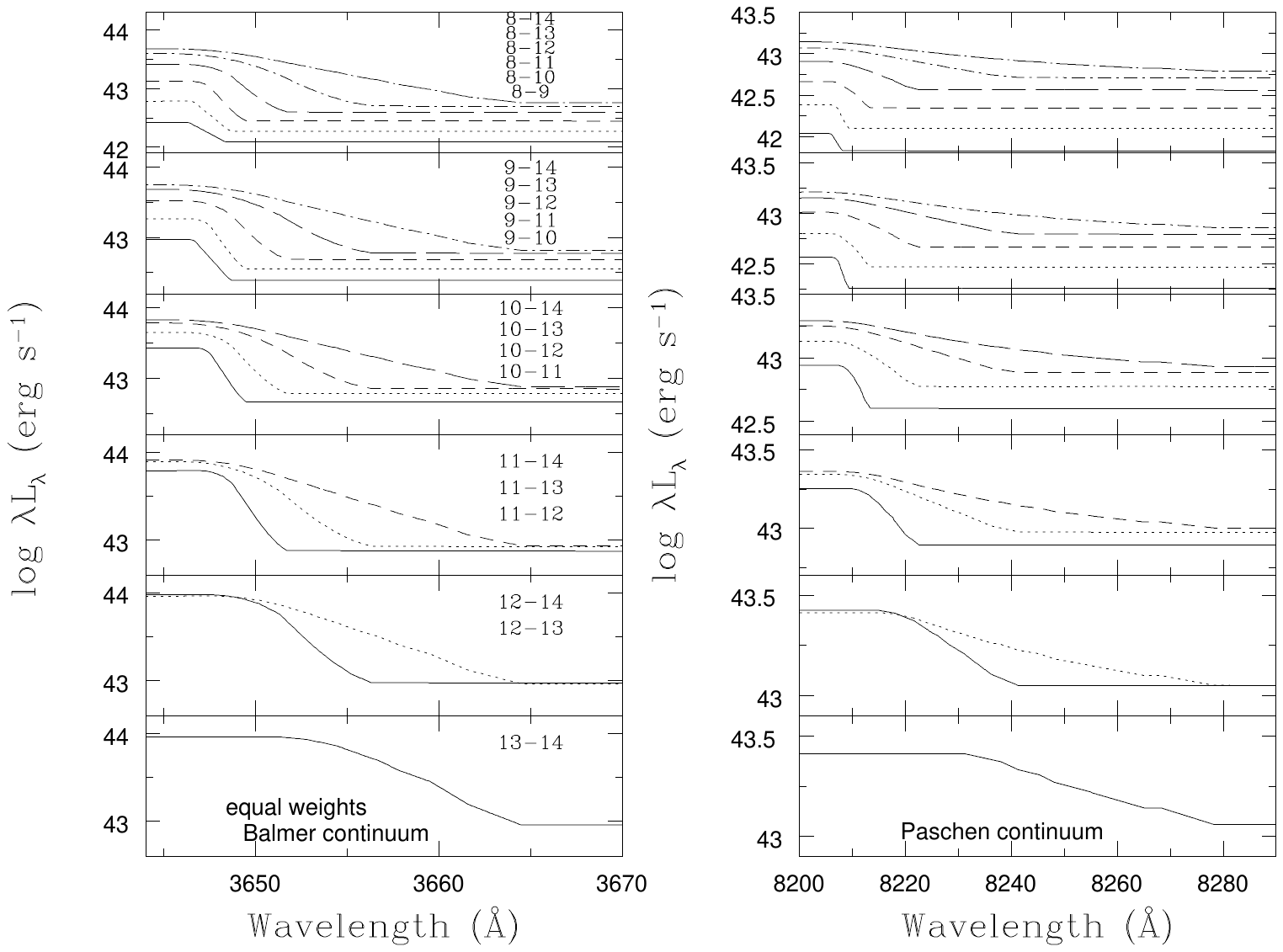}
%    \includegraphics[height=600pt]{Figures/plot_dens.pdf}
% original files are in balmer_sums.stk
    \caption{Model diffuse continuum spectral templates determined
      from summing over gas densities in the range $8 \leq \log n_{\rm
        H}$ (cm$^{-3}$) $\leq 14$. Each template comprises a simple
      average of the flux contributions from each slice in
      density). All assume 100\% coverage. Note the significant shift
      in jump location in the presence of high gas densities, and the
      smooth decline to longer wavelengths when a mixture of gas
      densities are available.  We include density contributions from
      gas with $\log n_{\rm H}$(cm$^{-3}$) = 14 to illustrate that
      even at high gas densities, the Balmer jump lies well shortward
      of 3700\AA\,.  }
    \label{const_dens}
\end{figure}

\subsubsection{A constant weighting in density scheme}
In Figure~\ref{const_dens} we show the shape of the DC only, in the
region spanning the Balmer jump (left) and Paschen jump (right), and
in the absence of any velocity broadening, for a range of summations
over the gas density-distribution, {\em all of which assume an equal
  weighting for each slice in density}. Note that the location of the
jump is governed by (i) our particular choice of weighting scheme, and
(ii) the minimum density contributing to the summation. Importantly,
because of its stronger dependence on jump location with density, the
intensity immediately longward of the Paschen jump declines more
slowly than that of the Balmer jump. In this scheme, in order for the
jump location to be significantly displaced from its vacuum
wavelength, the minimum gas density contribution must be high (e.g.,
compare Figure~\ref{const_dens}, panels 1 and 6). Conversely, if a
broad range in gas densities are available within the BLR, the DC
intensity longward of the jump declines smoothly toward longer
wavelengths and has an extended tail. Thus (in the absence of a
significant contribution from higher order Balmer and Paschen lines
longward of their respective jumps) the shape of the DC contribution
in the vicinity of the jump is a diagnostic of the range in gas
densities present within the BLR.
%/rfs/XROA/mg159/PRESS/CLOUDY/c17.02/SERIAL/NH08/plot_L_R.pdf
% plot_L_R_log - corrected Hgamma 4340
\begin{figure}
    \centering
    \includegraphics[width=\columnwidth]{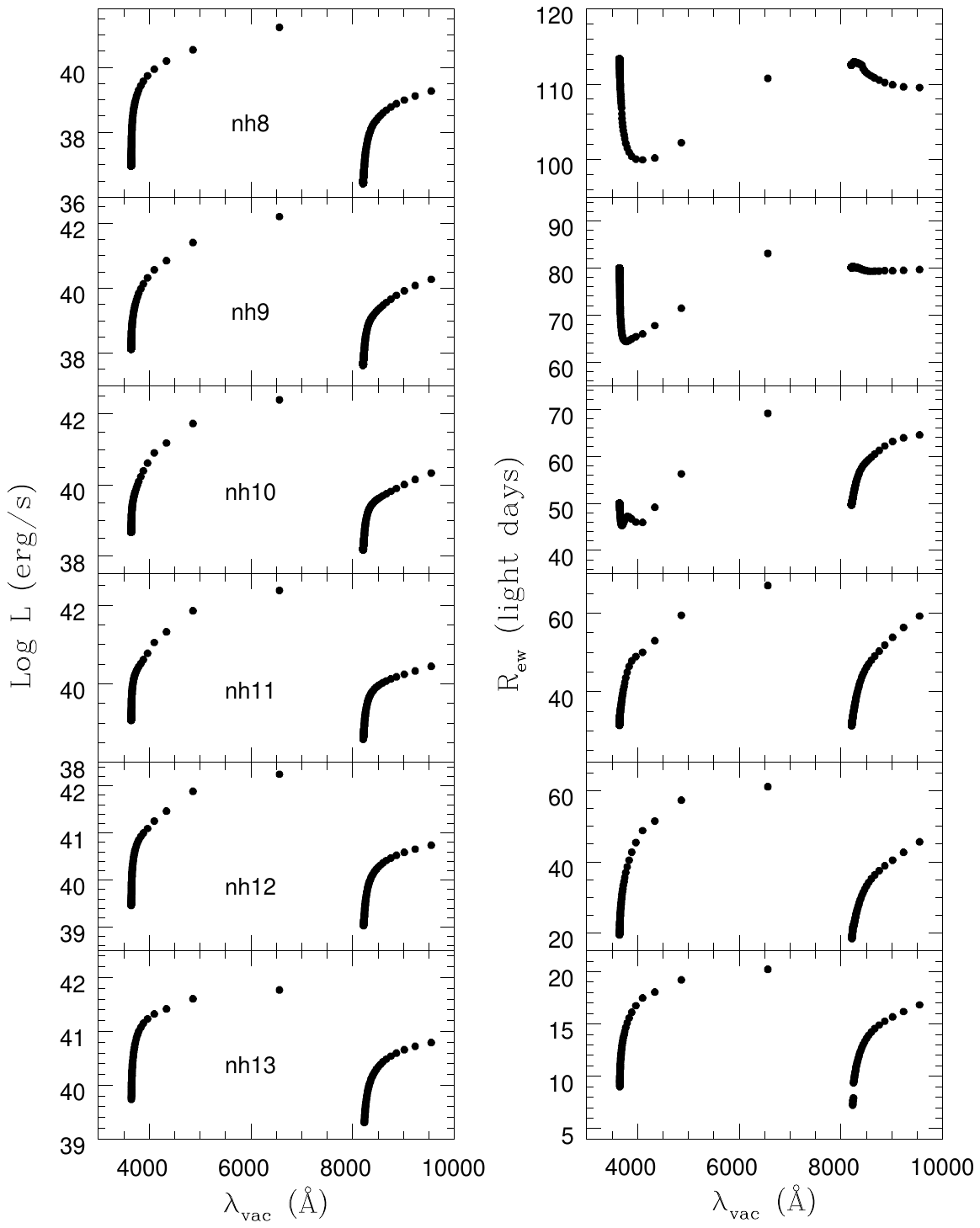}
    \caption{Left - Predicted emission-line luminosities for the
      Balmer and Paschen series as determined for constant density
      slices through the flux-density plane. From top to bottom -
      $\log~ n_{\rm H}({\rm cm}^{-3})=8,9,10,11,12,13$. Right -
      corresponding emissivity-weighted radii
      (light-days). Emission-line luminosities and emissivity-weighted
      radii for the higher order lines ($n_{\rm upper} >37$ -- Balmer,
      \, $n_{\rm upper} > 48$ -- Paschen) for densities in the range
      $8 < \log n_{\rm H}<13$, are here determined by
      extrapolation. Line luminosities decrease with increasing upper
      level. Emissivity-weighted radii also tend to decrease with
      increasing upper level, except at the lowest gas densities.  }
    \label{bal_pal_con}
\end{figure}

\subsection{Jump modification by higher-order emission lines}

Longward of their respective ionization edges, the Balmer and Paschen
continua may be significantly modified by the presence of blended
higher order Balmer and Paschen emission lines, as well as a smaller
(though still significant) contribution from higher order He~{\sc i}
and He~{\sc ii} lines. In order to estimate the contribution of these
lines to the DC spectrum in the vicinity of the jump we here adopt the
following scheme. First we estimate the relative strengths and mean
formation radius (emissivity-weighted radius) of individual emission
lines from our photoionisation model calculations. Next we perform a
spectral decomposition (here using multiple Gaussians) of one of the
strongest observed optical emission-lines (e.g., H$\beta$), to isolate
the broad emission line and estimate its intensity and width. Finally,
using this reference line as a template, we apply it to each of the
other lines in turn, scaling in width according to their mean
formation radius and under the assumption that the gas is virialised,
and scaled in intensity so that the relative strengths of the lines as
determined from photoionisation calculations are conserved.

Note, we have previously shown that in an LOC integration the higher
order Balmer lines form at smaller BLR radii (Korista and Goad 2004,
hereafter KG04). Thus, our expectation is that the higher order lines
in the Balmer and Paschen series will be broader than the reference
line. KG04 also showed that in the higher order lines the wings become
stronger relative to the core emission, broadening the base of the
lines. We do not account for this additional complication in this
work.

\begin{figure}
    \centering
    \includegraphics[width=0.75\columnwidth,angle=270]{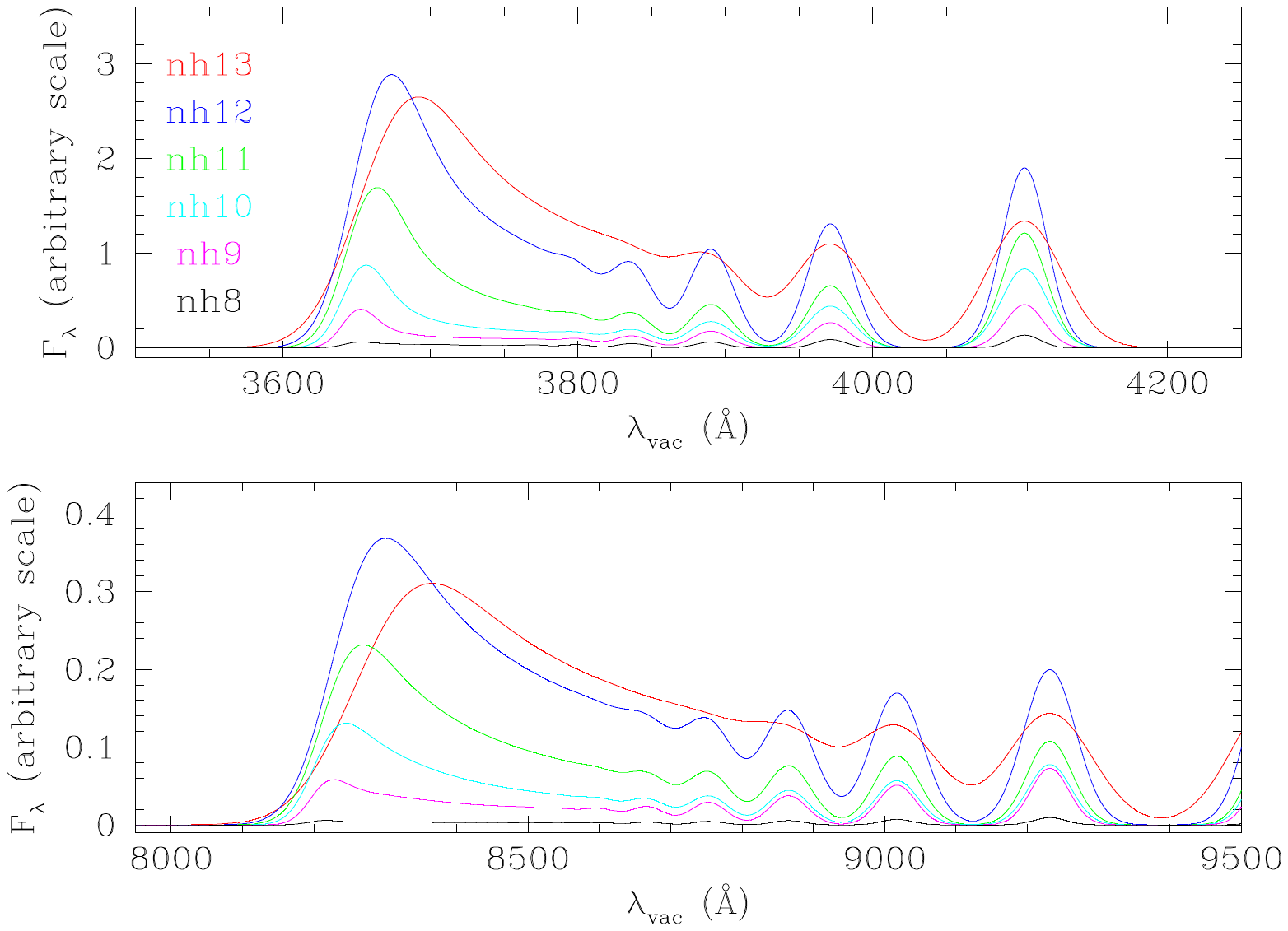}
    \caption{The effect of finite gas densities on the number of upper
      levels available for the Balmer series (upper panel) and Paschen
      series (lower panel). As the density increases, the number of
      upper levels available diminishes, the intensity of the pile-up
      relative to lower order lines decreases and the series limit
      shifts to longer wavelengths. Here line intensities are scaled
      relative to the predicted line intensity of the H$\beta$ line
      for gas densities of $\log n_{\rm H} ({\rm cm}^{-3})=13$ ($\log
      L=41.605$ erg~s$^{-1}$). Relative line widths are determined
      assuming that the gas is virialised and orbits a central black
      hole of mass $\log_{10} M_{\rm BH} = 7.692$M$_{\odot}$, as
      appropriate for NGC~5548 (e.g., Bentz \& Katz 2015). We assume
      the gas occupies an approximately flattened geometry with an
      inclination of $\approx$35 degrees.  This corresponds to a
      velocity width of 5760~km~s$^{-1}$ (FWHM) at a radial distance
      of 10 light-days from the central supermassive black hole.  }
        \label{plot_dens}
\end{figure}

\subsubsection{Higher-order emission-line strengths}

Estimates of the emission-line contributions of the lower order
Balmer, Paschen, and Helium lines (He~{\sc i} and He~{\sc ii}) in the
vicinity of the Balmer and Paschen jumps, are determined using
photoionisation calculations.  In Figure~\ref{plot_h_he}, we show
example radial surface emissivity distributions ($F(r)$) for selected
Balmer, He~{\sc i} and He~{\sc ii} lines. As for the DC bands we
integrate the emission line radial surface emissivity distributions
over the range in radii described in \S2.2 using the same power law
radial covering fraction dependence $dC(r)\propto r^{\gamma} dr$, with
$\gamma=-1.2$ to yield emission-line luminosities. Similarly, we
calculate using the same radial surface emissivity distributions, the
emissivity-weighted radius, $R_{\rm ew}$, corresponding to the
centroid of the distribution, representing the typical formation
radius of the line. However, since our {\sc cloudy} photoionisation
models predict Balmer and Paschen series emission line intensities for
bound-bound transitions up to n=48 (comprising 18 \emph{l}-resolved
levels and 30 collapsed levels) the strengths of the {\em higher
  level} transitions must be determined by extrapolation. Given that
the emission line luminosities and formation radii of the lower order
Balmer and Paschen lines vary smoothly as a function of wavelength
(KG04), we here choose to extrapolate to higher orders in $\log
\lambda$ -- $\log L$ space.

Predicted luminosities for low and higher order Balmer and Paschen
series lines are shown in Figure~\ref{bal_pal_con} (left-hand
panel). Also shown are the mean formation radii (the
emissivity-weighted radii) in light-days for each line
(Figure~\ref{bal_pal_con}, right-hand panel). As shown in KG04, the
Balmer and Paschen series lines are predicted to smoothly decrease in
strength from the lower order to higher order lines (see also Wills,
Netzer and Wills (1985), albeit for gas hydrogen densities of
$\sim\/10^{9}$~cm$^{-3}$).  Coupled with an ever decreasing mean
formation radius (for $\log n_{\rm H} > 10$~cm$^{-3}$), we therefore
expect the higher order Balmer lines in the vicinity of the Balmer
jump to be weaker, and substantially ($\approx 50$\%) broader than for
H$\beta$.

While the emission-line behaviour (luminosity and formation radius as
a function of wavelength) confirm our previous findings, integrating
over the constant density slices reveal some notable differences
(relative to an LOC integration) related to where the lines form. This
is particularly evident at low gas densities, as illustrated in
Figure~\ref{bal_pal_con} (right-hand panel), because this gas is
emissive over only a narrow range in incident hydrogen ionizing flux
$\Phi_{\rm H}$ (and correspondingly narrow range in radii). However,
since the contribution to the total light at low gas densities is
small (e.g., compare Figure~\ref{bal_pal_con} left, panels 1 and 6),
if a broad range in gas densities is available within the BLR, the
dominant effect will be a reduction in the mean formation radius
toward higher level transitions.

\subsubsection{A broad emission-line template : the case of NGC~5548}
To illustrate the contribution of the higher order lines to the
emitted spectrum in the vicinity of the Balmer and Paschen jumps we
start with a simple model for NGC~5548. Assuming a central black hole
mass of $\log (M_{\rm BH}/M_{\odot})=7.692$~ (e.g., Bentz \& Katz
2015), and a virialised velocity field, the Keplerian velocity at a
radial distance of 10 light-days is 5034~km~s$^{-1}$.  Assuming a
spherical BLR geometry and a Gaussian line profile, the corresponding
full width at half maximum (FWHM) is 10068~km/s. There have been
numerous campaigns monitoring the variability behaviour of the broad
optical emission-lines in NGC~5548. These provide a representative
delay for broad H$\beta$ of 10 days and a corresponding emission line
profile width of $\approx 5760$ km~s$^{-1}$ (FWHM).  These two
estimates can be reconciled if we instead assume that the BLR occupies
a largely flattened geometry with a line of sight inclination of
$\approx 35$~degrees\footnote{The mean formation radius for H$\beta$
in our model BLR is large ($R_{\rm ew} \approx 60$ lt-days) when
compared to that inferred from the measured emission-line
delays. Consequently, the template emission-line widths are
comparatively narrow and instead correspond to velocities
representative of their mean formation radius
(FWHM(H$\beta$)=37.5\AA\ $\equiv 2300$~km~s$^{-1}$). Thus in order to
match the measured line widths, the template must first be
broadened.}. Having found a representative line width at a radial
distance of 10 light-days, we next model each line assuming a Gaussian
line profile scaling in both intensity and width according to their
integrated luminosity and mean formation radius (luminosity-weighted
radius), assuming that the gas motion is dominated by the central
potential in which it resides (i.e., $v\propto 1/\sqrt r$), and such
that their integrated line luminosities are conserved.

%/rfs/XROA/mg159/PRESS/CLOUDY/c17.02/SERIAL/DC_DENSITY/NH08/plot_diff
\begin{figure}
    \centering   %\includegraphics[angle=270,width=\columnwidth]{Figures/plot_diff_nh12_b.pdf}
\includegraphics[angle=270,width=\columnwidth]{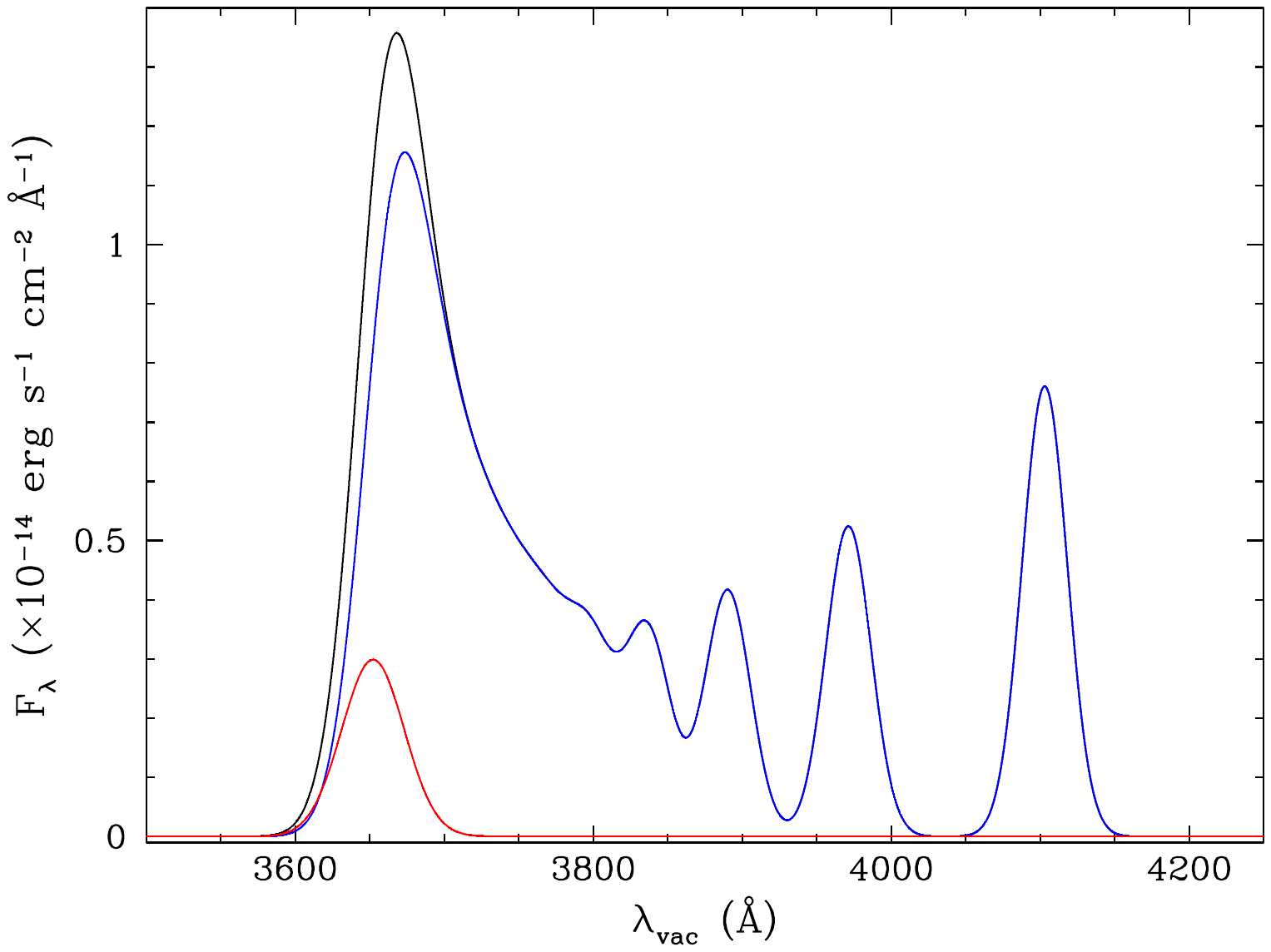}
    \caption{Unweighted (black) and down-weighted (blue) Balmer
      emission line pile-up for a hydrogen gas density of $n_{\rm
        H}=10^{12}$~cm$^{-3}$ (see \S2.2 for details). Shown in red is
      the total light transferred from the higher-n bound-bound
      transitions (black minus blue) to the continuum. This effect is
      greater in denser gas.}
    \label{plot_diff}
\end{figure}

%/rfs/XROA/mg159/PRESS/CLOUDY/c17.02/SERIAL/DC_DENSITY/NH08/plot_redist_5548_ba

\begin{figure}
    \centering
    \includegraphics[width=\columnwidth]{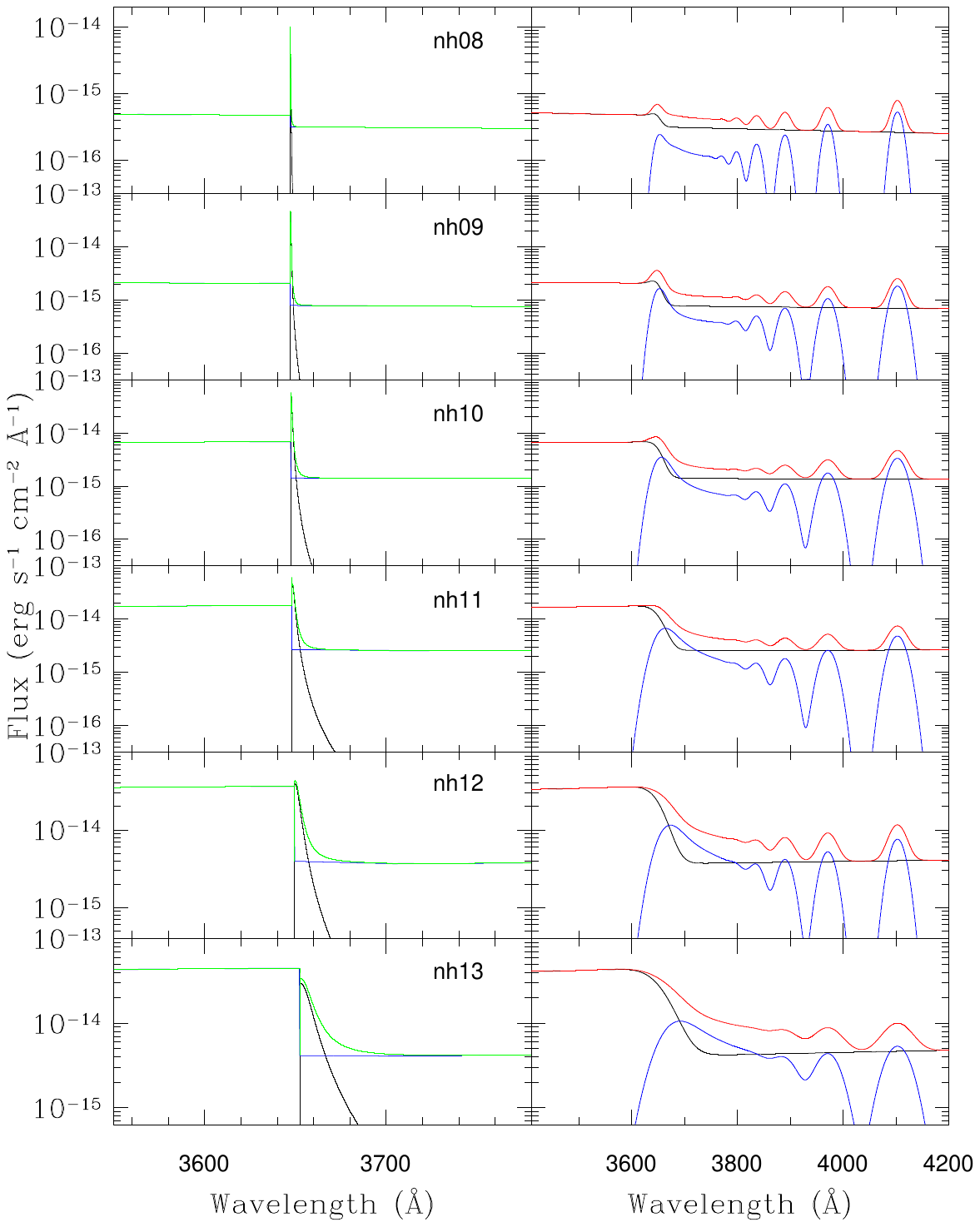}
    \caption{Left -- DC unbroadened (blue), redistributed pileup
      (black), and their sum (green) in the vicinity of the Balmer
      jump. Right -- summed components, Balmer DC$+$redistributed
      emission-line flux, broadened to match the velocity dispersion
      at the mean formation radius of the diffuse continuum short-ward
      of the jump (black), Balmer emission-line spectrum including
      pileup long-ward of the jump (blue), and their sum (red).}
    \label{plot_redist_ba}
\end{figure}

%/rfs/XROA/mg159/PRESS/CLOUDY/c17.02/SERIAL/DC_DENSITY/NH08/plot_redist_5548_pa

\begin{figure}
    \centering
    \includegraphics[width=\columnwidth]{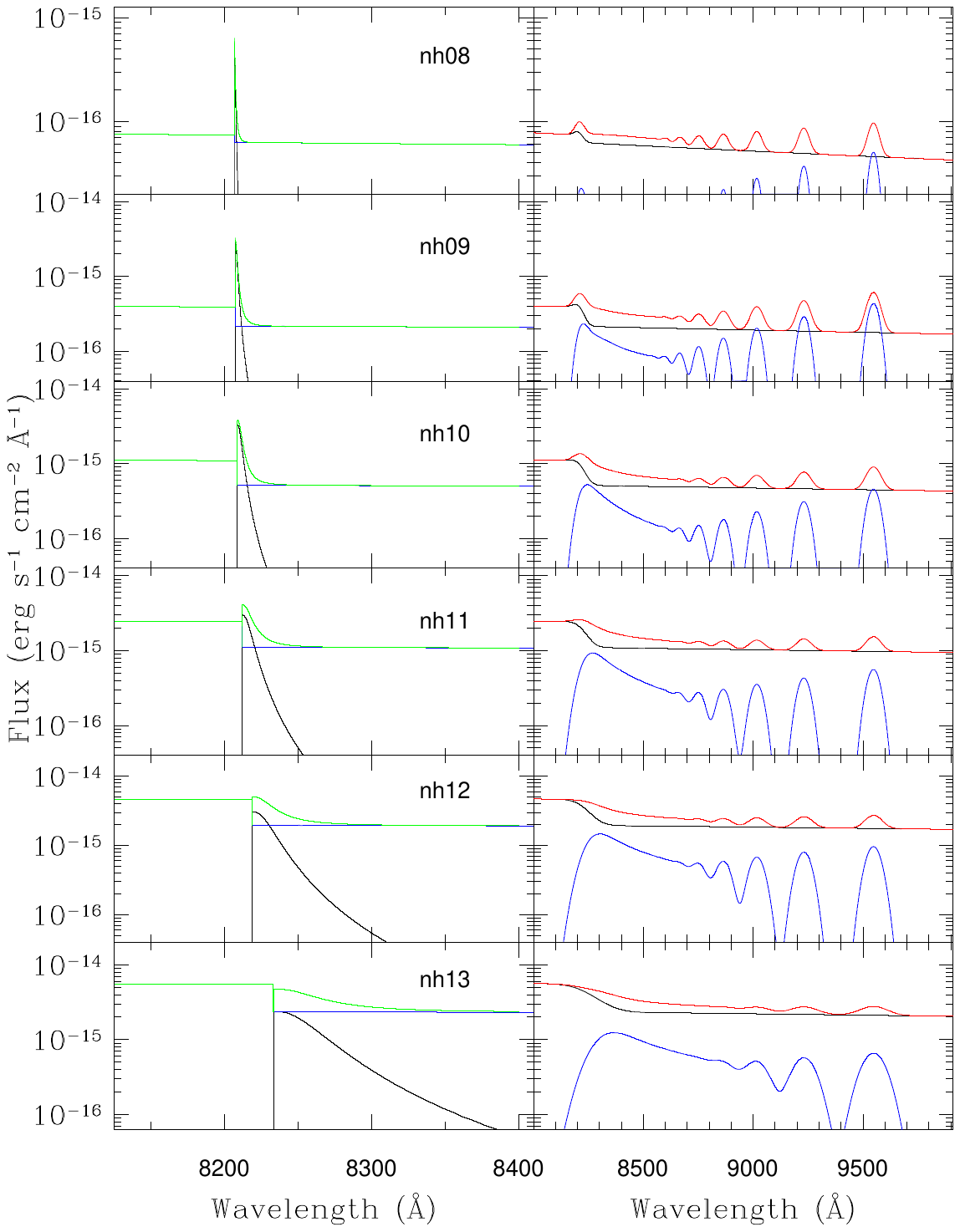}
    \caption{As for Figure~\ref{plot_redist_ba}, for the Paschen jump.}
    \label{plot_redist_pa}
\end{figure}

The resultant emission line spectrum in the vicinity of the Balmer and
Paschen jumps for each constant density slice through the flux density
($\log \Phi_{\rm H}$--$\log n_{\rm H}$) plane are shown in
Figure~\ref{plot_dens}.  As the gas density increases, the number of
available upper levels decreases. Thus the number of higher order
lines present drops steeply with increased density, and the series
limit shifts towards longer wavelengths. Furthermore, at fixed
density, the line intensities generally decrease with increasing $n$
(the number of upper levels available), for gas hydrogen densities
$\log n_{\rm H} \rm{(cm}^{-3}\rm{)}> 9$, and their mean formation
radius also decreases, consequently the higher order lines tend to be
both weaker and broader. Since line luminosity is a strong function of
density, the integrated emission-line spectrum for high gas densities
tends to be stronger, smoother, and shifted towards longer
wavelengths, and with fewer of the higher order lines resolved, e.g.,
compare the emission-line spectrum in Figure~\ref{plot_dens} for $\log
n_{\rm H} \rm{(cm}^{-3}\rm{)}=13$ (red line) and $\log n_{\rm H}
\rm{(cm}^{-3}\rm{)}=9$ (magenta line).

\subsubsection{Down-weighting and redistribution of the emission-line flux}

Thus far our summation of the higher order lines has ignored the
predicted de-population of the upper levels, which acts to lower their
contribution to the line emission, as described in \S2.1. This
emission is not lost, rather it is redistributed as recombination
continuum emission longward of the jump. Since we know the
luminosities of the higher order lines, and the reduction in the
statistical weights of the upper levels, we can reconstruct the
emission-line spectrum in the vicinity of the jump, after first
down-weighting their contributions to the total luminosity. A simple
integration of the difference spectrum, unweighted minus weighted,
then yields the integrated emission-line luminosity lost as a result
of this down-weighting scheme.  This we illustrate in
Figure~\ref{plot_diff} for a hydrogen gas density $n_{\rm
  H}=10^{12}$~cm$^{-3}$. Lines indicate the un-weighted spectrum
(black), the down-weighted spectrum (blue), and their difference
(red). This "missing light" (the integrated light in the difference
spectrum in Figure~\ref{plot_diff}), we redistribute into
recombination continuum emission just longward of the jump, using the
vacuum wavelengths and statistical weights for each line, and scaled
so that the energy redistributed as recombination continuum emission
is conserved. Importantly, we redistribute this light into the
recombination continuum {\it before any broadening\/} of the DC
spectrum is applied.

In Figures~\ref{plot_redist_ba}--\ref{plot_redist_pa} (left panel) we
show for each constant density slice, the DC spectrum (blue), shifted
to place the Balmer and Paschen jumps at their vacuum wavelengths
appropriate for the size of the model atom at that particular choice
in gas hydrogen density (see \S2.1), the redistributed emission-line
flux (black), and their sum (green). Also shown (right panel) is the
result of smoothing the summed components in the vicinity of the
jumps, adopting a Gaussian Kernel, of width (FWHM) appropriate to the
luminosity-weighted radius of the DC just blue-ward of the jump.  We
note that except at the lowest gas densities, as the order increases,
the mean formation radii of the lines converge to the same
luminosity-weighted radius as that found for the DC at wavelengths
just short-ward of the jump.

\begin{figure}
    \centering    
 \includegraphics[width=\columnwidth]{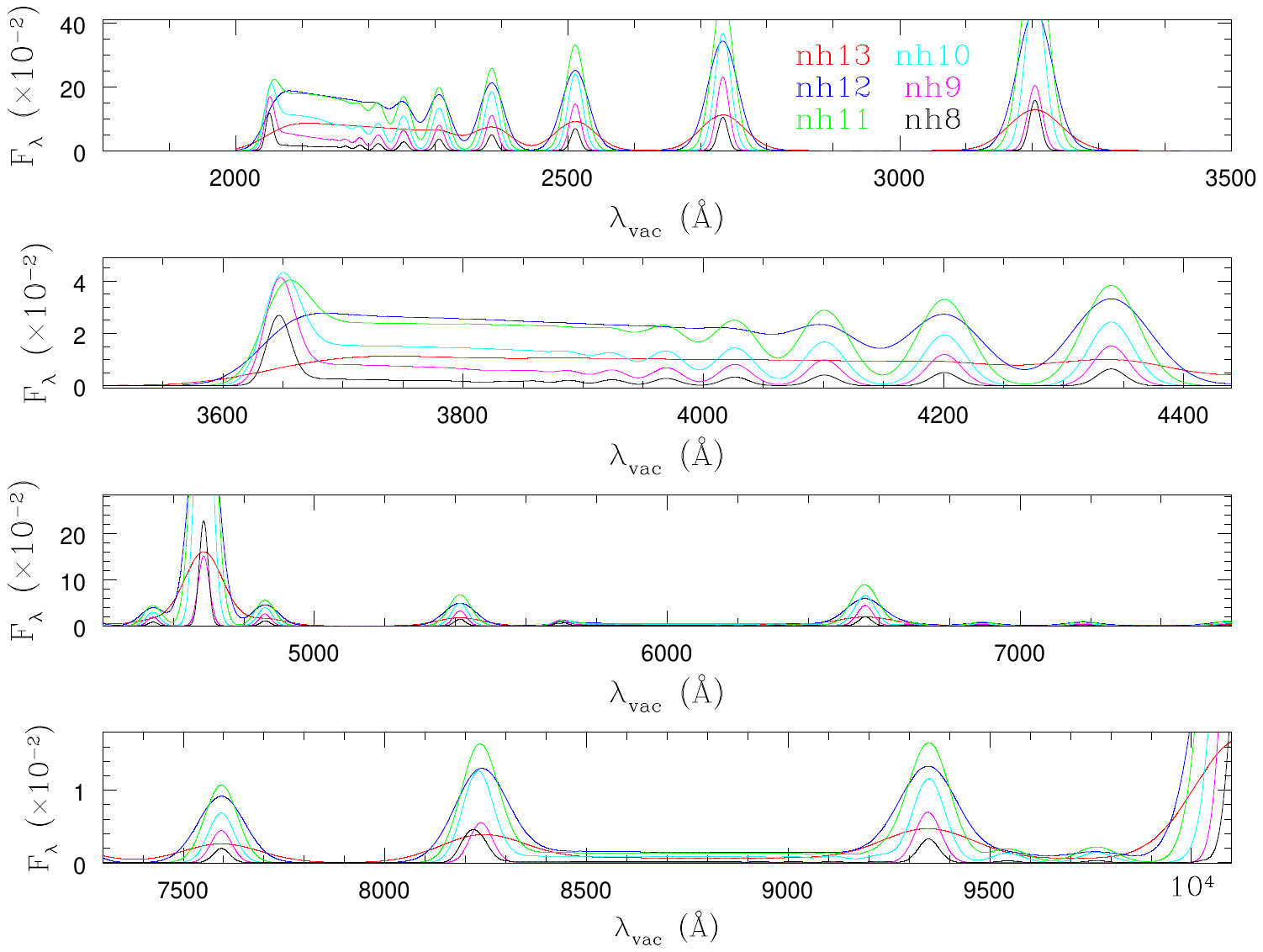}    
 \caption{The effect of finite gas densities on the predicted
   UV--optical He~{\sc ii} emission line spectrum. Emission-line
   fluxes have been placed on the same relative flux scale as the
   Balmer and Paschen lines shown in Figure~\ref{plot_dens}. The
   pileup of lines in the vicinity of their respective jumps for
   series ending n=3--6, have here been down-weighted (see \S2.1 for
   details). }
    \label{plot_dens_jul7_he2}
\end{figure}

\begin{figure}
    \centering   
    \includegraphics[width=\columnwidth]{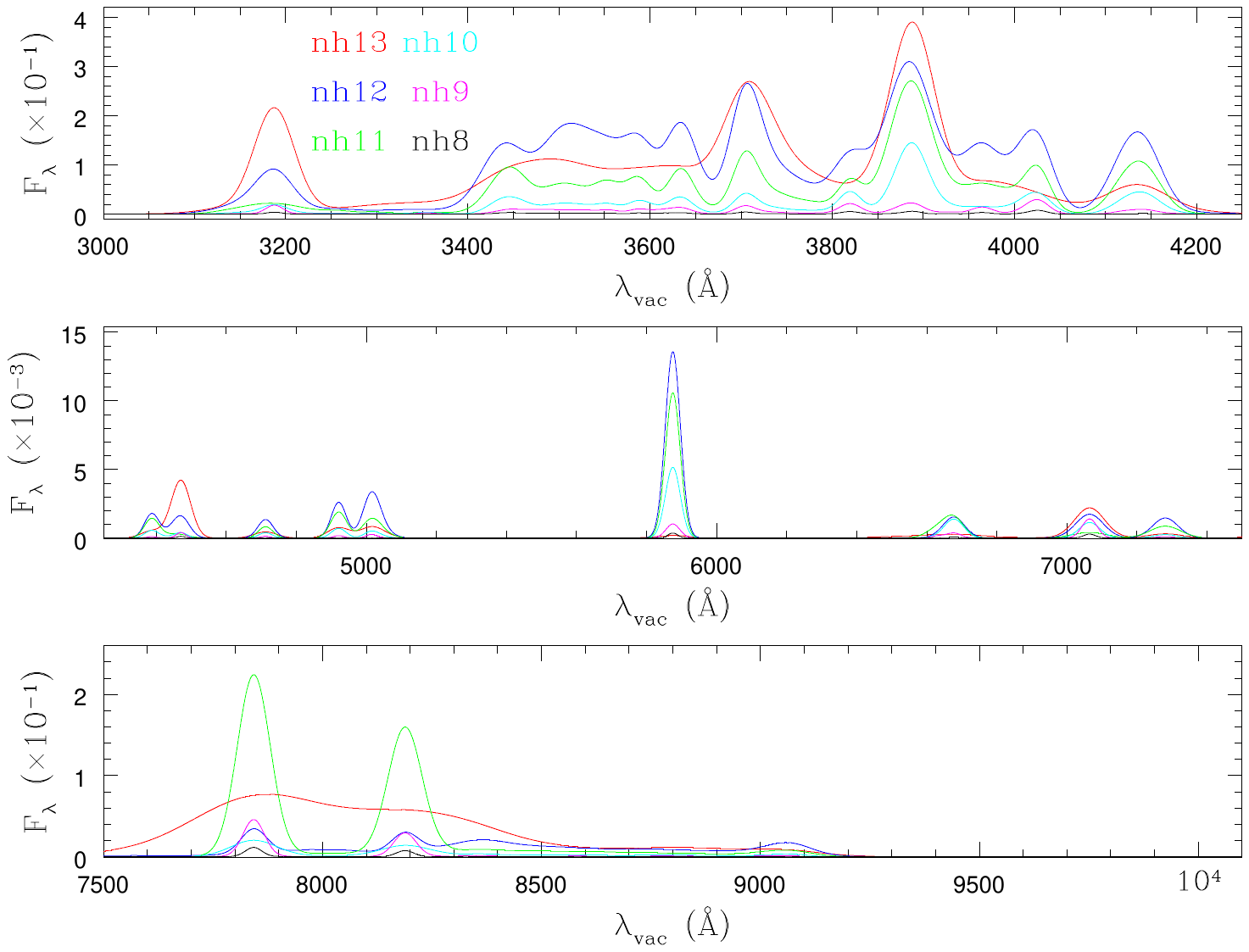}
    \caption{The effect of finite gas densities on the number of upper
      levels available for the UV--optical He~{\sc i} line
      series. Emission-line fluxes are on the same relative scale as
      the Balmer and Paschen lines (cf. Figure~\ref{plot_dens}). NB we
      only include those He~{\sc i} lines that are in the {\sc cloudy}
      output.}
    \label{plot_dens_jul7_he1}
\end{figure}

The final appearance of the jump depends upon jump location and
amplitude, and the amount of velocity broadening\footnote{In the
construction of the emission-line profile and jump location, we do not
include the effects of gravitational redshift and transverse doppler
shift, both of which will act to shift emission arising at small BLR
radii, to longer wavelengths. This effect will be largest for the
higher order lines arising in close proximity to the jump.} (a
consequence of black hole mass, mean formation radius and for a
flattened geometry, viewing angle), which acts to smooth out the jumps
and any additional sharp features resulting from redistributed line
emission. Since the mean formation radius is generally smaller at high
gas densities, the velocity broadening is larger, and the net effect
for this particular velocity field (a consequence of our chosen black
hole mass, geometry, and line of sight inclination), is to make the
jump appear to roll over at shorter wavelengths at higher gas
densities, than for lower gas densities.  However, both the size of
the jump and its span in wavelength is notably larger in the presence
of higher gas densities (Figure~\ref{const_dens}).

The final step in this process is to sum together the broadened
diffuse continuum template and the corresponding emission-line series
at each slice in density. This we illustrate in the right-hand panels
of Figure~\ref{plot_redist_ba}--\ref{plot_redist_pa} (solid red
lines).

\subsection{UV--Optical He~{\sc ii}}

Four major series contribute to the He~{\sc ii} emission-line spectrum
spanning the UV--optical. Transitions down to $n=3$, with a series
limit of ($\lambda_{\rm vac} = 2050.540$\AA), $n=4$ ($\lambda_{\rm
  vac}= 3645.404$\AA), $n=5$ ($\lambda_{\rm vac}=5695.943$\AA), and
$n=6$ ($\lambda_{\rm vac}=8202.158$\AA). The predicted He~{\sc ii}
emission line luminosities and mean formation radii for constant
density slices through the flux density plane are illustrated in
Figure~\ref{he2_form}. As for hydrogen, luminosities for the higher
order lines are determined via extrapolation from the lower order
lines, up to the highest available transition for that particular
density.

As He~{\sc ii} is hydrogenic, we can treat the level populations of
the upper levels in the presence of finite gas densities in a similar
fashion to those of hydrogen. Thus as for hydrogen we expect a pile-up
of lines just long-ward of each of the series limits, but reduced in
strength due to the effect of down-weighting, with the missing flux
redistributed as continuum flux just long-ward of their respective
jumps. In Figure~\ref{plot_dens_jul7_he2} we illustrate the He~{\sc
  ii} spectrum for constant density slices through the flux density
plane.

The He~{\sc ii} lines are substantially broader ($>$ 50\%) than
H$\beta$, owing to their smaller mean formation radius. This is a
consequence of their steeper power-law dependence of their radial
surface emissivity distributions ($F(r)\propto r^{-2}$, e.g.,
Figure~\ref{plot_h_he}, lower panel).

%/rfs/XROA/mg159/PRESS/CLOUDY/c17.02/SERIAL/DC_DENSITY/NH08/plot_dc_comp
\begin{figure}
  \centering
  \includegraphics[width=\columnwidth]{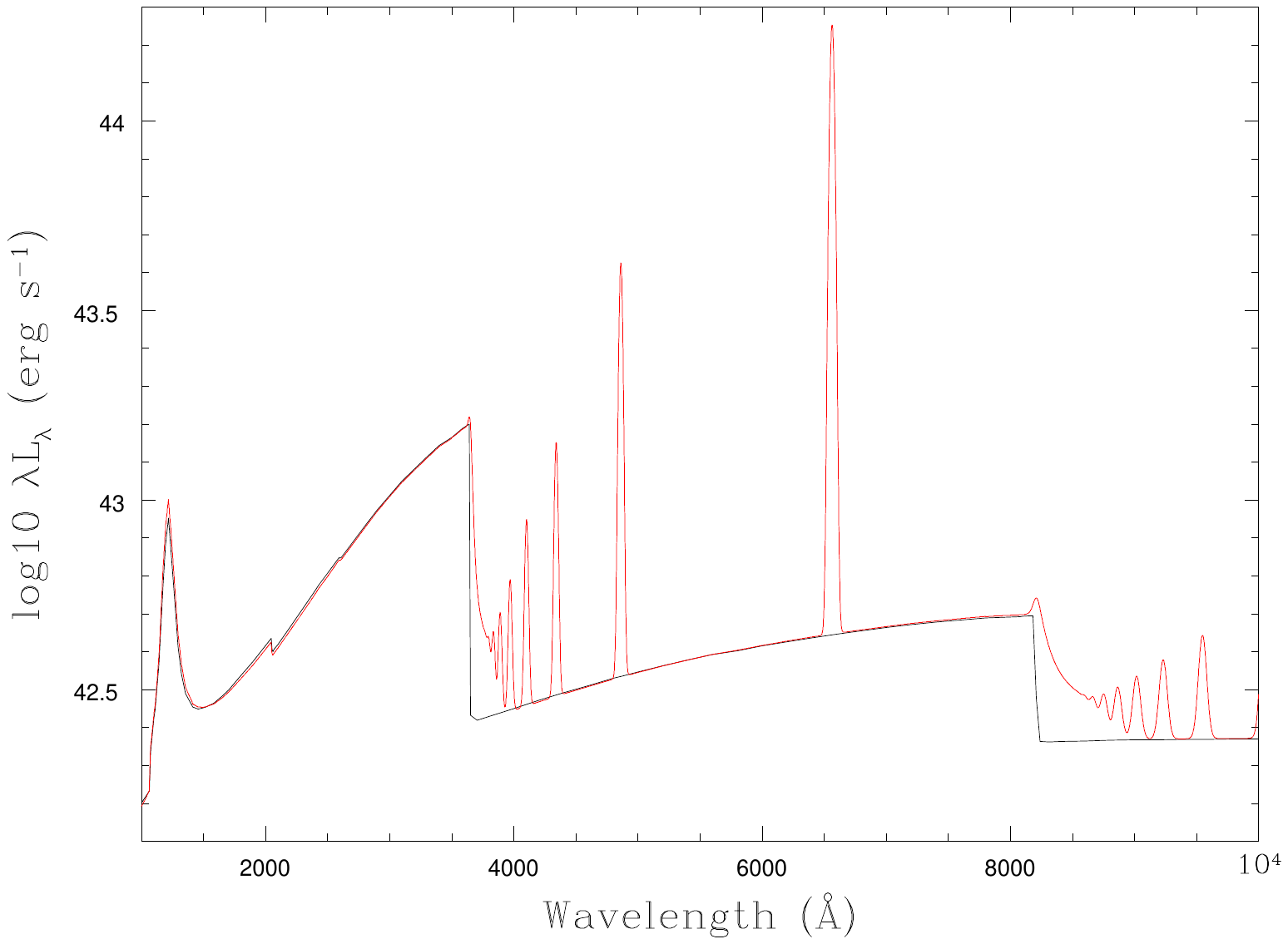}
    \caption{A comparison between the fiducial LOC model for NGC 5548
      ($8 < \log n_{\rm H} ({\rm cm}^{-3})< 12$) (Korista and Goad
      2019, their Figure~3), summed in 0.25 dex intervals in gas
      density (solid black line), and that constructed from the
      constant density slices (solid red line), including
      contributions from the Balmer and Paschen line series, a
      modified edge location (a consequence of finite gas densities),
      and redistributed flux (following our down-weighting scheme),
      all at the same resolution in gas hydrogen density.}
    \label{loc_comp}
\end{figure}

\subsection{UV--Optical He~{\sc i}}

There are several line series which contribute to the He~{\sc i}
spectrum spanning the UV--optical--IR. However, since He~{\sc i} is
non-hydrogenic we do not in this instance implement a down-weighting
scheme for the predicted higher order transitions in this line, and
instead simply extrapolate the lower order line strengths to higher
orders.
In Figure~\ref{plot_dens_jul7_he1} we illustrate the He~{\sc i}
spectrum for constant density slices through the flux density plane.

\subsection{A modified LOC DC spectral template}
Our original goal was to produce a modified DC template within the
context of an LOC model for the BLR. In the presence of cut-offs in
$U_{\rm H}c$, this can only be achieved by summing over the constant
density slices after first normalising to the number of models
contributing to each slice. In Figure~\ref{loc_comp} we show a
comparison between our fiducial LOC model of the DC contribution in
NGC~5548, and our modified template, created from summing the constant
density slices, and normalised as described above. In line-free
regions away from the jumps, there is good correspondence between the
overall shape of the DC spectrum calculated using these two
schemes. Of particular note is the slow ramp down in emission longward
of the Balmer and Paschen jumps.

Having validated our procedure for constructing a modified LOC DC
template from constant density slices through the $\Phi_{\rm
  H}$--$n_{\rm H}$ plane (Figure~\ref{loc_comp}), we next apply this
template to a spectral decomposition of the UV--Optical--IR spectrum
of the well-studied NLS1 Mrk~110.

\section{A spectral decomposition of Mrk~110}

To illustrate the spectral contribution of the pile up of higher order
emission lines in the vicinity of the Balmer and Paschen jumps,
accounting for both their mean formation radius and relative emission
line strengths, and thereby constrain the DC contribution to the
continuum emission, we here choose to model the UV-optical spectrum of
the nearby Narrow-Line Seyfert 1, Mrk~110, a source with a UV
luminosity ($\lambda1138$\AA) a factor $\sim 3\times$ larger than for
NGC~5548. Considering the factor $\approx$2 increase in luminosity
distance, $d_L=163$~Mpc (cf. 80.1~Mpc for NGC~5548), yields a received
flux which is $\approx 18\%$ smaller.

At a redshift of $z=0.03529$, Mrk~110 has been the subject of previous
broad emission-line (Peterson et~al. 1998, 2004; Zu et~al. 2011; Bentz
et al.\ 2013) and continuum reverberation mapping campaigns
(Vincentelli et~al. 2021). Consequently, the central black hole mass
is reasonably well-known, with ${\rm MBH}$ estimated to be $\approx
2.5\times 10^{7}$~M$_{\odot}$, and $L/L_{Edd} = 0.43$
(Meyer-Hofmeister and Meyer 2011), though MBH values as high as
$1.4\times 10^{8}$~M$_{\odot}$, based on the gravitational redshift of
the variable emission in the strong broad optical emission-lines, have
also been suggested (e.g., Kollastchny 2003). Continuum reverberation
mapping of this source indicates contributions to the delay spectrum
from at least two components which act on different timescales, a high
frequency component associated with disc reprocessing and a low
frequency component likely associated with the BLR (Vincentelli
et~al. 2021). Spectroscopically, Mrk~110 exhibits prominent Balmer
continuum emission, over and above that of the underlying disc
continuum, and broad emission-lines that are relatively narrow in
width. This affords easy identification of line-free continuum bands,
and increases the number of resolved lines in the vicinity of the
Balmer jump. Together, these make Mrk~110 the ideal test-bed for
spectral decompositions aimed at isolating, and ultimately
quantifying, the DC contribution to the continuum emission.

While there are numerous examples of AGN spectral decompositions in
the UV--optical, few utilize multi-wavelength observations spanning
more than a decade in wavelength, essential for constraining both the
underlying disc contributions, diffuse continuum emission, and the
longer wavelength toroidal (e.g., Korista and Goad 2019) and galaxy
contributions.  Here we use a modified version\footnote{We have
replaced the far UV spectrum of Mrk~110 with more recent UV HST/COS
(Costantini, PID~15699), and optical HST/STIS (Cackett, PID~15413)
spectroscopic data taken from the STScI/MAST portal. These spectra are
first de-reddened using the prescription given in Brown et al.\ (2019)
-- corrected for foreground dust extinction in the Milky Way using
dust maps from Planck (Planck Collaboration et al.\ 2011; Planck
Collaboration et al.\ 2014) and a Fitzpatrick (1999) model, and scaled
appropriately. Given the non-contemporaneous nature of the data used
in the construction of the broad-band SEDs, the spectral
decompositions should be considered as illustrative.} of the composite
spectrum of Mrk~110 from Brown et al.\ (2019,
DOI:10.17909/t9-3dbt-8734), covering the wavelength range $2\times
10^{-5}-36$ \micron\/, and more than sufficient for this purpose.  The
monochromatic flux at a wavelength of 2000\AA\ is $\lambda F_{\lambda}
\approx 4\times10^{-11}$~erg~s$^{-1}$~cm$^{-2}$ for this source.

\subsection{Spectral components}
%/home/m/mg159/DISC_LAGS/plot_16
% plot_16
\begin{figure}
    \centering
    \includegraphics[width=\columnwidth]{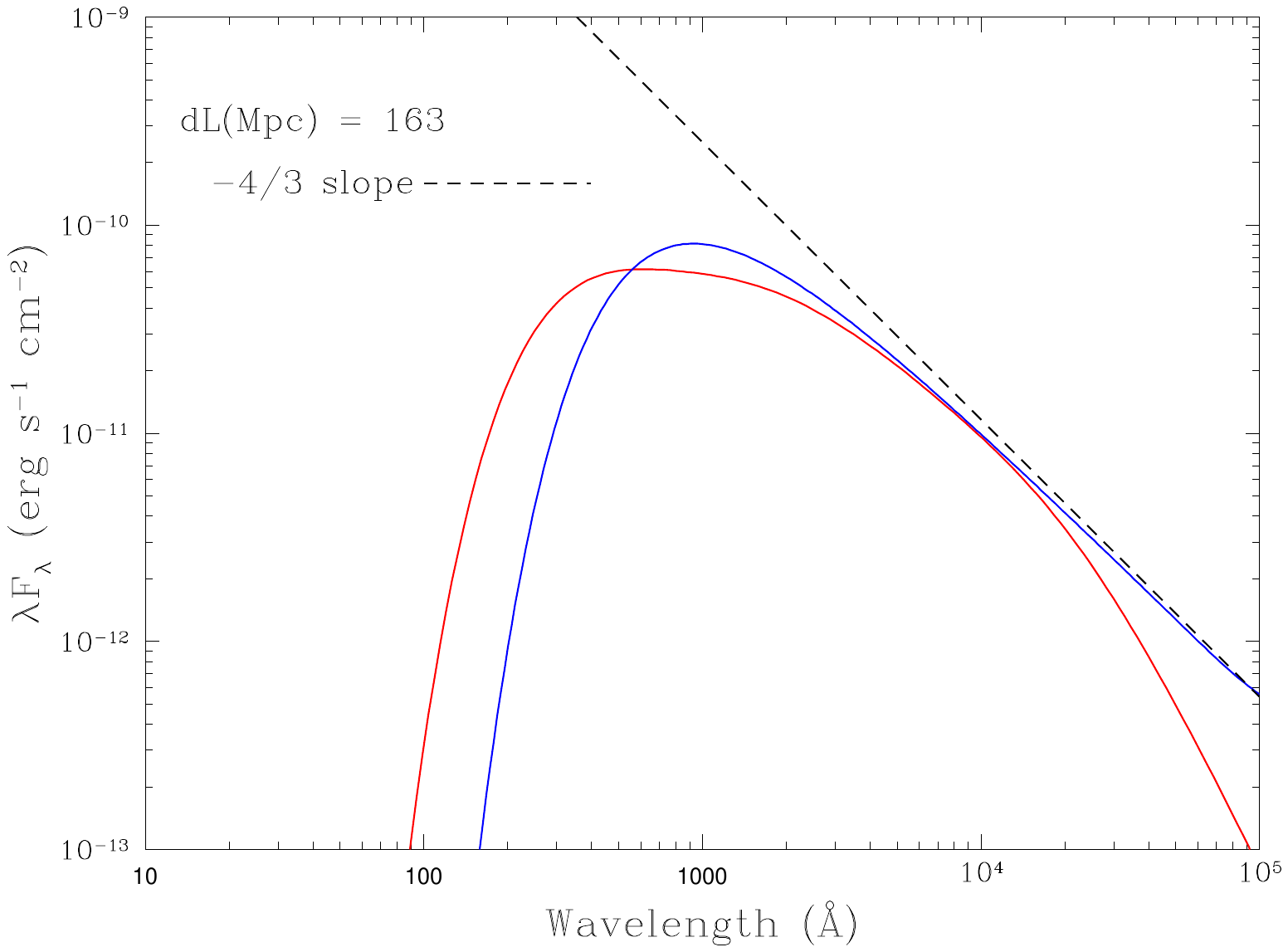}
    \caption{A comparison of multi-colour disc black-body spectral
      energy distributions for: (i) a standard infinite SS73 accretion
      disc (blue), and (ii) a finite accretion disc with a colour
      temperature correction (red). Disc model parameters for the
      finite disc -- ${\rm M}_{\rm bh}=10^{8}~{\rm M}_{\odot}$,
      $\dot{\rm {M}}=0.1~M_{\odot}~{\rm yr}^{-1}$, $R_{\rm
        min}=6~R_{\rm g}$, $T_{\rm min}=1800$~K ($R_{\rm
        max}=2230~R_{\rm g}$), have been chosen to be representative
      of values reported in the literature. The colour temperature
      correction acts to flatten and shift the peak of the standard
      MCD BB spectrum toward shorter wavelengths (e.g., Zdziarski
      et~al.\ 2022), while the finite disc size reduces the emission
      at wavelengths beyond $\sim 1$~$\mu m$.}
    \label{disc_colcor}
\end{figure}

The disc emission we model as a sum of black-bodies from a disc inner
radius $R_{\rm in} = 6~R_{g}$, where $R_{g}$ is the gravitational
radius GM/c$^2$, to an outer radius $R_{\rm out}$ set to the distance
at which the disc temperature is low enough for hot graphite grains to
condense ($T\approx1800$~K), and importantly we include a switch which
allows for a disc colour temperature correction using the formalism
presented in Done et al.\ (2012) and also discussed in Zdziarski et
al.\ (2022). The disc size and radial temperature profile $T(R)$ is
set by $M_{\rm BH}$ and $\dot{M}$, and for values appropriate to
Mrk~110, and in the absence of any colour temperature correction,
$T(R)\propto R^{-3/4}$\, at UV--optical wavelengths\footnote{Since
$T(R) \propto \left ( GM\dot{M}/R^{3} \right )^{1/4}$ (approximately),
the overall shape of the standard accretion disc 0.1--1 microns SED is
determined by the ratio $\dot{M}/(x^3 M^{2})$, where x =
$R_{in}/R_g$.}. The colour temperature correction (Done et~al.\ 2012;
Zdziarski et al.\ 2022) acts to flatten the SED at far-UV energies,
redistributing UV continuum flux to EUV energies (e.g.,
Figure~\ref{disc_colcor}). We note that the adopted spectral model of
the underlying thermal continuum spectrum does not consider
contributions from the warm corona, although these are not usually
important for wavelengths in this source $\lambda > 1000$~\AA\/ (see
e.g., Juranova et al. 2024).

We include in our spectral decomposition: a UV iron template (e.g.,
Tsuzuki et~al. 2006; Mejıa-Restrepo et al.\ 2016), but here modified
as described below, an optical Fe~{\sc ii} template (e.g., Vestergaard
and Wilkes 2001), our modified LOC diffuse continuum and hydrogen
emission line template, individual LOC templates for He~{\sc ii} and
He~{\sc i}\footnote{While the relative strength of the DC and hydrogen
emission line contributions is not particularly sensitive to the shape
of the ionizing continuum, their strengths relative to the higher
ionization Helium lines are. Thus, we keep the He~{\sc ii} and He~{\sc
  i} templates separate from the DC$+$broad hydrogen emission line
composite so that they can be scaled independently during the fitting
process.}, and an estimate of the (oft-neglected) toroidal
contribution as described in KG19: a combination of thermal emission
and scattered continuum light from graphite grains with size
distributions (larger than ISM grains) similar to those found in the
Orion Nebula, along with the thermal continuum emitted by the gas,
plus diffuse continuum contributions from a model narrow-line
region. For the host galaxy spectral contribution we use an 11~Gyr
solar metallicity model from Bruzual and Charlot (2003), broadened to
the resolution of HST/STIS G430L and G750L with flux normalization
appropriate for the HST aperture ($F_{\lambda5100} =6.4 \times
10^{-17}$~erg~s$^{-1}$~cm$^{-2}$~\AA$^{-1}$).\\

\subsection{DC and emission-line templates for Mrk~110}

Since it desirable that our modified DC template $+$ emission lines be
as widely applicable as possible, we once again model the line
emission for constant density slices through the flux-density plane
with Gaussian profiles, with widths determined by the mean formation
radius of each line under the assumption that the BLR gas is
virialised.  If the BLR comprises gas with a broad range in density,
then for overlapping lines that are coincident in wavelength, even if
similar in strength, differences in their mean formation radii will
result in non-Gaussian profiles.  Furthermore, since the LOC template
is formed by a weighted sum of the spectra generated for the
individual slices in gas density, the resultant emission line profile
even for well-isolated lines will be non-Gaussian, displaying
comparatively broader wings and narrower cores, as the higher density
gas is mainly confined to smaller BLR radii and lower density gas to
larger BLR radii.

Assuming $R_{\rm BLR} \propto L_{\rm ion}^{1/2}$, and broadly similar
SEDs, the factor $\sim3\times$ larger luminosity of Mrk~110
c.f.\ NGC~5548, results in a BLR which is a factor $\sim{(3)}^{1/2}$
larger ($\approx$200--300 lt-days at the outer radius, where robust
grains can survive).  On the other hand, black hole masses that differ
by an order of magnitude have been reported for Mrk~110, with virial
mass estimates favouring low black hole mass, and spectropolarimetric
measurements (e.g. Afanasiev et~al. 2019) and SED fitting favouring
high black hole mass. Profile fitting that includes the effects of
gravitational redshift (e.g., Kollatschny 2003) also favour large
black hole masses ($M_{\rm grav}=1.4\pm0.3\times 10^{8}~M_{\odot}$),
and may be reconciled with virial mass estimates if a low line of
sight inclination is assumed ($i\sim 21$~deg).  Given the large range
in black hole mass estimates published in the literature, as a
starting point, we here adopt the low virial mass estimate from Bentz
and Katz 2015, $M_{\rm vir} = 1.959\times 10^{7}~M_{\odot}$, and for
which (ignoring any differences in viewing angle) the broad
emission-line template for Mrk~110 will be a factor $\approx 2$
narrower than for NGC~5548.

%/mg159/KIRK_EXCEL/plot_fe2_col_update.pdf
\begin{figure}
    \centering
\includegraphics[width=\columnwidth]{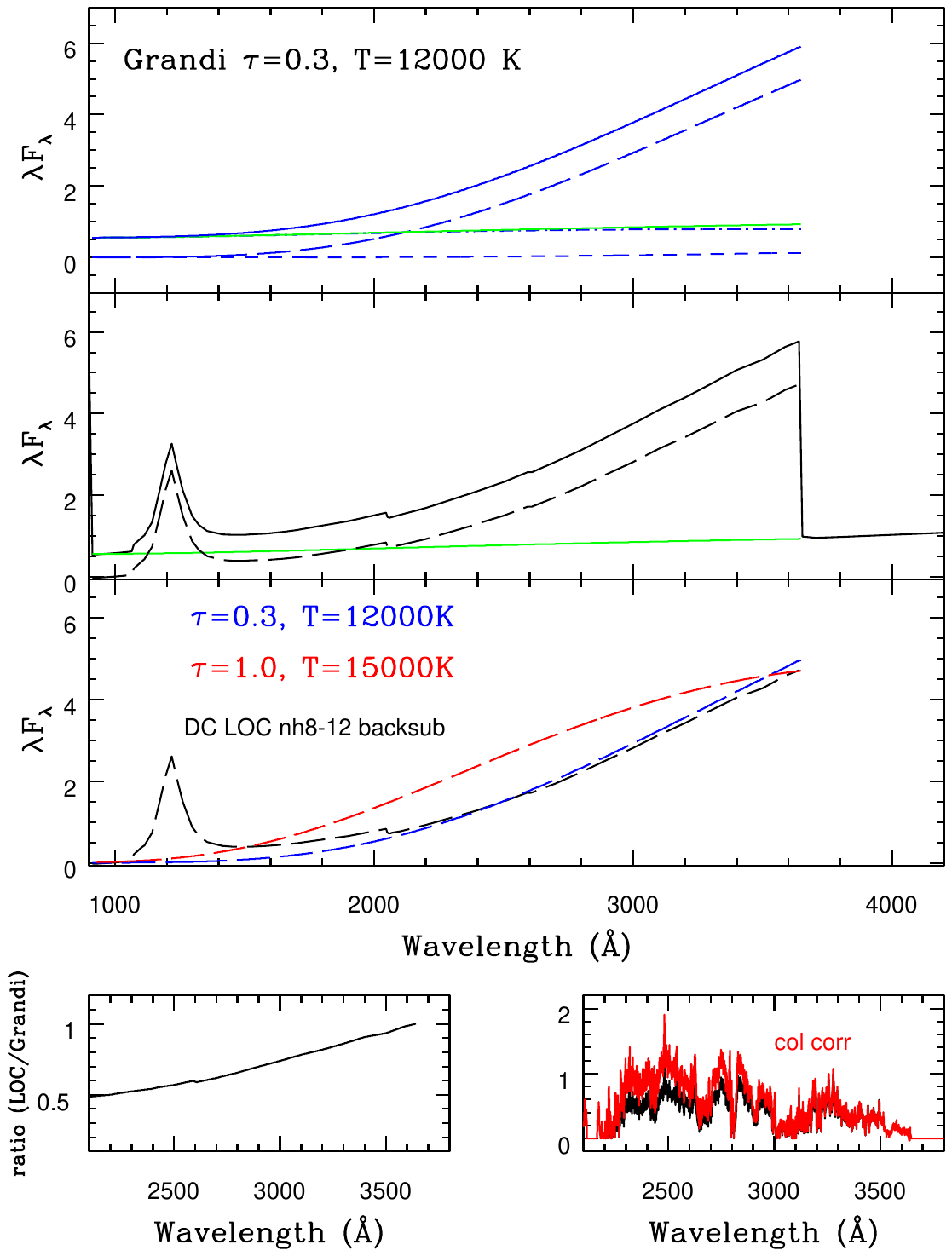}
    \caption{Upper panel - Estimated Paschen continuum (short-dashed
      line), free-free emission $+$ electron scattering (dot-dashed
      line) and their sum (solid green line), with optical depth and
      temperature chosen to approximately match the background
      emission of the LOC diffuse continuum component from Korista and
      Goad (2019). The corresponding Balmer continuum for these same
      parameters is indicated by the long dashed line, and their sum
      by the solid blue line.  Panel 2 - DC template from Korista and
      Goad (2019), and the corresponding Balmer continuum (long dashed
      line) following background subtraction (see text: solid green
      line).  Panel 3 - background subtracted LOC Balmer continuum
      (dashed black line), and Grandi (1982) Balmer continuum,
      assuming typical values for the threshold optical depth
      ($\tau=1.0$) and temperature (T=15000K) from the literature. The
      ratio of the {\sc cloudy}-predicted LOC BaC to the Grandi BaC
      (panel 4), provides a colour correction to the estimated UV
      Fe~{\sc ii} contribution. Panel 5, the Fe~{\sc ii} template from
      Tsuzuki et~al.\ (2006) before (solid black line) and after
      (solid red line) applying a colour-correction.}
    \label{fe2_colcor}
\end{figure}

Our model H$\alpha$ broad emission-line profile for Mrk~110 is a sum
of Gaussians centred on a common wavelength, whose strengths and
widths are based upon our LOC model line emissivity and a virial
velocity field. It has a FWHM$\approx$ 28.4\AA\/, consistent with the
mean formation radius of this line, and is approximately Gaussian
being somewhat broader in the line wings and narrower in the line core
than a single Gaussian profile.  By contrast, the measured H$\alpha$
broad emission line profile in Mrk~110 can be approximated by a single
Gaussian of width $\approx$ 44\AA\ (FWHM), ignoring the red wing.
Thus, in order to match the H$\alpha$ emission line profile, the
template profile must be broadened by convolving with a Gaussian with
kernel width $\sigma \approx [(\rm
  {FWHM}_{H\alpha(Mrk~110)})^{2}-(FWHM_{H\alpha(LOC)})^{2}]^{1/2}/2.354$
(e.g., Vestergaard and Wilkes 2001), yielding a kernel width of
$\sigma \approx 14.3$\AA\ ($ = 33.6$\AA\, FWHM).  Template broadening
is here performed in log-$\lambda$ space, so that a constant velocity
broadening is applied to each of the H, He~{\sc i} and He~{\sc ii}
templates in turn. We note that this will introduce a small additional
broadening at the location of the jumps (which are already quite
smooth) unless the Ba$+$Pa lines are treated separately from the DC
component.

\subsection{A modified UV iron template}

Typically UV and optical Fe~{\sc ii} templates employed in spectral
decomposition are determined from observations of strong UV Fe~{\sc
  ii} emitters (e.g., I~ZW~1). Isolating the UV Fe~{\sc ii} template
normally involves first estimating the background continuum, a
combination of disc continuum emission (often represented by a single
powerlaw), and Balmer continuum emission (e.g., Wills, Netzer and
Wills 1985; Vestergaard \& Wilkes 2001; Veron-Cetty et~al.\ 2004;
Tsuzuki et~al. 2006; Bruhweiler \& Verner 2008; Kovacevic
et~al.\ 2010; Popovic et~al.\ 2019; Park et~al.\ 2022). In most
previous work the Balmer continuum has been modelled following the
prescription of Grandi (1982), e.g., Tsuzuki et~al.\ (2006). However,
as reported in Mejia-Restrepo et al.\ (2016), and confirmed here,
typical parameters chosen for modelling the Balmer continuum are in
fact a poor representation of the diffuse continuum typically
associated with BLR clouds.  In Figure~\ref{fe2_colcor} we illustrate
the discrepancy between the Balmer continuum predicted by Grandi
(1982) using a parameterization commonly adopted in the literature --
a threshold optical depth $\tau=1.0$, and temperature $T=15000$~K, and
that which is more characteristic of our LOC photoionisation model
(panel~3 -- red and black dashed lines respectively).

In removing the other contributions to the diffuse continuum within
the Balmer continuum (a combination of the Paschen continuum and
weaker higher-order free-bound continua, free-free emission and modest
electron scattering) as predicted by photoionisation models
contributing to the LOC DC template (KG2019, black dashed line), the
standard parameterisation from Grandi (1982) over-predicts the Balmer
continuum contribution in the vicinity of the main UV Fe~{\sc ii}
emission (dashed red line). Indeed, the Balmer continuum within the
LOC DC template -- a composite spectrum of BLR clouds -- is better
characterized by lower electron temperature, $T\approx\/12000$~K, and
lower threshold optical depth, $\tau=0.3$ (dashed blue line).  Since
the UV Fe~{\sc ii} strength is underestimated at the shortest
wavelengths, we here apply a colour-correction (lower-left panel),
dividing through by the ratio of the relative strengths of the
LOC/Grandi continuum (Figure~\ref{fe2_colcor}, lower left panel). The
semi-empirical UV Fe~{\sc ii} template (from Tsuzuki et~al. 2006) and
corresponding colour-corrected version are shown bottom right (black
and red lines respectively). Given the uncertainty in the amplitude
and colour of the Balmer continuum applied in previous estimates of
the UV Fe~{\sc ii} strength (and in particular in the vicinity of the
Balmer jump), here we introduce one additional correction; we break
the colour-corrected UV Fe~{\sc ii} template into two components
centred at a wavelength of 3026\AA, and scale each component
independently.  For wavelengths shortward of 2200\AA\, we use an
appropriately scaled version of the UV Fe~{\sc ii} template from
Vestergaard and Wilkes (2001).
%\twocolumn
\subsection{Estimating the DC contribution via spectral fitting}

With the major spectral components identified, we next perform a
spectral decomposition of Mrk~110, to place limits upon the diffuse
continuum contribution to the total flux spectrum, and importantly,
constrained by the strength of the broad Balmer and Paschen
emission-lines.  The combined DC $+$ Balmer and Paschen emission-line
template we scale to match the measured strengths of the broad
component in the strongest optical emission-lines in each series
(e.g., H$\alpha$, H$\beta$, see Appendix~\ref{decomp} for
details). He~{\sc i} and He~{\sc ii} emission-line templates are
scaled independently and in a similar fashion using the strong broad
emission-lines of He~{\sc ii} $\lambda$4686 and He~{\sc i}
$\lambda$5876, in order account for uncertainties in the relative
number of high energy (54~eV) photons when compared to those near
1~Ryd for this source and which may differ substantially from that for
NGC~5548.  Appropriate scale factors have here been determined by
matching the predicted broad emission-line strengths with
observations. This amounts to a global scaling of the cloud covering
fraction.  Once scaled, the strengths of these components remain fixed
throughout the fitting process.

In Table~\ref{line_ews} we indicate the predicted emission-line
strengths of the major UV/optical emission lines for our two LOC
integrations (columns 4 \& 5) alongside the measured EWs (column 3)
for the major UV--optical emission-lines in Mrk~110.  Quoted line EWs
are measured relative to a reference band, here chosen to be the
incident continuum flux at 1215\AA\, and assuming full source coverage
at the outer radius), so that they might easily be compared with
similar EW contours on the flux density plane (e.g., KG04). For
Mrk~110, the measured line EWs are {\em unusually\/} large. For
example, EW(Ly$\alpha$)> 200\AA\, while typical values in AGN span
50--150\AA\,, though we note that the EW(Ly$\alpha$) may be reduced by
up to $\sim$10\% if the Rayleigh scattering feature beneath Ly$\alpha$
is significant. Such large EWs suggest large global covering fractions
are required ($\sim$70\%, falling to $\sim$40\% in the presence of
significant microturbulent velocities). We finally note that the
ionising SED in Mrk~110 may simply have a larger value in
$\rm{\Phi_H}$/$\lambda\/F_{\lambda1215}$ than was adopted in the
photoionisation models presented here.

\begin{table}
    \centering
    \begin{tabular}{lrr|rrr}\hline
    Line ID   & Flux ($\times 10^{-14}$)   & EW(\AA) & EW(\AA)     &  EW(\AA)         \\
              &  (erg~s$^{-1}$~cm$^{-2}$)  &         & ($0$~km~s$^{-1}$) &   ($100$~km~s$^{-1}$) \\ \hline
  Ly$\alpha$ 1216  & 868.9                & 214.1  &  273.5  &   407.2  \\
  C~{\sc iv} 1550  & 581.7                & 143.3  &  206.8  &   294.5  \\
  He~{\sc  ii} 1640 & 46.2                & 11.4   &  36.0   &    38.5  \\
  He~{\sc ii} 4686 & 8.7                  &  2.2   &  4.3    &    5.6   \\
  He~{\sc i} 5876  & 4.8                  & 1.2    &  2.2    &    3.3   \\
  H$\beta$  4861   & 40.6                 & 10.0   &  16.1   &   29.0   \\
  H$\alpha$ 6563   & 186.7                & 46.0   &  62.0   &   87.2   \\ 
  Mg~{\sc ii} 2800 & 74.2                 & 18.3   &  46.4   &   85.1   \\ \hline
\end{tabular}

\caption{Columns 2--3, rest-frame broad emission line fluxes and
  equivalent widths measured relative to the continuum flux at
  $\lambda$1215\AA\,. Quoted EWs for the Balmer lines include all
  components associated with a particular line excluding the narrow
  component. The measured rest-frame continuum flux at 1215\AA\,
  $F_{\lambda}$(1215)$=4.059\pm0.016 \times
  10^{-14}$~erg~s$^{-1}$~cm$^{-2}$~\AA$^{-1}$ has been determined from
  a linear fit to line free continuum bands bracketing
  Ly$\alpha$. This likely includes $\approx 10\%$ contribution from
  the DC, and thus measured EWs referenced to the incident continuum
  could be as much as 10\% larger. Model EWs (Columns 4--5) have been
  calculated relative to the incident continuum flux at $\lambda
  1215$\AA\, and are quoted assuming full coverage of the ionizing
  continuum source and for two values of the microturbulent
  velocity. }
\label{line_ews}
\end{table}

We include scaling factors for the MCD BB continuum to allow for
uncertainties in the source distance and inclination, a
color-corrected UV Fe~{\sc ii} template (\S3.3), an optical Fe~{\sc
  ii} template (though weak), and a component for the Toroidal
Obscuring Region (hereafter, TOR) which at near-IR wavelengths (0.8-3
microns) is dominated by thermal radiation from hot graphite grains
near their sublimation temperature. We note, however, that this
spectral template is the full thermal and scattered light spectrum
from gas embedded with grains similar to those found in the harsh UV
radiation environment of the Orion nebula (Korista and Goad 2019).  We
also include in our fit an 11~Gyr solar metallicity galaxy template
(Bruzual and Charlot 2003) that is fixed in strength, and a bespoke
narrow emission-line template for the strongest Balmer and Paschen
lines that is fixed in strength, adopting Case B recombination
emission line ratios, and assuming a narrow H$\beta$ emission line
strength which is 10\% of the measured strength of narrow [O~{\sc
    iii}] 5007\AA\/, as determined via spectral decomposition (see
Appendix~\ref{decomp} and Table~\ref{mega_tab}). The strong broad UV
emission lines have been determined independently via spectral
fitting, adopting a locally linear fit to line-free continuum bands to
isolate the strong broad UV emission lines. This process will tend to
overestimate the underlying disc continuum contribution. We account
for this discrepancy as part of the overall fit.  Finally, we broaden
the 3 emission line templates; the combined DC+Ba+Pa template, and
He~{\sc i} and He~{\sc ii} templates, in logarithmic wavelength space
using a Gaussian convolution kernel of unit area, but whose width is
free to vary. For sources in which the optical Fe~{\sc ii} line widths
are narrow (FWHM $<$ 3200~km~s$^{-1}$), UV and optical Fe~{\sc ii}
emission-lines widths are comparable, and tend to be narrower than
either H$\beta$ or Mg~{\sc ii} (Le \& Woo 2019; Kovacevic et
al.\ 2010). For the UV and optical Fe~{\sc ii} templates, we here
smooth with a Gaussian kernel of fixed width ($\approx 10$\AA), noting
that the optical Fe~{\sc ii} strengths in Mrk~110 are relatively weak.

\begin{figure}
    \centering
\includegraphics[width=\columnwidth]{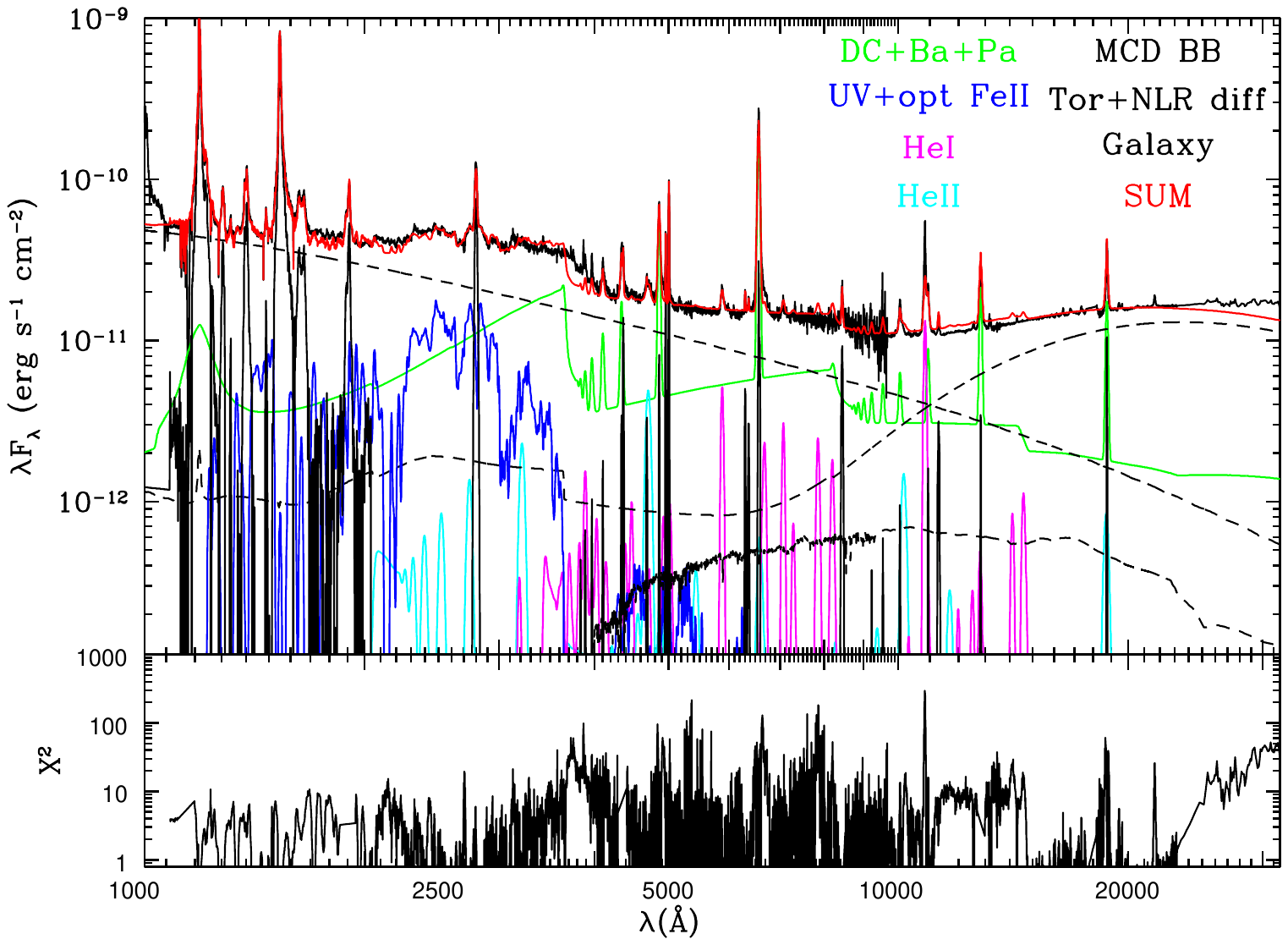}
    \caption{An example spectral decomposition of the UV--Optical--IR
      spectrum of Mrk~110. Upper panel -- Mrk~110 scaled input SED
      (black-dashed), MCD BB ($R_{\rm min}=6$~$R_{g}$, $R_{\rm max}$
      set to radius at which $\rm{T_{min}} = 1800$~K), a combined DC
      $+$ Balmer $+$ Paschen line template (green), He~{\sc i}
      (magenta), He~{\sc ii} (cyan), a colour-corrected UV Fe2
      template and an optical Fe~{\sc ii} template (blue), scaled
      diffuse (NLR$+$TOR, black dashed), and fixed (in strength)
      galaxy (black dashed) and narrow emission line templates (black
      solid) which are not part of the fit. The summed components are
      indicated in red. An indication of the quality of the fit is
      shown in the lower panel. Zoomed in versions of the spectral
      decomposition may be found in Appendix~\ref{zoom}. The
      microturbulent velocity is here set to $v_{\rm
        turb}=0$~km~s$^{-1}$.}
    \label{plot_global_vt0}
\end{figure}

With all components specified, we fit the rest-frame UV-optical-IR SED
of Mrk~110 by minimizing $\chi^{2}$ using a combination of Simulated
Annealing, following Vanderbilt and Louie's implementation of the
Metropolis-Hastings algorithm (Vanderbilt and Louie 1984), and
down-hill simplex methods (Press et~al. 1992), optimizing over 9 free
parameters (Table~\ref{model_params}).  In
Figure~\ref{plot_global_vt0} we show an example global fit to the
broad band SED of Mrk~110. The upper panel indicates the MCD BB disc
emission (black dashed line), the summed DC+Ba+Pa spectrum (green), UV
and optical Fe~{\sc ii} templates (blue), He~{\sc ii} (cyan), He~{\sc
  i} (magenta), galaxy template (black dashed line) and NLR$+$toroidal
diffuse continuum contributions (black dashed line). Expanded views of
this model fit can be found in Appendix~\ref{zoom}. The best fit model
parameters are presented in Table~\ref{model_params}. Note that our
preferred solution requires a larger MBH and smaller $\dot{\rm M}$,
than the values commonly reported for this source (e.g., $\log ({\rm
  M}_{\rm BH}/{\rm M}_{\odot}$)=7.4~, and $\dot{\rm M}=0.24~{\rm
  M}_{\odot}$~yr$^{-1}$, corresponding to $L/L_{\rm Edd}=0.43$;
Meyer-Hofmeister and Meyer 2011).

\begin{table}
    \centering
    \begin{tabular}{lrr}\hline
Parameter         &  \multicolumn{2}{c}{Best fit value} \\
                  & \multicolumn{1}{c}{$v_{\rm turb}$} & \multicolumn{1}{c}{$v_{\rm turb}$} \\
                  &   \multicolumn{1}{c}{($0$~km~s$^{-1}$)}  &  \multicolumn{1}{c}{($100$~km~s$^{-1}$)}        \\ \hline
  $^{\dagger}$ DC$+$Ba$+$Pa scale factor      & $0.69$ &  $0.38$ \\
  $^{\dagger}$ He~{\sc ii} scale factor  & $0.55$ &  $0.42$  \\
  $^{\dagger}$ He~{\sc i} scale factor   & $0.59$ &  $0.40$ \\
$\log_{10}$ M$_{\rm BH}$ (M$_{\odot}$)       &  7.77 & 7.97 \\
Tor$+$NLR DC~ SF    & $-$0.16 & $-$0.16 \\
UV Fe~{\sc ii} SF (2200--3600\AA) & 0.71  & 0.76 \\
Opt Fe~{\sc ii} SF (4250--7000\AA) & $-$1.48 & $-$1.87 \\ 
$\log_{10}$ $\dot{{\rm M}}$ (${\rm M}_{\odot}~{\rm yr}^{-1}$)    &   $-$0.78 &  $-$1.09 \\
$\log_{10}$ FWHM smoothing kernel (\AA) & 1.50 & 1.51 \\
Disc luminosity scale factor & 0.82 & 0.95  \\ 
$\log_{10}$ UV line scale factor & $-$1.59 & -1.75  \\
$\log_{10}$ Fe~{\sc ii} correction factor & $-$0.24 & 0.00 \\
\\
\hline
\\
\multicolumn{3}{c}{Diagnostics}\\
\\
\hline
Torus $F_{\lambda}($2.35$\mu$m) (erg~s$^{-1}~$cm$^{-2}$~\AA$^{-1}$) & 5.53e-16 & 5.63e-16 \\
UV Fe~{\sc ii} flux (erg~s$^{-1}$~cm$^{-2}$~)  & 5.27e-12 & 6.36e-12 \\
Opt Fe~{\sc ii}  flux (erg~s$^{-1}$cm$^{-2}$) & 5.17e-14 & 2.08e-14 \\
DC/(disc$+$DC) (1630-3600\AA) &  27.1\% & 12.6\% \\
DC/(disc$+$DC) 3600\AA & 56.4\% & 31.7\% \\
DC/(disc$+$DC) 4200\AA & 22.1\% & 9.8\% \\
DC/(disc$+$DC) 5100\AA & 30.4\% & 13.8\% \\
DC/(disc$+$DC) 8000\AA & 51.6\% & 27.3\% \\
% data are stored in flux_measurements.sdf
F(DC 3600/F(DC 4200) & 5.41 & 5.03 \\
F(disc$+$DC 3600)/F(disc$+$DC 4200) & 2.12 & 1.55 \\
$^{\ddagger}$F(BaC)/F(H$\beta$) & 11.74 & 4.87 \\
$\log_{10}$(F(H$\beta$)/$F_{\lambda}$(DC4850\AA) & 2.65 & 2.99 \\
\hline
    \end{tabular}
    \caption{Upper panel -- Parameter estimates obtained from the
      spectral decomposition of Mrk~110. Lower panel -- Diagnostic
      fluxes and their ratios.
    \newline $^{\dagger}$ Indicates parameter fixed during the fitting
    process.\\ $^{\ddagger}$ Indicates a lower limit to the flux
    ratio, since the BaC extends from 1--0.25~Ryd. }
    \label{model_params}
\end{table}

\section{Discussion}

\subsection{Initial findings}

Our multi-component spectral decomposition
(Figure~\ref{plot_global_vt0}, upper panel) is able to reproduce the
gross features of the UV--Optical--IR spectrum of Mrk~110 with just a
handful of components. Inclusion of a significant DC component acts to
flatten the spectrum through the UV--optical, allowing for a steeper
underlying disc continuum than might otherwise be derived, and {\it
  negating the requirement for substantial intrinsic reddening in the
  host galaxy}.  Furthermore, if the disc is finite in size, and host
galaxy contributions through optical--IR are weak, significant DC
contributions can fill in the substantial emission gap likely to be
present at 1 micron, a region in which disc contributions are rapidly
declining and toroidal contributions rapidly rising.

A key result from this study is that the DC covering fraction that
importantly is first constrained by matching photoionisation model
predictions of the strength of the broad optical-IR recombination
lines of hydrogen with observations, is found to be $\approx 70$\%
(Table~2, column~2, $v_{\rm turb}=0$~km~s$^{-1}$). This is
$\sim$50--100\% larger than that found for NGC~5548 ($\sim$35--50\%)
using the same power-law radial covering fraction dependence, and
toward the upper end of the predicted range in covering fraction
($33\% \leq c_{\rm f} \leq 80$\%) found for NGC~5548 based on fitting
the strength and variability behaviour of the strongest UV
emission-lines in that source (e.g., Korista and Goad 2000).  When
compared to the underlying disc continuum contributions, and
integrating over a band-pass spanning 1630--3600\AA\ , the DC
contributes $\approx$ 27\% of the total continuum emission (disc$+$DC)
over the same band-pass. The DC contribution modelled here, is
$\approx$ 25\% larger ($\lambda F_{\lambda} (3600) =
2.05\times10^{-11}$~erg~s$^{-1}$~cm$^{-2}$) than that found by
Vincentelli et~al. (2021), using a scaled version of the original
DC-only template from Korista and Goad (2019), while also adopting a
far smaller value of MBH and larger $\dot{M}$.

Note that when fitting the major optical--IR broad emission-lines only
their widths have been allowed to vary, their strengths remain
fixed. While our model emission-line profiles can be non-Gaussian for
well-isolated lines, they are symmetric, and therefore cannot
reproduce the observed extended red-wing emission present in H$\alpha$
and H$\beta$ (e.g., Figure~\ref{fit_jul_06}).  Fits to the
emission-line profiles therefore tend to favour solutions that match
either the line cores or the line wings.  Even so, our best fit model,
requires a larger black hole mass than that adopted in the initial
stages of construction of the emission-line template with an estimated
MBH near the upper end of the range of published values for this
source ($\sim 10^{8}$M$_{\odot}$), and supported by the substantial
smoothing kernel width required to match the widths of the strong
optical and IR emission-lines. Such high black hole masses are not
unreasonable for this source, and more closely align with MBH
determinations based on profile fitting that includes the extended red
wings of the broad optical emission-lines while also accounting for
the effects of extreme gravity, gravitational redshift and transverse
doppler shift, $\log (M_{\rm BH}/M_{\odot}$)=8.1 (Kollastchny 2003),
\S3.2, and see also Goad, Korista and Ruff (2012).

Evidence for possible GR effects in the form of extended red-wing
emission is clearest in the strong optical Balmer lines of H$\alpha$
and H$\beta$\footnote{Enhanced red-wings may also be present in the
higher order Balmer lines. However, these lines tend to be weaker,
more highly blended, and the line continuum contrast poorer, making
such features harder to discern.}, with a more prominent red-wing in
H$\beta$ than in H$\alpha$, that is, the Balmer decrement is not
constant across the line profile (see e.g.,
Appendix~\ref{decomp}). This is not unexpected, as higher order Balmer
lines form at smaller BLR radii on average (see \S2.5, and KG04),
where GR effects are stronger. A multi-component fit to the broad
H$\beta$ emission-line profile (Figure~\ref{fit_jul_06}) indicates a
red-wing component centred 20\AA\ redward of the narrow emission-line
core, a similar shift to the red-ward displacement of the entire broad
He~{\sc ii} 4686 emission-line. Since He~{\sc ii} primarily forms at
smaller BLR radii (measured emission-line delays relative to the
optical continuum at 5130\AA\ in Mrk~110 are 3.5~days for He~{\sc ii},
23.5~days for H$\beta$, and 32.5 days for H$\alpha$; Kollatschny
2003), the bulk of the emission for this line will arise within radii
subject to strong gravity\footnote{If strong GR effects (gravitational
redshift and transverse Doppler shift) are responsible for the strong
line asymmetry observed in the broad wings of the optical
recombination lines, and for the measured shift in location of broad
He~{\sc ii}~4686 to longer wavelengths, such effects likely also play
a role in both shifting and broadening the jump locations and may act
to reduce the deficit in the vicinity of the jump. Such effects are
beyond the scope of this paper.}.

While we set out to address the shift in location and slow ramp down
in emission at the location of the Balmer jump, by considering the
effect of finite gas densities on jump location and the pile-up of
higher order lines, a substantial systematic mismatch between model
predictions and observations in the vicinity of the Balmer jump
nevertheless remains.

\subsection{Microturbulent velocities within the BLR gas}

Despite adding significant additional complexity to our model DC
template, a substantial deficiency in emission in the vicinity of the
Balmer jump remains. One way of reducing this deficit is to increase
the line-continuum contrast.  A physical process not yet considered
that acts precisely in this fashion is the presence of microturbulence
(Bottorff et~al. 2000). Microturbulent velocities (locally
extra-thermal line widths) act to reduce the line optical depths
thereby increasing the line escape probabilities. Consequently, line
cooling is enhanced at the expense of continuum cooling. This effect
is greater at the higher gas densities and incident photon fluxes
where the line optical depths are largest.  Since the higher-order
Balmer and Paschen lines are emitted preferentially at higher gas
densities, we expect significant enhancement of the emission-line
strengths in the vicinity of the jumps. While line flux ratios are not
particularly sensitive to the degree of microturbulence, the strengths
of permitted lines are selectively enhanced. Additional contributions
may also arise via continuum pumping of the upper levels of
transitions (see Bottorff et al.\ 2000).

%/rfs/XROA/mg159/PRESS/CLOUDY/c17.02/SERIAL/DC_DENSITY/NH08/MKN110
\begin{figure}
    \centering
    \includegraphics[width=\columnwidth]{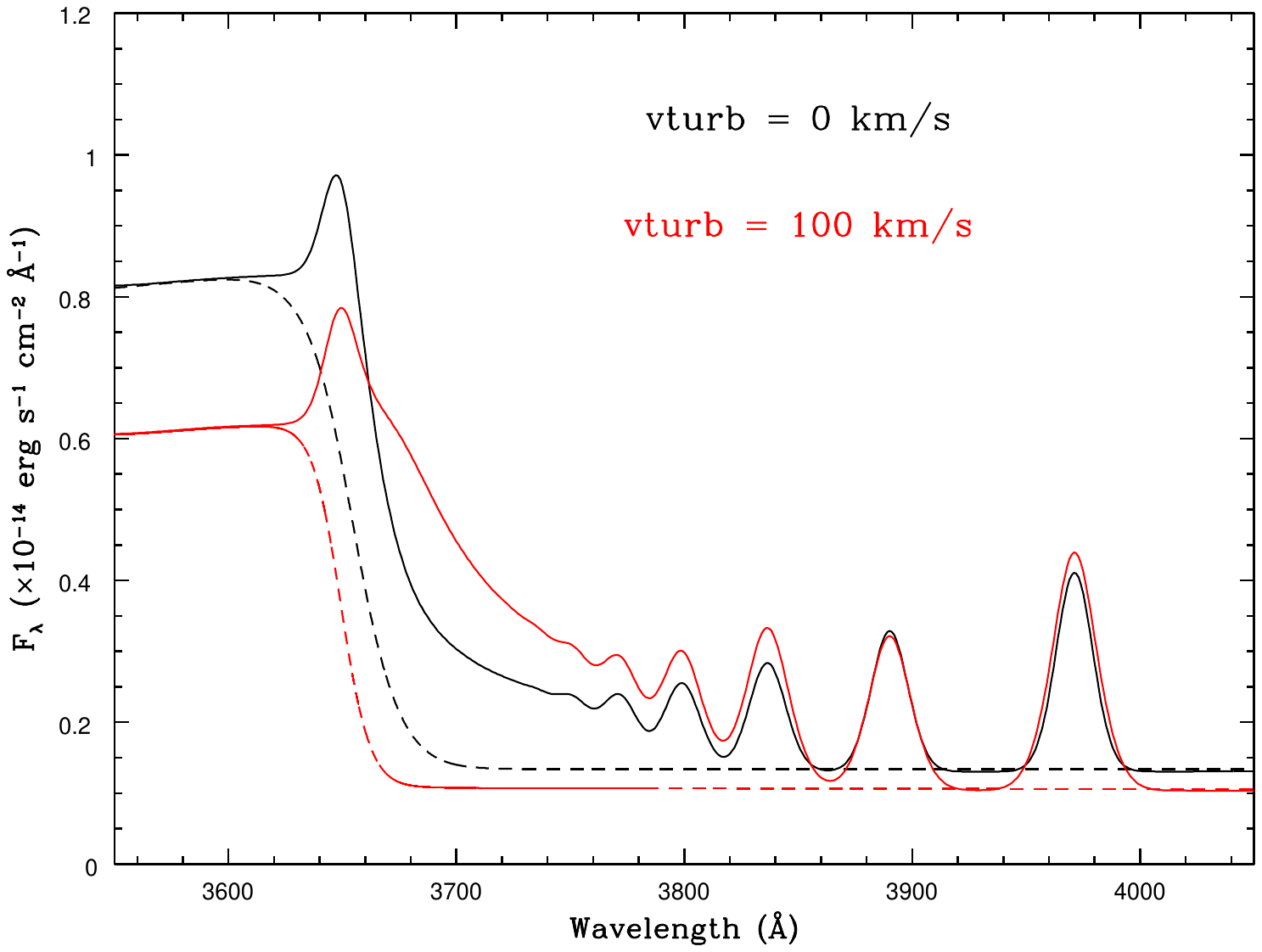}
    \caption{A close up view of the Balmer jump illustrating the
      pile-up of higher order lines long-ward of the jump. The Balmer
      jump location has been modified as described in \S2. In black we
      show the summed DC $+$ higher order Balmer lines for our
      fiducial LOC model ($6 \leq \log U_{\rm H}c$ (cm~s$^{-1}$)$~\leq
      11.25$, and $8 \leq \log n_{\rm H} {\rm (cm}^{-3}\rm{)} \leq
      12$. In red we illustrate the same LOC model but with a
      microturbulent velocity of 100~km~s$^{-1}$.  The dashed lines
      indicate the DC component only. Microturbulent velocities reduce
      the DC emission and increase the line emission, increasing the
      line--continuum contrast.}
    \label{vturb_edge}
\end{figure}

To investigate the effect of microturbulence on the appearance of the
DC template, we recompute our LOC models using photoionisation model
grids spanning the same seven orders of magnitude range in $\Phi_{\rm
  H}$, $n_{\rm H}$, but now with the inclusion of microturbulent
velocities, adopting a value of 100~km~s$^{-1}$, similar to that found
in magnetohydrodynamic simulations addressing the stability of broad
line region clouds (Krause et~al. 2012). Microturbulent velocities of
this order of magnitude have been suggested in order to match the
shape and strength of the UV Fe~{\sc ii} spectrum in AGN (Netzer and
Wills 1983; Baldwin et al.\ 2004; Sarkar et~al. 2021). We implement
this local line broadening mechanism in an {\em ad hoc} fashion,
without reference to a particular physical cloud model.

In the presence of microturbulence, the EW contours on the
flux-density plane of the Balmer and Paschen continuua open up, and
shift towards higher $\Phi_{\rm H}$--$n_{\rm H}$. This results in
lower DC luminosities on average ($\approx$26\% for our fiducial LOC
integration), and smaller mean formation radii (a reduction of
$\approx$10\% across the whole delay-spectrum).  A comparison of the
DC emission from constant density slices in the flux--density plane,
and an LOC summation, with and without microturbulence, suggests that
the effect is largely colour-independent; in the presence of
microturbulence, the DC contribution is reduced, but its spectral
shape remains approximately constant, with a small ($\sim$30\%)
reduction in the height of the Balmer and Paschen jumps (e.g.,
Figure~\ref{vturb_edge}).

Similarly, for the Balmer and Paschen series lines, the EW contours on
the flux--density plane shift toward higher $\Phi_{\rm H}$--$n_{\rm
  H}$. The line escape probabilities are significantly enhanced for
the majority of lines, including many of the higher order lines. Line
luminosities are significantly enhanced $\sim$ 0.3--0.4 dex, while the
mean formation radii of the majority of lines is significantly reduced
with a much weaker wavelength-dependence. Reduced DC contributions and
enhanced line emission increase the line--continuum contrast long-ward
of the jumps allowing for a more gradual decline toward longer
wavelengths.  In Figure~\ref{vturb_edge} we show combined
DC$+$emission-line templates in the vicinity of the Balmer jump for
the case of zero microturbulence (black) and $v_{\rm
  turb}=100$~km~s$^{-1}$ (red). Dashed lines show the DC component
only.  By contrast, the presence of microturbulence has a far more
modest effect ($\approx $20\% increase) on the strongest lines of
He~{\sc i} and He~{\sc ii}, and little to no effect on the strength of
the higher order lines of He~{\sc ii}.

\begin{figure}
    \centering
\includegraphics[width=\columnwidth]{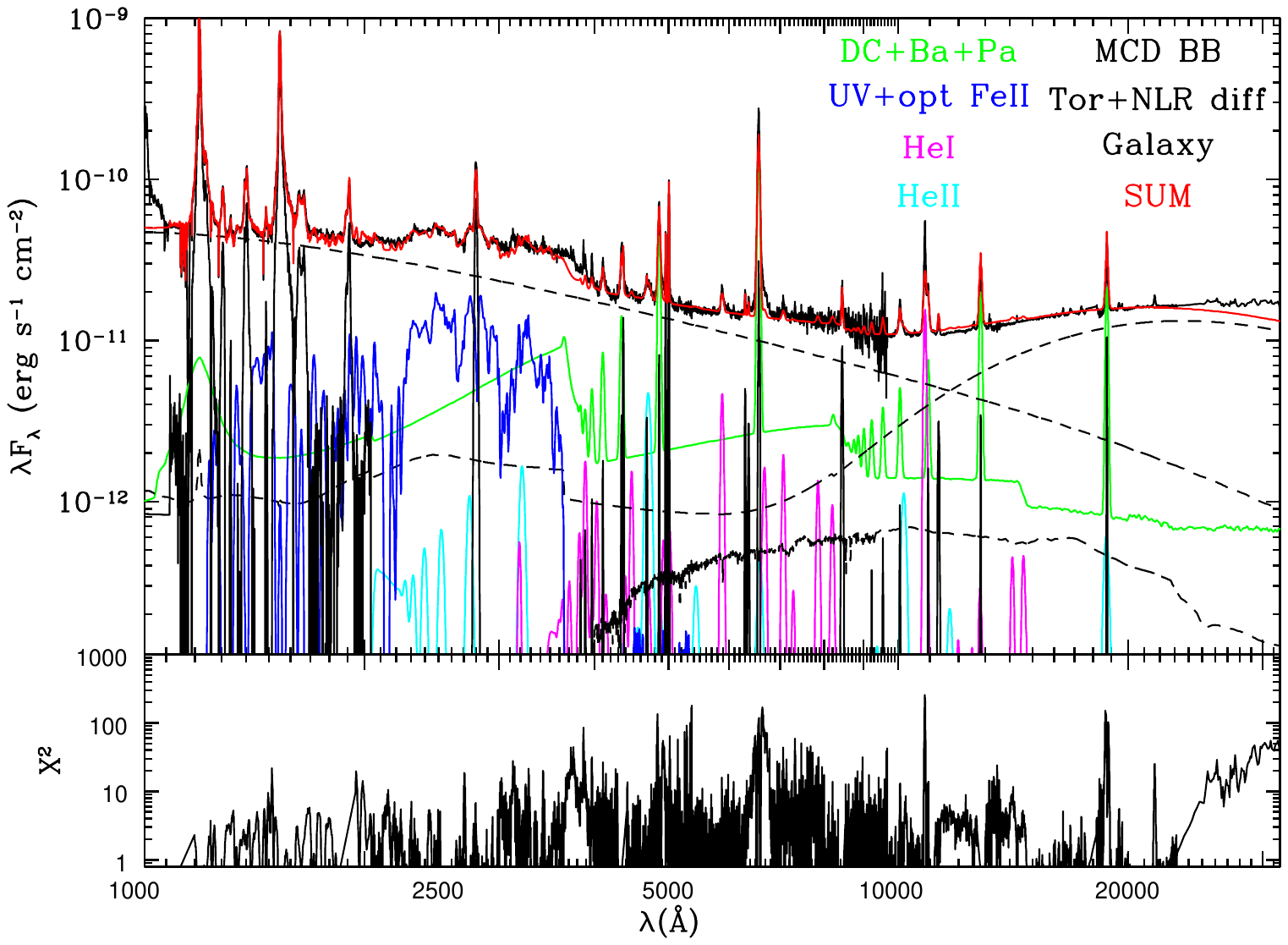}
    \caption{As for Figure~\ref{plot_global_vt0} but now using a
      combined DC$+$Ba$+$Pa template with microturbulent velocities
      $v_{\rm turb}=100$~km~s$^{-1}$.  The increased line to continuum
      contrast suppresses the DC contribution to the total light,
      requiring increased contributions from the underlying disc.}
    \label{plot_global_vt100}
\end{figure}

In Figure~\ref{plot_global_vt100} we show a spectral decomposition of
the rest-frame UV-Opt-IR spectrum of Mrk~110 but now substituting the
previous combined DC$+$Ba$+$Pa template (with zero microturbulence)
with that calculated for microturbulent velocities of
100~km~s$^{-1}$. The increased line continuum contrast reduces the
covering fraction required to match the observed emission line
strengths of the major optical recombination lines to $<$40\%
(Table~\ref{model_params}, column 3).  The DC contribution is
therefore significantly reduced, as is the height of the Balmer jump,
which also has a slower ramp down toward longer wavelengths. The
reduction in DC emission must be compensated for by an increased
contribution from the disc requiring larger M and/or larger $\dot{M}$
(e.g., Table~\ref{model_params}), and in such a fashion that the
underlying disc contribution preserves its shape (since the shape of
the DC spectrum is largely insensitive to the degree of
microturbulence). Our fits including microturbulence tend to favour
larger $M$ rather than $\dot{M}$, likely due to the deficit in
emission at 1 micron in the presence of suppressed diffuse continuum
contributions; a larger $M$ results in a flatter spectrum through the
UV-optical-IR, and the spectrum doesn't cut-off as sharply.

%/scratch/ngts/mg159/OPTIMIZERS/TURB/AMOEBA/ANCHOR/NOCOL/ONCE

\subsection{Discussion of the various dependencies.}

The shape of the underlying incident continuum depends on MBH and
$\dot{M}$, which for the standard Shakura-Sunyaev disc model a fixed
spectral shape requires ($\dot{M}/M^{2}$)$=$constant. For fixed MBH, a
larger $\dot{M}$ shifts the peak emission to shorter wavelengths, and
the UV-optical spectrum becomes somewhat bluer and especially
brighter. For a fixed $\dot{M}$, a larger MBH shifts the peak (in
$\lambda L_\lambda$) of the disc emission to longer wavelengths, with
the consequence that the spectrum flattens through the UV--optical and
becomes brighter at especially longer wavelengths. To further
complicate matters, the colour temperature correction of the disc
emission shifts the peak emission to shorter wavelengths and flattens
the spectrum through mainly the UV spectral region (\S3.1,
Figure~\ref{disc_colcor}).  At the longer optical-to-near-IR
wavelengths, the disc emission steepens to approximately its canonical
slope $\lambda L_\lambda \propto \lambda^{-4/3}$ due to the locations
of the Wien blackbody peaks given an approximate $r^{-3/4}$ dependence
in the disc effective temperature. Then beyond $\sim 1 \mu\/m$ or so
the spectrum steepens further toward the Rayleigh-Jeans slope $\lambda
L_\lambda \propto \lambda^{-3}$, as the minimum effective temperature
of a finite disc is encountered. The host galaxy, if sufficiently
bright within the entrance aperture, will act to flatten the observed
AGN spectrum through the optical-to-near-IR spectral region. Beyond
$\sim 1 \mu\/m$, emission from the dusty torus due to hot grains will
initially flatten and then dominate the observed spectrum, with the
location of the peak emission near $\sim 3\mu\/m$ (in $\lambda
L_\lambda$ vs.\ $\lambda$ space) due to thermal emission by hot,
likely larger graphite, grains.

The spectral energy distribution of the DC is on average flat in
$\lambda L_{\lambda}$ through the UV--optical. Increased
microturbulent velocities which act to increase line cooling at the
expense of continuum cooling is largely colour-independent, the lines
become stronger and the DC is suppressed, but the SED of the DC
remains largely the same. A significant DC component will thus act to
flatten the spectrum through the UV—optical, and even out to $\sim
1\mu\/m$ in the absence of significant host galaxy contribution.  For
our spectral decompositions, the inclusion of both a disc colour
temperature correction {\bf and} DC component, pushes the best fit
solution to low black hole mass and large $\dot{M}$; a steeper
incident continuum is required to accommodate the flattening of the
spectrum in the far UV induced by the disc colour temperature
correction and the flattening through the UV--optical arising from a
significant DC component.  Higher mass solutions (approaching $\sim
10^{8}$ $M_{\odot}$), which are more in line with both historical mass
estimates of this source (from the gravitational redshift of the
optical emission lines), and more recently reported values from the
analysis of gravitationally shifted X-ray lines (Reeves et al. 2021),
do not require a disc colour temperature correction. The flattening of
the spectrum can be accomplished simply by increasing MBH and the
inclusion of a significant DC component. Our best fit solutions, found
in the absence of any disc colour temperature correction are shown in
Figures~\ref{plot_global_vt0}, \ref{plot_global_vt100}. Best fit
parameters are presented in Table~\ref{model_params}. We note that
comparable solutions are recovered with the disc colour temperature
correction implemented, and the impact of the spectral flattening
reduces the inferred MBH by about a factor of $\approx$5.  {\it The
  complex interplay between the various contributory components
  illustrate the ambiguity of assigning black holes masses and
  accretion rates via SED fitting alone}. Nevertheless, based on the
measured emission-line strengths we can provide broad limits on the
contribution of the DC emission to the total light in line free
continuum bands at optical—IR wavelengths. These are indicated in the
lower half of Table~\ref{model_params} and equate to 32--56\% of the
total light shortward of the Balmer jump and 27--52\% just shortward
of the Paschen jump. Thus even in the presence of significant
microturbulent velocities, the DC contributes a significant fraction
of the total light through the UV—optical.

\subsection{Disc atmosphere's impact on SED near free-bound jumps}
An alternate route for reducing the flux deficit in the vicinity of
the jumps is to posit the presence of a disc atmosphere.  Previous
studies indicate that significant hydrogen, helium and metal line
opacities act to modify the emergent disc spectrum, especially in the
vicinity of the series limits for hydrogen and helium (e.g., Hubeny et
al.\ 2001, their Figure~13). Metal line opacities reduce and flatten
the emergent spectrum through the far-UV relative to that for an MCD
BB and display prominent bound-free edges. The increase in emission
longward of the jump relative to that for a standard MCD BB, may act
to fill-in the emission gap. Such effects are beyond the scope of this
contribution.

\subsection{Energetics}

The UV--optical--IR spectrum of Mrk~110 is far richer in complexity
than is often recognised, with contributions from multiple line series
and free-bound continua of hydrogen and neutral and singly-ionised
helium, many of which are coincident in wavelength (e.g.,
Figure~\ref{plot_global_vt0}).

Also apparent is that a physically motivated spectral decomposition
gives a far better estimate of the underlying disc continuum, and
thereby DC contribution, than studies that attempt to isolate the
latter using a linear fit to designated line-free continuum
bins.\footnote{Searches for DC continuum contributions to the
inter-band continuum delays in other AGN have met with mixed
results. In low-z AGN, the continuum inter-band delays appear to
correlate with the delays of the broad emission-lines (e.g., Wang et
al.\ 2023) as would be expected if the DC makes a significant
contribution to the continuum inter-band delays (Li et al.\ 2021;
Netzer et al.\ 2022). While in high-z samples, no such correlation is
found (Sharp et al.\ 2024). We suggest that part of the reason for the
mixed results is the absence of a physical model for isolating the
underlying disc contribution. A local linear fit to designated "line
free" continuum bands will likely grossly underestimate the DC
contribution.} In the {\it absence of microturbulence}, our best-fit
model indicates a DC contribution which at the location of the Balmer
and Paschen jumps is approximately equal in strength to the underlying
disc emission, similar to that found for models of NGC~5548 (e.g.,
Korista and Goad 2001, 2019). Integrated across a band-pass spanning
1360--3600\AA\ the DC contributes $\approx$ 27\% of the total
(disc$+$DC) continuum emission and exceeds the UV Fe~{\sc ii} emission
over the same band-pass by as much as $\sim$ 50\%.  By contrast,
significant microturbulent velocities increase the line--continuum
contrast suppressing the DC contribution to the total light (as low as
13\% of the total light in the 1360--3600\AA\ band-pass) and due to
the enhanced line strengths, reduces the gas covering fraction
requirements to $<$40\% (see Tables~\ref{model_params}, and
Figure~\ref{plot_global_vt100}).  Moreover, the observed similarity in
shape between the Paschen, Brackett and Pfund DC emission ($\sim
4000$\AA\, -- 2.5$\micron$), and host galaxy contribution, underlines
its thermal nature. Notably, in the presence of a finite disc, the DC
contribution fills in the emission gap near 1 micron, a region in
which disc contributions to the total light are steeply declining, and
dust contributions steeply rising. If host galaxy contributions are
small, there is little else that can flatten the spectrum in this
region.

%/rfs/XROA/mg159/DC_TEMPLATE/FITTING/plot_spec.pdf
\begin{figure}
  \centering
  \includegraphics[width=0.7\columnwidth,angle=270]{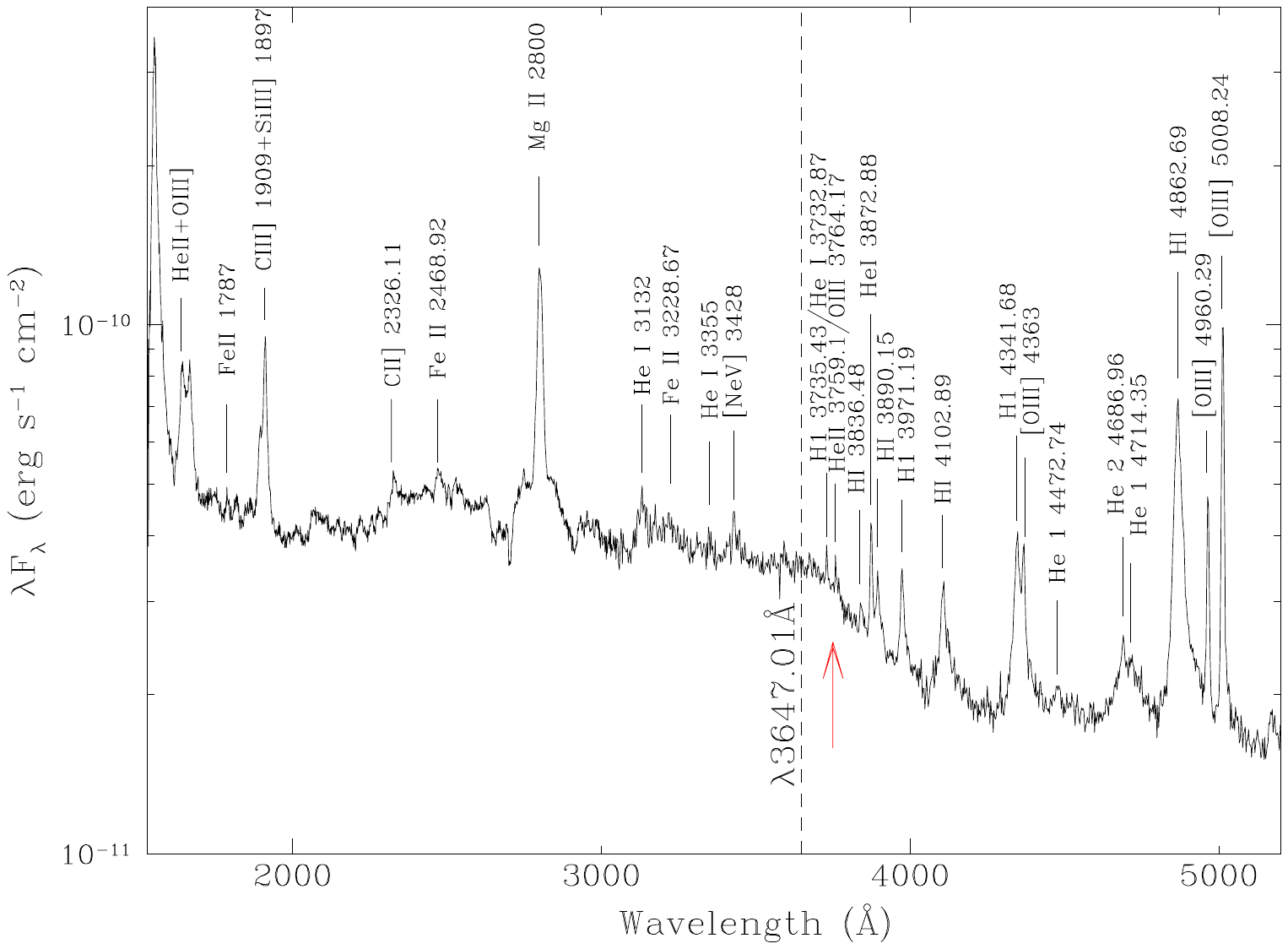}
    \caption{A close up view of the UV-optical spectrum of Mrk~110
      taken with HST/STIS(G230L), illustrating the prominent blend of
      Balmer continuum emission and UV Fe~{\sc ii}, a significant
      Balmer jump with a slow ramp down to longer wavelengths. The
      most prominent emission-lines have been identified. The dashed
      vertical line indicates the vacuum wavelength location of the
      Balmer jump in the low density limit, 3647.01 \AA\/. The
      apparent location of the jump ($\approx \lambda$3750\AA) is
      indicated by the red arrow. Also of note is the Fe~{\sc ii} line
      at $\lambda$1787\AA, a key indicator of microturbulent
      velocities (e.g., Baldwin et al. 1996) or some other
      extra-thermal line broadening mechanism.}
    \label{plot_spec}
\end{figure}

The DC contribution constrained by the emission-line strengths of the
Balmer and Paschen series lines, provides a surprisingly good match to
the data, even though the adopted line profiles -- comprising a
superposition of Gaussian profiles of variable strength and width --
were not modelled on the detailed line shape of individual lines
present in the spectrum. Close up views of selected parts of the SED
can be found in Appendix~\ref{zoom}.  These reveal that the majority
of Balmer and Paschen series lines are well fit by our scaled
DC$+$emission-line template, except in the extended red wing of
H$\beta$ and H$\alpha$ (e.g., Figure~\ref{zero_turb}1, panels 2 and
3), both of which exhibit strong red wing asymmetries. Such line
asymmetries may also be present in the higher order Balmer lines, but
these lines become weaker, broader, and more highly blended as the
order increases, and such low contrast red wing asymmetries cannot
easily be discerned.

\subsection{Limits on the gas density}
Modifications to the hydrogen model atom in the presence of finite gas
densities were included in an effort to address the observed later
than expected decline in the Balmer jump emission at UV
wavelengths. As we have shown (\S2.1), increased gas density reduces
the number of upper electron energy levels available and reduces the
ionization potential, shifting the jump to longer wavelengths. At the
same time, emission lines from higher order transitions are removed,
and their energy redistributed as DC emission just long-ward of the
shifted jump. Summing over a broad range in gas densities then results
in Balmer continuum emission which declines toward longer wavelengths,
with the starting point, termination, and rate of decline, dependent
on the range of gas densities available, and the degree of Doppler
broadening.

%/MG159/PHI_NHPLOTS/ratios_noturb.pdf
\begin{figure}
    \centering
    \includegraphics[width=\columnwidth]{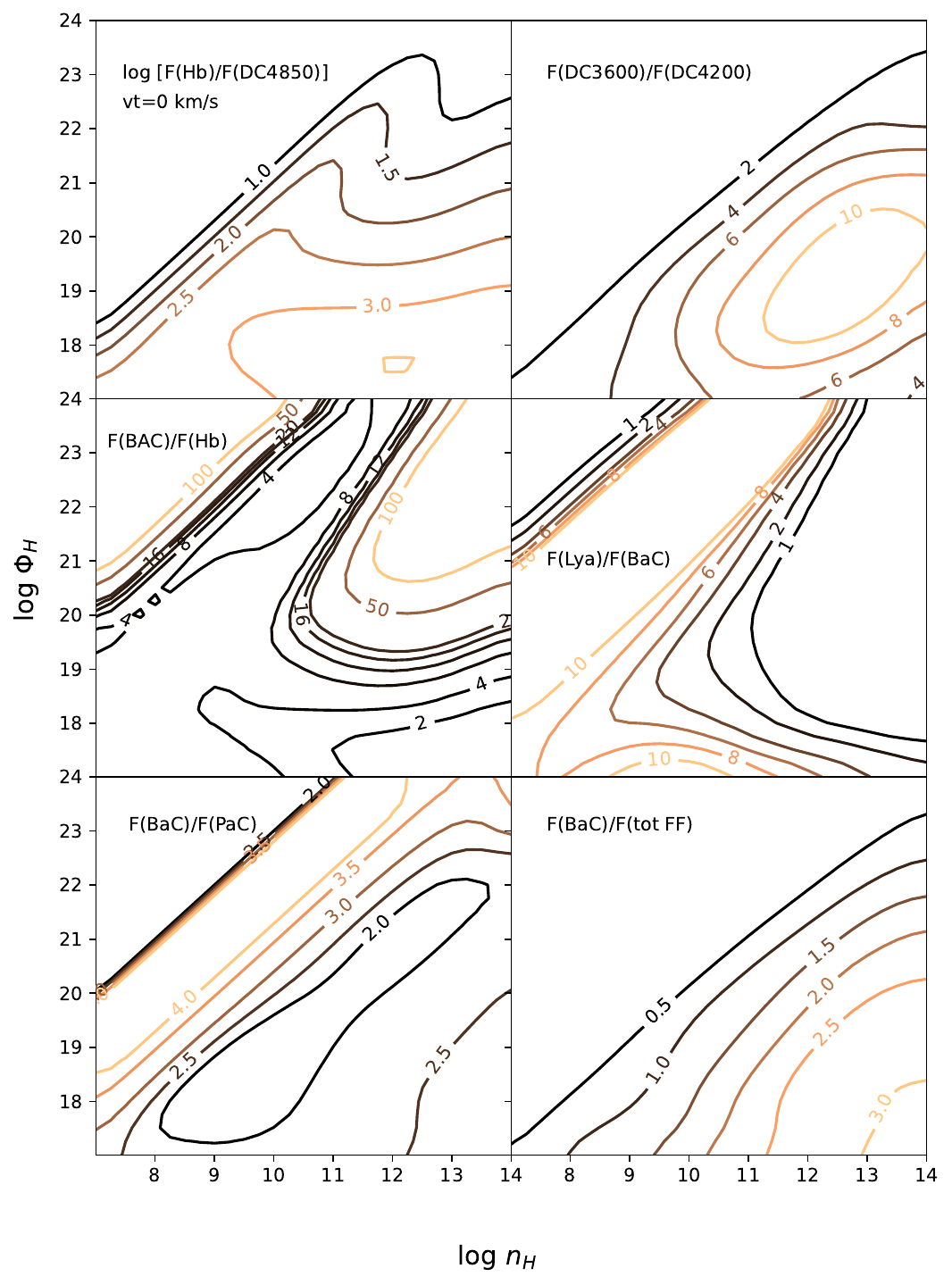}
    \caption{Diffuse continuum diagnostics for standard photoionised
      gas with no microturbulence: Upper left -- the ratio
      F(H$\beta$)/F$_{\lambda\/4850}$(DC); the values in units of
      Angstroms are contoured logarithmically for this panel.  Middle
      left -- the ratio of the integrated Balmer continuum flux to the
      H$\beta$ emission line flux.  Lower left -- the ratio of the
      integrated Balmer to Paschen continuum fluxes.  Upper right --
      the ratio of the diffuse continuum flux ($\lambda F_{\lambda}$)
      at 3600\AA\, to that at 4200\AA.  Middle right -- the ratio of
      Ly$\alpha$ emission line flux to the integrated Balmer continuum
      flux.  Lower right -- the ratio of the integrated Balmer
      continuum flux to that in the free-free continuum.}
    \label{diag_noturb}
\end{figure}

\begin{figure}
    \centering
    \includegraphics[width=\columnwidth]{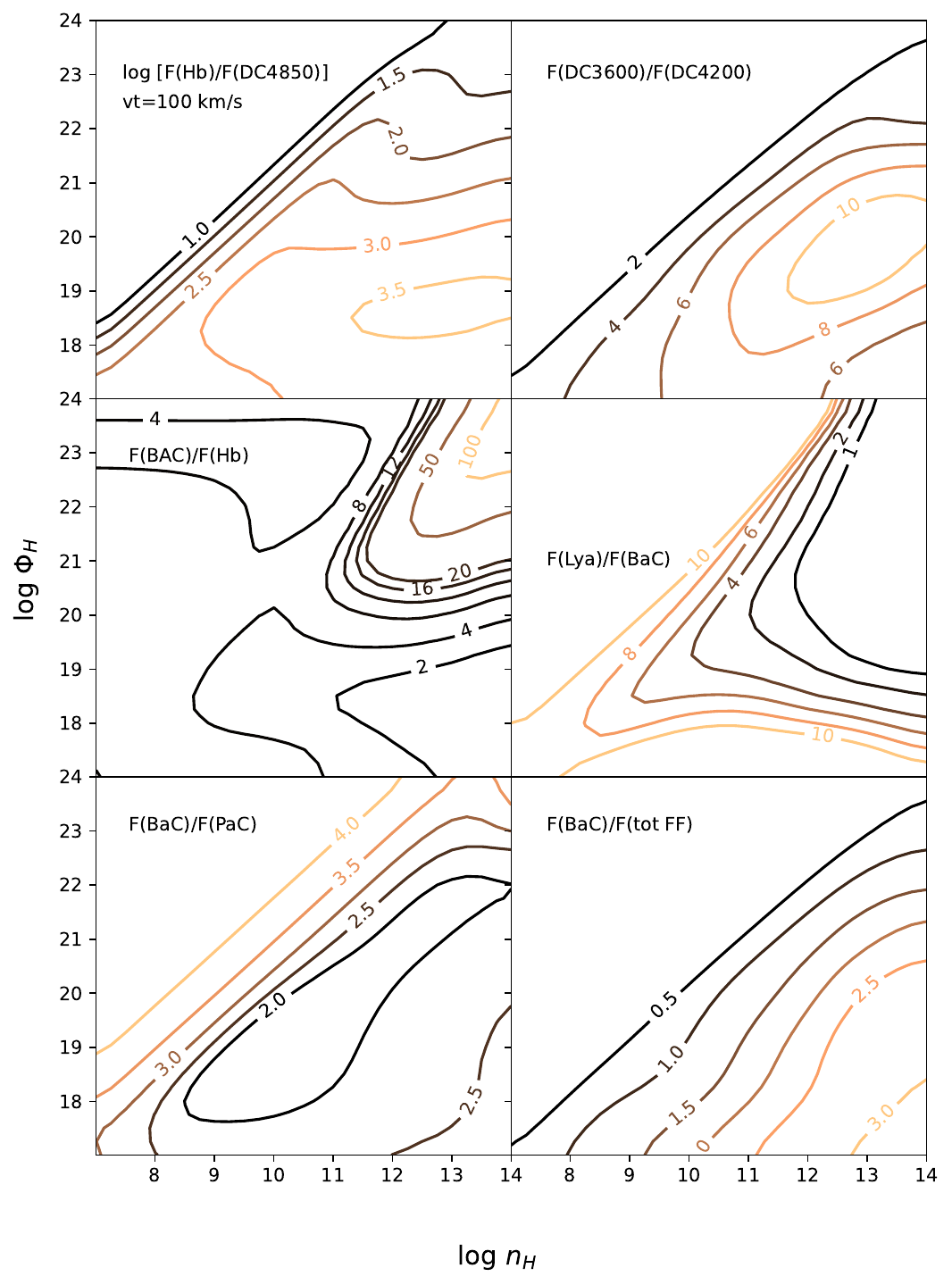}
    \caption{Same as Figure~\ref{diag_noturb} except that the gas has
      microturbulence with a characteristic width of $v_{\rm turb}=
      100$~km~s$^{-1}$.}
    \label{diag_turb}
\end{figure}

However, even at the upper end of gas densities thought appropriate to
the BLR, the shift in location of the Balmer jump, and the pile up of
lines, are insufficient to reproduce the slow ramp down in emission
out to wavelengths approaching 3750\AA\,, as is seen in Mrk~110. To
illustrate the magnitude of the discrepancy, we show in
Figure~\ref{plot_spec} an HST(STIS) spectrum for Mrk~110 spanning the
region of the Balmer jump. To guide the eye, we highlight the expected
location of the Balmer jump for the commonly adopted model atom
(dashed vertical line), and identify many of the major emission
features. As seen in Figure~\ref{plot_spec} and confirmed by our model
fitting (e.g., Figure~\ref{plot_global_vt0}), the turn-down in the
Balmer emission in the vicinity of the jump appears to start at longer
wavelengths, beyond $\approx 3700$\AA\, and there is thus a deficit in
flux in the model fit just long-ward of the predicted location of the
Balmer jump. Moreover, this discrepancy in jump location is not
peculiar to Mrk~110. Many other AGN in which the Balmer continuum is
prominent, display a jump location at wavelengths that are longer than
expected even for gas densities associated with the BLR (see e.g.,
Brown et~al.\ 2019).

One possible solution to this issue is to restrict the range in BLR
gas densities available. However, while greater contributions by high
density gas can shift the jump location to longer wavelengths, the
height of the jump tends to be largest at the highest gas densities
(e.g., Figure~\ref{plot_redist_ba}), and the number of available
levels is much diminished, reducing the number of lines contributing
to the pileup long-ward of the jump. The net effect is a large jump
but with a more gradual decline (e.g., lower right panel of
Figure~\ref{plot_redist_ba}). Although we cannot exclude the presence
of even higher density gas, for the Balmer continuum emission to reach
wavelengths as large as 3700\AA, would require gas densities as large
as $n_{\rm H}\sim 5\times 10^{15}$~cm$^{-3}$. Such gas will contribute
little to the observed line emission, and therefore is not associated
with the BLR, but if present will act to reduce the DC contribution
from the BLR.

Although it is possible that the line strengths of the higher order
lines have been underestimated, where such lines can be clearly
identified, our model fit performs well in simultaneously matching
their observed line strengths\footnote{We note that {\sc cloudy}
photoionisation predictions for the emission-line strengths of
H$\gamma$ and Pa$\beta$, lines that originate from the same upper
level (n=5), appear anomalously high with respect to the general trend
in decreasing line strength with increasing order.}.  Thus we have no
clear evidence that the predicted line strengths of higher order lines
from extrapolation of the general trend in line strength with
increasing order are erroneous.

Finally, we note that the inclusion of somewhat higher density gas in
our LOC model (e.g., $n_{\rm H} = 10^{13}$~cm$^{-3}$) would enhance
the diffuse continuum relative to the hydrogen (and most other) line
emission. Thus in the presence of such dense gas the DC from the BLR
would be substantial even in the presence of microturbulence.

\subsection{Diagnostic flux ratios to assist in spectral fitting of UV--near-IR AGN SEDs}

As a service to the community we show in
Figures~\ref{diag_noturb}--\ref{diag_turb} diagnostic flux ratios
which can be used to estimate the contribution of the BLR diffuse
continuum emission in spectral decompositions of the UV--Optical--IR
spectra of AGN.  These include: the ratio of the H$\beta$
flux\footnote{We caution here that the hydrogen Balmer and Paschen
line emission from dense, optically thick clouds in the broad-line
regions of AGN historically have been systematically under-predicted
by modern photoionisation codes (e.g., Netzer 2020), although we would
estimate factors $<$~2 when including gas densities up to
$10^{12}-10^{13}$~cm$^{-3}$. We note that for the model described
here, the flux ratio F(Ly$\alpha$)/F(H$\beta$) lies between 17--21,
compared to a measured flux ratio of $\approx 21$ (e.g.,
Table~\ref{mega_tab}), suggesting in this case more than sufficient
Balmer line emission.} to the value of the specific flux in the
diffuse continuum (F$_{\lambda\/4850}$) just beneath (upper left), the
ratio of the diffuse continuum flux for two representative continuum
bands (3600\AA\ and 4200\AA) bracketing the Balmer jump (upper right),
the ratio of the integrated Balmer continuum (0.25--1~Ryd) to the
H$\beta$ flux (middle--left), the ratio of the Ly$\alpha$ flux to that
in the integrated Balmer continuum (middle--right), the ratio of the
integrated Balmer to Paschen (0.11--0.25~Ryd) continuum fluxes
(bottom--left), and the ratio of the integrated Balmer continuum to
integrated free-free emission (lower--right). These diagnostic flux
ratios are presented for two cases: standard photoionised ``clouds''
with zero microturbulence (Figure~\ref{diag_noturb}) and those with
microturbulent velocities set to 100~km~s$^{-1}$
(Figure~\ref{diag_turb}). These likely span the range of extrathermal
local line widths that may affect the emitted BLR spectra of
AGN\footnote{The flux ratio contours for models with log(U$_{\rm H}$c)
= log($\Phi_{\rm H}$/$n_{\rm H}$) > 12~cm~s$^{-1}$ may be ignored, as
this gas is overionised (and those clouds near this boundary are
likely thermally unstable) and contributes little to the observed
broad emission line spectra of AGN in any case. We also note that gas
clouds with properties similar to those toward the lower right corner
of the density--ionizing photon flux plane (very low U$_{\rm H}$) are
also unlikely to exist, emitting {\em strongly} in Na~{\sc i}
(5891~\AA\/, 5897~\AA\/), Ca~{\sc ii} (3934~\AA\/, 3969~\AA\/,
8500~\AA\/, 8544~\AA\/, 8664~\AA\/), O~{\sc i} (8446~\AA\/), and in
the UV resonance lines of Fe~{\sc ii}.}.

For example, by measuring the integrated flux of broad H$\beta$
following removal of the narrow line component, and then referring to
the top--left panels of Figures~\ref{diag_noturb}, \ref{diag_turb} the
specific flux level in the diffuse continuum F$_{\lambda}(\rm{DC})$
beneath H$\beta$ may be constrained. As the 4000--8000\AA\/ diffuse
continuum is expected to be fairly constant in F$_\lambda\/$
vs.\ $\lambda$, then an estimate for the specific flux in the diffuse
continuum at 3600~\AA\/ (and thus the magnitude of the Balmer jump)
may then be bounded using the information within the upper--right
panels of the two figures.  Since the EW contours for larger
microturbulent velocities shift upwards on the flux density plane, the
fiducial cloud ($\log \Phi_{\rm H}$ $(\rm {photons~cm^{-2}~s^{-1}})$ =
19.5, $\log n_{\rm H}$ (cm$^{-3}$) = 11.0) gives a larger value for
this ratio in the presence of significant microturbulence, counter to
what is seen in the LOC composite-cloud DC spectrum, highlighting the
difference between single cloud and ensemble model predictions.

\begin{figure}
    \centering
    \includegraphics[width=\columnwidth]{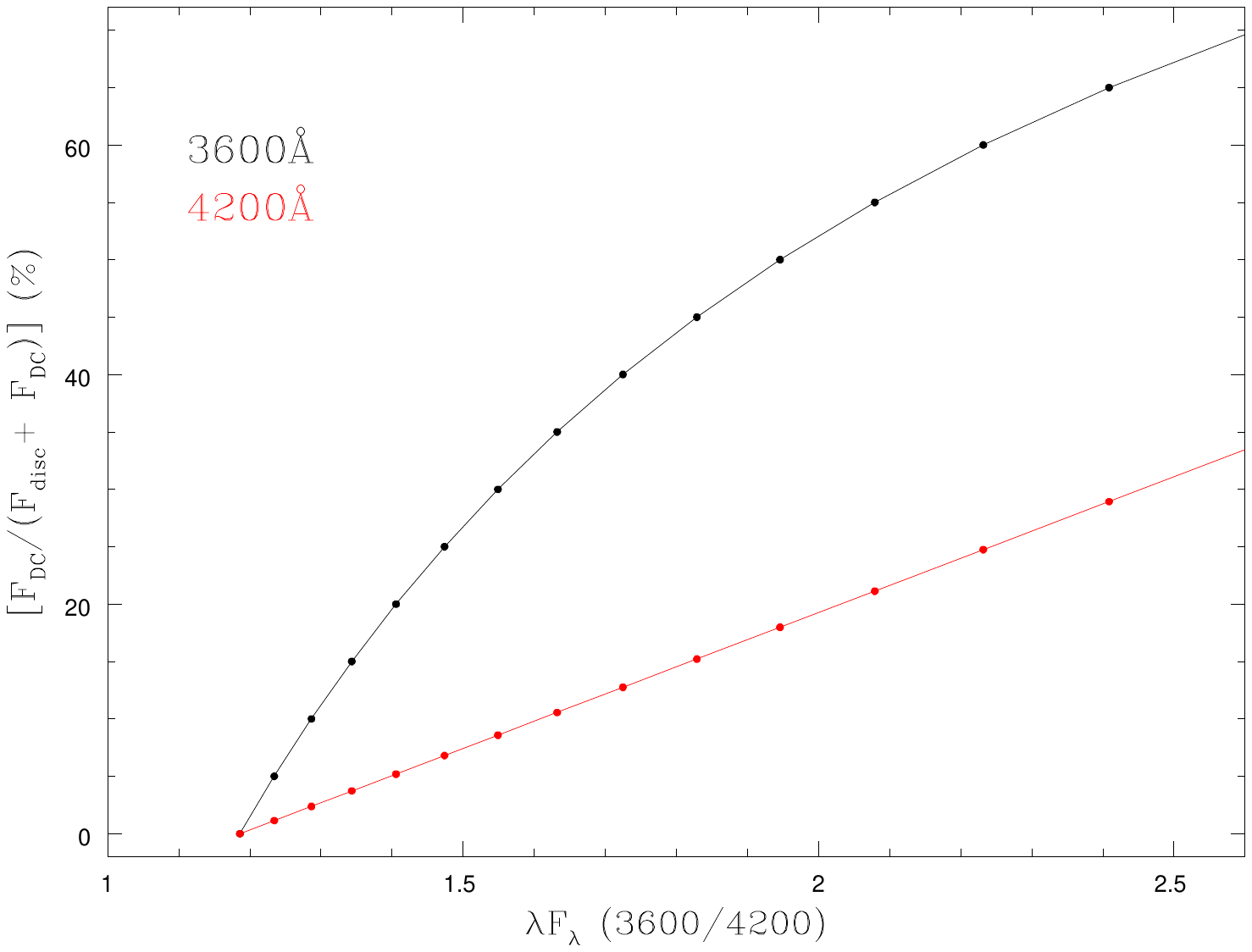}
    \caption{The predicted DC percentage contribution to the total
      nuclear light [$F_{DC}/(F_{disc}+F_{DC})$] determined at
      wavelengths of 3600\AA\, (black) and 4200\AA\, (red) as
      functions of a measured flux ratio $\lambda
      F_{\lambda3600}$/$\lambda F_{\lambda4200}$ in the total nuclear
      light (disc$+$DC).}
    \label{frac_dc}
\end{figure}

A related ratio of fluxes $\lambda F_\lambda\/$ at 3600\AA\/ and
4200\AA\/ is one measured in {\em total nuclear} light, after
correcting for host galaxy contribution. Fe~{\sc ii} and other line
emission are expected to be weak at both locations, and so this flux
ratio should be a reasonably good, relatively model-independent,
measure of the strength of the Balmer jump along with the overall
strength in the diffuse continuum arising from the BLR. We plot in
Figure~\ref{frac_dc} the expected percentage contribution of the
BLR-dominated DC to the total light at these two wavelengths as
functions of a measured nuclear flux ratio $\lambda
F_{\lambda3600}$/$\lambda F_{\lambda4200}$. The behaviour of the BLR
DC contribution functions at both wavelengths is nearly identical for
both sets of assumptions about the magnitude of the microturbulent
velocity.  For comparison, the flux ratio expected from the underlying
central thermal continuum source will have values {\em less than}
$(3600/4200)^{-4/3} \approx 1.23$. Slightly larger values might be
incurred in the presence of local heating above the standard viscous
one. Our accretion disc spectral models have a flux ratio value of
just under 1.2, as can be seen in Figure~\ref{frac_dc} where the DC
contribution goes to zero. Thus, {\bf a measured flux ratio value in
  signifiant excess then indicates a significant contribution of the
  BLR DC to the UV--Optical spectrum.} In the case of Mrk~110, the
measured flux ratio of 1.9 from Figure~\ref{plot_spec} suggests a
contribution of the DC from the BLR to the total nuclear light near
the Balmer jump and 4200~\AA\/ of at least 40\% and 15\%, respectively
(see e.g. Figure~\ref{frac_dc}), allowing for potential small
contributions ($\sim$ few percent) from weak BLR line emission and
other light sources including scattered light from the torus plus the
thermal continuum from the NLR (e.g., Figures \ref{plot_global_vt0},
\ref{plot_global_vt100}).  If the measured flux ratio value (1.9) is
dominated by DC emission from the BLR plus the underlying thermal
continuum, then comparison to the values noted in
Table~\ref{model_params} for our two spectral models of Mrk~110 argues
for BLR microturbulent velocities substantially lower than 100
km~s$^{-1}$.

%/scratch/ngts/mg159/OPTIMIZERS/TURB/AMOEBA/ANCHOR/NOCOL/ONCE/plot_frac

Next, the two middle panels then gauge the strength of the integrated
Balmer continuum, based on measurements of the broad H$\beta$ and
Ly$\alpha$ fluxes. Finally, the strength of the Paschen and free-free
continua relative to the Balmer continuum may be gauged from the lower
two panels. We suggest that the coordinates (log~n$_H = 11$,
log~$\Phi_H = 19.5$, near log~U$_H \approx -2$) might be
characteristically representative of the composite {\em thermal}
diffuse continuum emission from the gas within the broad emission line
regions of AGN (Table~\ref{fiducial}).

\begin{table}
    \centering
    \begin{tabular}{lr|lr}
 &     & \\ \hline
\multicolumn{4}{c}{$v_{\rm turb}=0$~km~s$^{-1}$}  \\ \hline
$\log$[F(H$\beta$)/F(DC4850)] (\AA) &  2.53  &  F(DC3600)/F(DC4200) & 7.64 \\  F(BaC)/F(H$\beta$) & 16.88   & F(Ly$\alpha$)/F(BaC) & 1.05 \\   
F(BaC)/F(PaC)      & 1.87   &  F(BaC)/F(totFF) & 1.44 \\   
& & & \\ \hline
\multicolumn{4}{c}{$v_{\rm turb}=100$~km~s$^{-1}$}  \\ \hline
$\log$[F(H$\beta$)/F(DC4850)] (\AA) & 3.13   & F(DC3600)/F(DC4200) & 8.17\\   
F(BaC)/F(H$\beta$) & 4.07   &  F(Ly$\alpha$)/F(BaC) & 2.29\\   
F(BaC)/F(PaC) & 1.70   &  F(BaC)/F(totFF) & 1.42\\   

\hline \end{tabular}
    \caption{Diagnostic ratios for a fiducial cloud model with parameters $\log \Phi_{\rm H}$ $(\rm {photons~cm^{-2}~s^{-1}})$ = 19.5, $\log n_{\rm H}$(cm$^{-3}$) = 11.0.}
    \label{fiducial}
\end{table}

Guided by the above diagnostic flux ratios (Figures~\ref{diag_noturb}
and \ref{diag_turb}), we suggest that spectroscopic analyses of UV to
near-IR spectra of AGN should adopt a sum over
hydrogen\footnote{Hydrogen free-bound emission will dominate over that
of neutral and singly-ionized helium, which often have nearly
coincident wavelength thresholds as those in hydrogen in any case.}
free-bound continua using a characteristic value in electron
temperature (e.g., T$_{\rm fb} \approx 12000$~K) and a modest
characteristic optical depth ($\tau \approx 0.3$) at the hydrogen
Balmer threshold\footnote{Photoionization models suggest smaller
threshold optical depths for the Paschen, Brackett and Pfund continua,
$\approx 0.6 \tau(\rm{BaJ})$.}. A free-free continuum with a somewhat
higher characteristic electron temperature (e.g., T$_{\rm ff} \approx
19000$~K) should also be included; it dominates over free-bound
continuum emission longward of the Paschen jump. Its integrated flux
in a composite BLR DC spectrum should be comparable to that in the
Balmer continuum (Figures~\ref{diag_noturb}, \ref{diag_turb}). See
Grandi (1982) and Draine (2011) for discussions of how these may be
implemented within a spectral fit. The inclusion of free-electron and
neutral hydrogen scattering contributions is important to the diffuse
continuum from the BLR at wavelengths mainly shortward of
$\sim$2000~\AA\/. However, even here the scattered light is likely to
be generally modest, a few to $\sim$10\%, in contribution. The neutral
hydrogen scattering feature may manifest in extended wings centred on
the Ly$\alpha$ broad emission line.  We note that the above parameters
result in a surprisingly good fit to the DC template presented in
KG19.

Comparing Figures~\ref{diag_noturb}, \ref{diag_turb}, it is clear that
the primary impact of introducing extra-thermal line widths to the
line emitting gas is to increase the emission line to diffuse
continuum flux ratio. Thus, to match a measured amount of broad
emission line flux, the underlying diffuse continuum will be weaker in
the presence of increasing local line widths above thermal. It has
little impact on the overall SED shape or on relative contributions of
various emission processes to the diffuse continuum.

In summary, the presence of Paschen and higher-order free-bound
continua as well as a free-free continuum are required elements in
addition to the Balmer continuum. The strength and overall shape of
the diffuse continuum emanating from the BLR {\it are not
  arbitrary\/}, but must be related to the strengths of the observed
broad emission lines. Additionally, the magnitude of its presence is
in part revealed by a measurement of the nuclear flux ratio $\lambda
F_{\lambda3600}$/$\lambda F_{\lambda4200}$.

\section{Summary and conclusions}

A major goal of this work was to provide an improved broad
emission-line region diffuse continuum template, for use in spectral
decompositions of the UV-Optical-IR flux spectra and
wavelength-dependent delay spectra of AGN, and in that respect we have
been largely successful. An accurate DC template represents a critical
first step toward quantifying the BLR DC contributions to the measured
interband continuum delays in disc reverberation mapping experiments.
Our new and improved LOC DC template accounts for the finite range in
gas densities likely to be present within the BLR, and includes
contributions from both the major optical-IR hydrogen recombination
emission lines, and the pile-up of higher order lines longward of
their respective jumps emitted from the same gas. By fitting model
emission line strengths to observations, we can provide limits on the
BLR gas covering fraction, the DC contribution to the flux spectrum,
and its impact on measured interband continuum delays.

Our spectral modelling (comprising model and empirical templates) was
deliberately focused on Mrk~110, a type 1 AGN with emission-lines that
are sufficiently narrow that nearly line-free continuum bands can be
readily identified. However, given the likely lack of sensitivity of
the DC emission to the incident ionizing spectrum, {\em such templates
  may be usefully applied to spectral decompositions of other AGN.}

Key results from this study are as follows:

\begin{enumerate}
    \item{The inclusion of finite gas densities within the BLR reduces
      the number of electron energy levels available in the model
      atom, lowering the ionization potential and thereby shifting the
      location of the prominent Balmer and Paschen jumps to longer
      wavelengths. This shift in wavelength is a factor $\sim$5.1
      larger for the Paschen jump than the Balmer jump.}
    \item{The reduction in available bound electron energy levels with
      increasing gas number density reduces the number of higher order
      lines contributing to the pile-up longward of the jump.}
    \item{The lower statistical weights of energy levels near a
      discrete series threshold, reduces the emission resulting from
      discrete transitions with the emission appearing instead as
      recombination continuum emission.}
    \item{The net effect of (i)--(iii) is a shift in the location of
      the major jumps, and a shallower (rather than abrupt) decline in
      the recombination continuum contributions at wavelengths
      longward of the jump as is observed.}
    \item{The shape of the jump may be further modified by bulk motion
      (ie., Doppler broadening) of the gas, and our model includes the
      effects of Doppler broadening on both lines and continua, with
      the degree of broadening determined by their respective mean
      formation radius.}
    \item{The BLR diffuse continuum, which at wavelengths longward of
      $\sim2000$\AA\, is dominated by thermal emission, acts to
      flatten the SED from the near-UV to the near-IR (e.g., relative
      to that expected from a pure thermal disc spectrum), negating
      the requirement for significant intrinsic reddening within the
      host AGN.}
    \item{In the presence of a finite disc and a weak host galaxy
      contribution, the absence of a significant DC component results
      in a substantial emission gap at 1 micron. Hence, if DC
      contributions are not correctly accounted for, contributions
      from other components, e.g., the hot dust, may be significantly
      overestimated.}
    \item{The ratio of the total nuclear light ($\lambda
      F_{\lambda3600}/\lambda F_{\lambda 4200}$), provides a good
      indication of the strength of the DC contribution.}
\end{enumerate}

While the spectral decomposition of Mrk~110 provides a reasonable fit
to the data and useful limits on the DC contribution to the flux and
delay spectrum, the deficit in flux just longward of the Balmer jump
remains problematic. Microturbulent velocities, which enhance the
line--continuum contrast mitigates some, but not all, of the observed
deficit, while at the same time, significantly reducing the DC
contribution. {\bf The measured ratio of the total nuclear light
  $\lambda F_{\lambda3600}/\lambda F_{\lambda 4200}$ of 1.9 in
  Mrk~110's spectrum would suggest a substantial contribution by the
  BLR DC.} We further speculate that the presence of a disc atmosphere
and/or a hydrogen bound-free opacity feature in the disc spectrum may
act to fill in some of the remaining emission gap.

Finally, we note that current theoretical models of UV Fe~{\sc ii} do
not alleviate this problem, requiring significant enhancement of the
predicted Fe~{\sc ii} contributions in the vicinity of the Balmer
jump. This may point to current limitations in specifying the numerous
energy levels present in the model atom for Fe used in current
photoionisation calculations, which we here suggest may be a fruitful
avenue for further progress.

%%%%%%%%%%%%%%%%%%%%%%%%%%%%%%%%%%%%%%%%%%%%%%%%%%
\section*{Data Availability}

The publicly available data described here may be obtained from the
MAST archive at \url{https://dx.doi.org/10.17909/t9-3dbt-8734}.  Data
generated in support of this research will be shared on reasonable
request to the corresponding author.

\section*{Acknowledgements}
This research is based in part on observations made with the NASA/ESA
Hubble Space Telescope obtained from the Space Telescope Science
Institute, which is operated by the Association of Universities for
Research in Astronomy, Inc., under NASA contract NAS 5–26555. These
observations are associated with programs PID~15699 (Costantini),
PID~15413 (Cackett).

We thank the referee for providing comments and suggestions which led
to improvements in clarity of the work presented here. We thank Gary
Ferland for his continued development and support of the
photoionisation code, {\sc cloudy}.



%%%%%%%%%%%%%%%%%%%%%%%%%%%%%%%%%%%%%%%%%%%%%%%%%%

%%%%%%%%%%%%%%%%% APPENDICES %%%%%%%%%%%%%%%%%%%%%

\appendix

\section{Isolating the broad and narrow emission-line contributions}

\label{decomp}
To create a template profile for broad H$\beta$ we first isolate this
component by fitting the region in the vicinity of this line with
multiple Gaussians, having first subtracted a linear fit to the local
background continuum, using line free continuum regions blue-ward of
broad He~{\sc ii}~$\lambda4686$\AA\, and red-ward of the broad He~{\sc
  i} line at $\lambda$5876\AA. We iteratively fit Gaussians to the
emission lines, starting with the strongest components, only adding
more components if required by the data. We use the strong narrow
[O~{\sc iii}]~$\lambda$5007\AA\, line as a template for isolating the
narrow line contributions to H$\beta$ and He~{\sc
  ii}~$\lambda$4686\AA, fixing the strength of narrow H$\beta$ to 10\%
of [O~{\sc iii}]~$\lambda$5007\AA, as appropriate for narrow line
region conditions in AGN, and keeping their widths tied.

Our fit includes the two He~{\sc i} lines at $\lambda$4923.31\AA\, and
$\lambda$5017.08\AA\, (vacuum wavelengths), with their strengths tied
to 20\% of the intensity of the longer wavelength He~{\sc i}
$\lambda$5876\AA\, line, and widths scaled to be $\approx$30\% larger,
in accordance with their smaller mean formation radius, and assuming a
virialised velocity field. We also include the Na~D lines (vacuum
wavelengths $\lambda$5891.58\AA\, and $\lambda$5897.56\AA\,), with
their intensities and wavelength separations tied. Their intensities
we fix to 7.7\% of He~{\sc i}~$\lambda$5876\AA\,, and their widths are
set equal to that for He~{\sc i}~$\lambda$5876\AA\,, in accordance
with predicted LOC model line intensities and mean formation radii.

\begin{figure}
    \centering
    \includegraphics[width=\columnwidth]{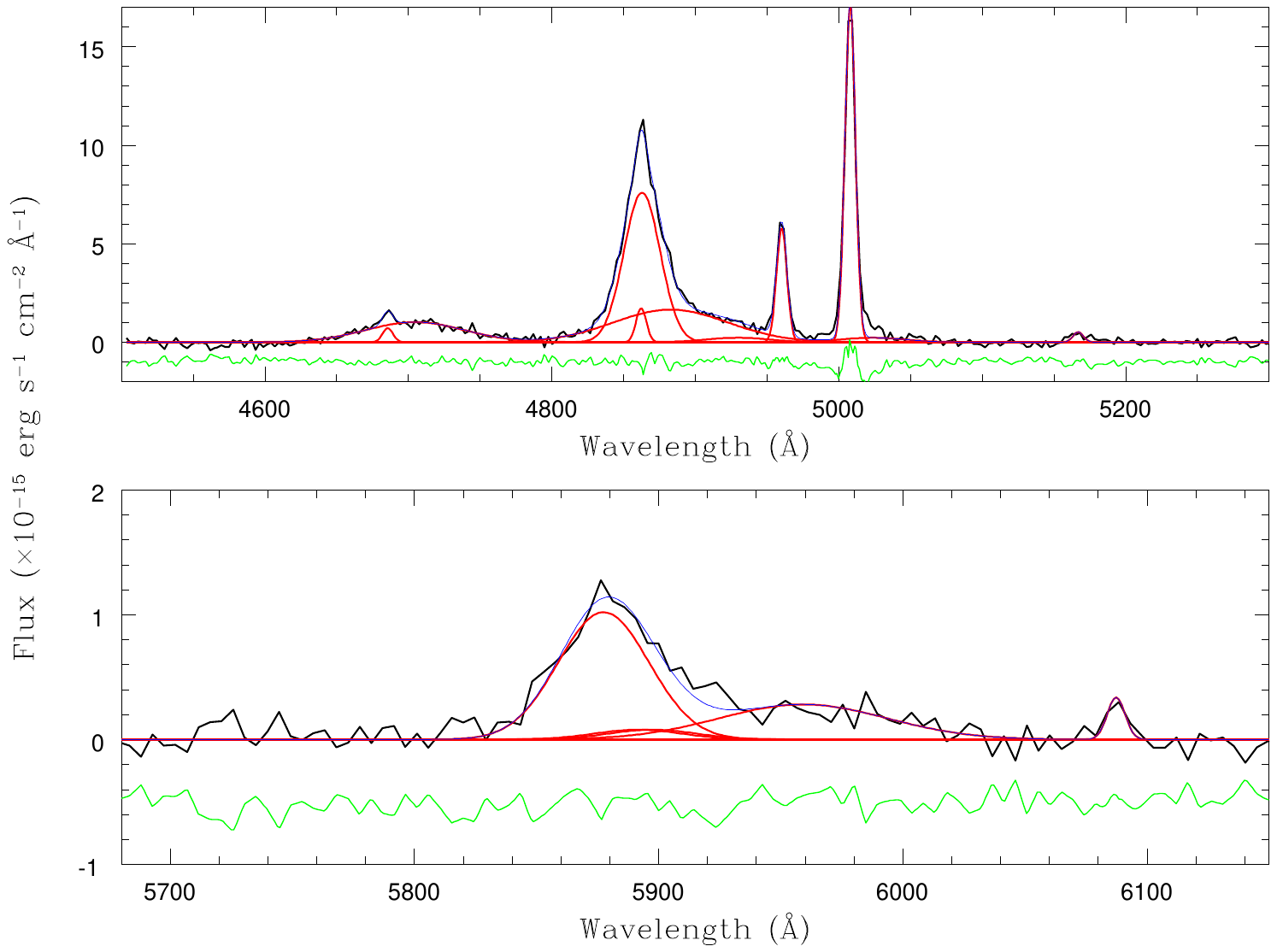}
    \caption{Model fit to the continuum subtracted spectrum of Mrk~110
      in the vicinity of broad H$\beta$. The spectrum has first been
      shifted to the vacuum wavelengths based on a measurement of the
      line centre of the strong forbidden [O~{\sc iii}] line at
      5008.24\AA. Individual components, here modelled as Gaussians
      are illustrated in red, the summed fit in blue. Residuals
      (data$-$model) are shown in green and offset for clarity. The
      broad He~{\sc i} lines ($\lambda_{\rm vac}$=4923.31\AA,
      $\lambda_{\rm vac}$=5017.08\AA) are tied in wavelength and their
      flux tied to $\approx$20\% of the He~{\sc i} 5876\AA\, line
      flux. Narrow emission line contributions are tied in width using
      the width of the longer wavelength component of [O~{\sc iii}],
      with their strengths free to vary. For narrow H$\beta$ we
      additionally fix its strength to 10\% of narrow [O~{\sc
          iii}]~$\lambda$5007, a compromise between photoionisation
      model predictions (F(H$\beta$)$\sim 5$\% F([O~{\sc
          iii}]~$\lambda 5007$), and results from model fits in which
      this component is left free to vary (F(H$\beta$) $\approx 14$\%
      F([O~{\sc iii}]~$\lambda 5007$). We do not include optical
      Fe~{\sc ii} in the fitting process as it is known to be weak in
      this source.  For the fit to the longer wavelength He~{\sc i}
      5877.26\AA (vacuum) spectral region, we include a contribution
      from the Na D doublet at 5891.58, 5897.56\AA\, (vacuum). Their
      strengths are here set to 7.7\% of the He~{\sc i} line
      intensity, and their widths are found to be similar to He~{\sc
        i}~5876\AA\, , consistent with their similar mean formation
      radius ($\approx$50 light-days) found from our LOC model
      integrations.  }
    \label{fit_jul_06}
\end{figure}

Fitting a single broad component to broad H$\beta$ results in
significant residuals in the wings of broad H$\beta$.  Our preferred
fit requires an additional broad component which significantly reduces
the residuals in the broad wings of H$\beta$ (see
Figure~\ref{fit_jul_06}, Table~\ref{mega_tab}, for
details). Individual components are shown in Figure~\ref{fit_jul_06}
along with the summed fit (blue) and residual spectrum (green), the
latter offset for clarity. The integrated flux in broad H$\beta$ (the
sum of components 4 \& 5) is $4.056\times
10^{-13}$~erg~cm$^{-2}$~s$^{-1}$, and the profile indicates an
extended red tail. While some portion of the red tail may be
attributed to Fe~{\sc ii}, mainly m42, $\lambda$4924\AA\, and
$\lambda$5018\AA\,, this pair of emission lines must be very weak in
this AGN, given the weakness in the third transition of m42,
$\lambda$5169\AA. We remark also that a similar (though relatively
weaker) extended red-wing is also seen in the profile of broad
H$\alpha$. Thus, we have also investigated using the broad H$\alpha$
profile as a template for broad H$\beta$. However, the red wing in
broad H$\alpha$, though extended, is much weaker than that in broad
H$\beta$, and consequently, fitting the H$\alpha$ template to broad
H$\beta$ while able to fit the core of the profile, does not
significantly reduce the fit residuals in the red wing. Similarly,
H$\delta$ has broader wings compared to H$\beta$ (as predicted by
photoionisation calculations, e.g., Korista and Goad 2004).
Regardless of the exact form adopted for the broad H$\beta$ template
the sum of the broad and semi-broad components
(Figure~\ref{fit_jul_06}) used in the construction of a template
profile, is for illustrative purposes only.

\subsection{Intrinsic reddening}
The Balmer decrement (H$\alpha$/H$\beta$) estimated via spectral
decomposition and often used as a reddening indicator when compared to
Case B values, is here measured to be $< 4.6$ for Mrk~110. However,
photoionisation models of BLR clouds are far from Case B. In
particular, our LOC model predictions {\it in the absence of any
  intrinsic reddening\/} predict a Balmer decrement of $\approx 4$,
close to the value estimated above.

%/rfs/XROA/mg159/DC_TEMPLATE/FITTING/FIT_MULTI/DISC_FIT/MKN110/SA/plot_ha
\begin{figure}
    \centering
    \includegraphics[width=\columnwidth]{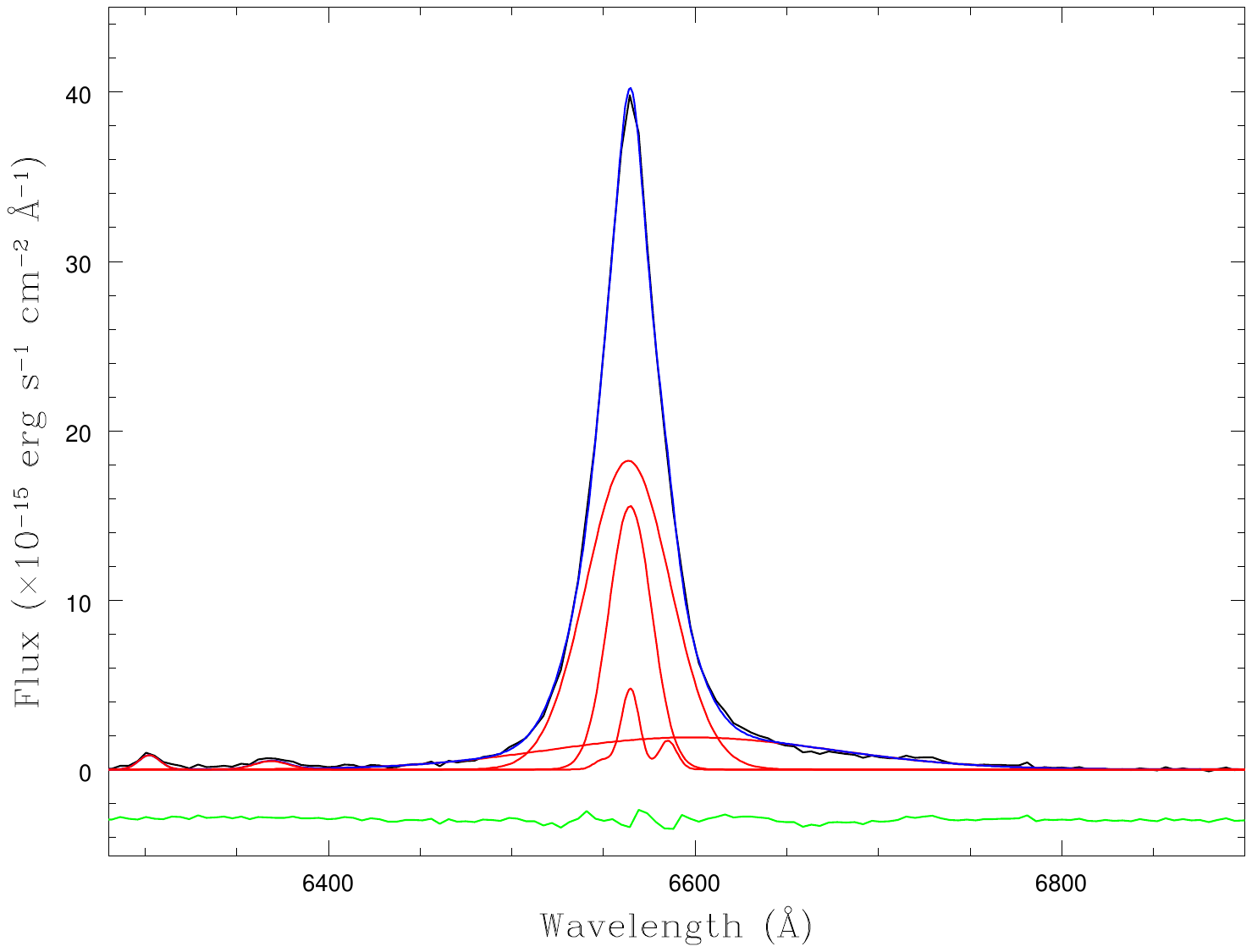}
    \caption{Spectral decomposition of the optical spectrum of Mrk~110
      in the vicinity of H$\alpha$. The fit includes a narrow
      H$\alpha$ $+$ N~{\sc ii} template. Individual components are
      shown in red, the summed fit is shown in blue, and fit residuals
      in green. The red wing evident in H$\beta$ and He~{\sc ii} 4686
      though present, is much weaker in H$\alpha$ in a relative sense
      than in H$\beta$.}
    \label{fit_ha}
\end{figure}

%/rfs/XROA/mg159/DC_TEMPLATE/FITTING/FIT_MULTI/DISC_FIT/MKN110/AMOEBA/plot_fit_c4
\begin{figure}
    \centering
\includegraphics[width=\columnwidth]{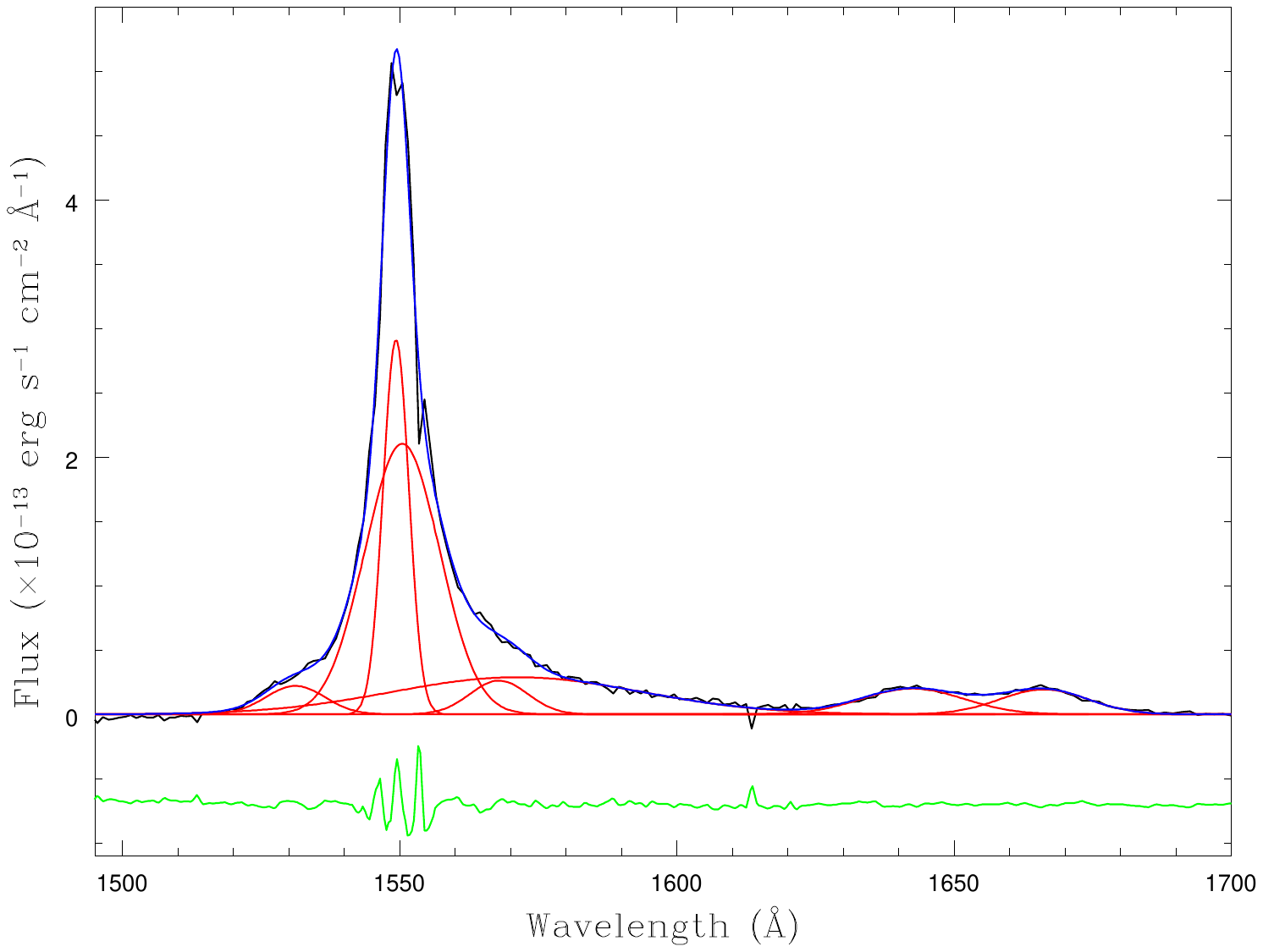}
    \caption{Spectral decomposition of the continuum subtracted
      HST/COS spectrum of Mrk~110 in the vicinity of C~{\sc iv} 1549.
      Individual components are shown in red, the summed fit is shown
      in blue, and fit residuals in green.}
    \label{fit_c4}
\end{figure}

\section{UV--optical line series for He~{\sc ii}}
\label{he2}
In Figure~\ref{he2_form} we show the predicted luminosities for the
He~{\sc ii} emission-line series, for constant density slices through
the flux--density plane (\S2.6).

\setcounter{figure}{0}

\begin{figure}
    \centering
    \includegraphics[width=\columnwidth]{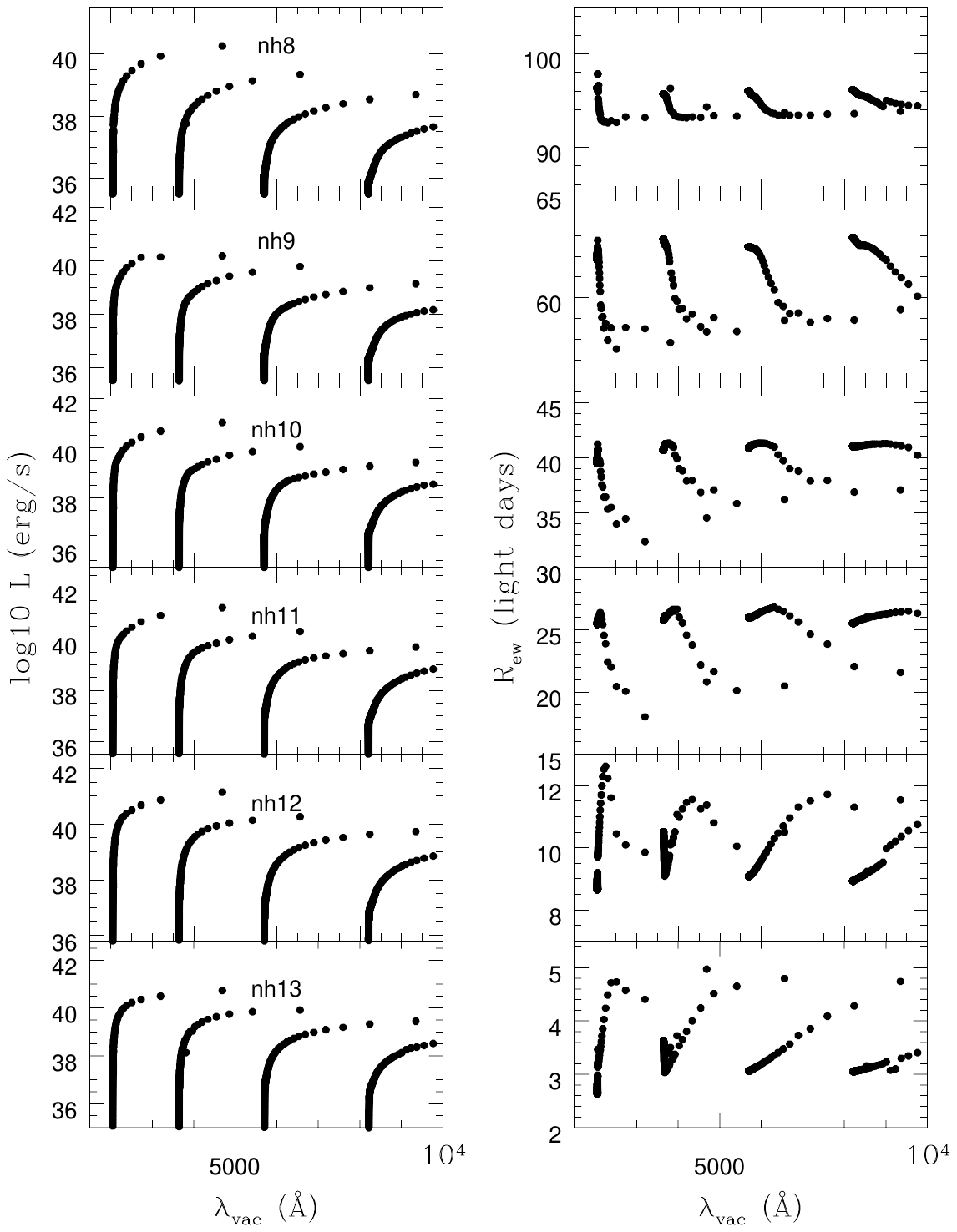}
%    {Figures/plot_L_R_jul7_he2_extrap.pdf}
    \caption{Left - Predicted luminosities for the He~{\sc ii}
      emission-line series (transitions down to n=3,4,5 and 6), as
      determined for constant density slices through the flux-density
      plane. From top to bottom - $\log~ n_{\rm H}({\rm
        cm}^{-3})=8,9,10,11,12,13$. Right - corresponding
      emissivity-weighted radii (light-days). Emission-line
      luminosities and emissivity-weighted radii for the higher order
      He~{\sc ii} lines for densities in the range $8 \leq \log n_{\rm
        H}$ (cm$^{-3}$)~$ \leq 13$, are here determined by
      extrapolation. Line luminosities decrease with increasing upper
      level. Emissivity-weighted radii show a more complex behaviour,
      with mean formation radii generally increasing with increasing
      {\it n} for low gas densities, and generally decreasing with
      increasing {\it n} for high gas densities.  NB The intensities
      of the higher order lines are determined first by extrapolation
      up to the series limit, and then down-weighted in accordance
      with the scheme described in \S2.1.}
    \label{he2_form}
\end{figure}

\section{The effect of turbulence on spectral decompositions}
\label{zoom}

In Figures~\ref{zero_turb}1--\ref{turb_100}2 we show close-up views of
the spectral decomposition of Mrk~110 for vt=0 km~s$^{-1}$ and vt=100
km~s$^{-1}$ microturbulent velocities, so that they may be directly
compared. In both cases no colour temperature correction has been
applied to the MCD BB emission.

\onecolumn
\begin{table}
    \centering 
    \begin{tabular}{rrrrr}
    \hline
    \multicolumn{1}{c}{Component} &
    \multicolumn{1}{c}{Central Wavelength} & 
    \multicolumn{1}{c}{ FWHM } & 
    \multicolumn{1}{c}{ Peak Flux ($\times 10^{-13}$) } & 
    \multicolumn{1}{c}{ Flux ($\times 10^{-12}$) }  \\ 
 & \multicolumn{1}{c}{(\AA)}  & \multicolumn{1}{c}{(\AA)}     & \multicolumn{1}{c}{ (erg~s$^{-1}$~cm$^{-2}$~\AA$^{-1}$)} & \multicolumn{1}{c}{(erg~s$^{-1}$~cm$^{-2}$)}  \\ \hline
narrow Ly$\alpha$  & 1215.978 $\pm$ 0.028  & 3.814 $\pm$ 0.097  & 7.976 $\pm$ 0.2305 & 3.237$\pm$ 1.554 \\
broad Ly$\alpha$   & 1216.364 $\pm$ 0.068  &  11.161 $\pm$ 0.401   & 3.949 $\pm$ 0.1928   & 4.6913 $\pm$ 0.1244 \\
red wing Ly$\alpha$ & 1222.118 $\pm$ 0.680 & 45.285 $\pm$ 1.553  & 0.8291 $\pm$ 0.04733 & 3.998 $\pm$ 1.479\\
nv doublet    & 1238.820 $\pm$ 0.000  & 2.798 $\pm$ 0.678  & 0.2163 $\pm$ 0.06970   & 0.6444 $\pm$ 0.1923 \\
nv doublet  & 1242.800 $\pm$ 0.000  & 2.798 $\pm$ 0.678  & 0.2174 $\pm$ 0.06407 & 0.6477 $\pm$ 0.2063 \\
Si~{\sc ii}   & 1304.826 $\pm$ 0.587  & 11.163 $\pm$ 1.384  & 0.3099 $\pm$ 0.03299 & 3.681 $\pm$ 0.3935 \\
\hline
\\
blue shoulder C~{\sc iv}  & 1531.195 $\pm$ 1.453  & 12.504 $\pm$ 3.013  & 0.2217 $\pm$ 0.04337 & 0.2951$\pm$ 0.1018 \\
   narrow C~{\sc iv}  & 1549.336 $\pm$ 0.050  &  5.504 $\pm$ 0.189  & 2.915 $\pm$ 0.1476   & 1.708 $\pm$ 0.1364 \\
broad   C~{\sc iv}    & 1550.474 $\pm$ 0.221  & 15.846 $\pm$ 1.120  & 2.105 $\pm$ 0.1608   & 3.551 $\pm$ 0.1734 \\
red shoulder C~{\sc iv}  & 1567.891 $\pm$ 1.430  & 12.236 $\pm$ 2.895  & 0.2627 $\pm$ 0.06455 & 0.3422 $\pm$ 0.1443 \\
red wing C~{\sc iv}   & 1571.164 $\pm$ 4.241  & 52.928 $\pm$ 5.050  & 0.2891 $\pm$ 0.04924 & 1.629 $\pm$ 0.2956 \\
broad   He~{\sc ii}   & 1642.537 $\pm$ 2.559  & 21.620 $\pm$ 6.140  & 0.2007 $\pm$ 0.02539 & 0.4619 $\pm$ 0.1284 \\
   O~{\sc iii}]        & 1666.212 $\pm$ 2.426  & 18.760 $\pm$ 5.182  & 0.1937 $\pm$ 0.03006 & 0.3868 $\pm$ 0.1211 \\
   Mg~{\sc ii} & 2799.621 $\pm$ 0.088 & 22.454 $\pm$ 0.088 & 0.3103 $\pm$ 0.0025 & 0.7416 $\pm$ 0.0060\\
\\
\\
\hline
    \multicolumn{1}{c}{Component} &
    \multicolumn{1}{c}{Central Wavelength} & 
    \multicolumn{1}{c}{ FWHM } & 
    \multicolumn{1}{c}{ Peak Flux ($\times 10^{-15}$) } & 
    \multicolumn{1}{c}{ Flux ($\times 10^{-15}$) }  \\ 
 & \multicolumn{1}{c}{(\AA)}  & \multicolumn{1}{c}{(\AA)}     & \multicolumn{1}{c}{ (erg~s$^{-1}$~cm$^{-2}$~\AA$^{-1}$)} & \multicolumn{1}{c}{(erg~s$^{-1}$~cm$^{-2}$)}  \\ \hline
narrow He~{\sc ii} & 4685.685 $\pm$ 0.859 &  8.461 $\pm$ 0.093 & 0.719 $\pm$ 0.136 & 6.477 $\pm$ 1.225 \\
broad He~{\sc ii} & 4704.666 $\pm$ 1.951 & 80.605 $\pm$ 4.043 & 1.018 $\pm$ 0.050 & 87.310 $\pm$ 4.091\\
narrow H$\beta$ & 4862.434 $\pm$ 0.309 &  8.461 $\pm$ 0.093 & 1.730 $\pm$ 0.015 & 15.580 $\pm$ 0.218 \\
semi-broad H$\beta$ & 4862.997 $\pm$ 0.212 & 29.741 $\pm$ 0.737 & 7.590 $\pm$ 0.175 & 240.30 $\pm$ 10.44 \\
v. broad H$\beta$ & 4882.901 $\pm$ 2.778 & 93.974 $\pm $7.060 & 1.653 $\pm$ 0.188 & 165.3 $\pm$ 9.4 \\
narrow [O~{\sc iii}] & 4960.375 $\pm$ 0.100 &  8.461 $\pm$ 0.093 & 5.815 $\pm$ 0.141 & 52.370 $\pm$ 1.360 \\
narrow [O~{\sc iii}] & 5008.089 $\pm$ 0.036 &  8.461 $\pm$ 0.093 & 17.300 $\pm$ 0.149 & 155.8 $\pm$ 1.5\\
broad He~{\sc i} & 4929.784 $\pm$ 6.542 & $^{\dagger}$51.600 $\pm$ 0.000 & 0.227 $\pm$ 0.017 & 12.450 $\pm$ 0.946 \\
broad He~{\sc i} & 5023.784 $\pm$ 6.542 & $^{\dagger}$52.600 $\pm$ 0.000 & 0.227 $\pm$ 0.017 & 12.690 $\pm$ 0.964\\
broad He~{\sc i} & 5877.536 $\pm$ 2.411 & 43.754 $\pm$ 5.868 & 1.020 $\pm$ 0.078 & 47.510 $\pm$ 6.197\\
broad Na~D & $^{\dagger}$5891.600 $\pm$ 0.000 & $^{\dagger}$41.900 $\pm$ 0.000  &  0.079 $\pm$ 0.006 &  3.503 $\pm$ 0.266\\
broad Na~D & $^{\dagger}$5897.600 $\pm$ 0.000 &  $^{\dagger}$41.900 $\pm$ 0.000 &  0.079 $\pm$ 0.006 &  3.503 $\pm$ 0.266 \\
narrow [Fe~{\sc X}] & 6087.404 $\pm$ 2.645 &  8.461 $\pm$ 0.093 & 0.338 $\pm$ 0.156 & 3.046 $\pm$ 1.407 \\
unknown & $^{\dagger}$5167.000 $\pm$  0.000 & 8.461 $\pm$ 0.093 & 0.515 $\pm$ 0.117 & 4.640 $\pm$ 1.052\\
unknown & 5959.182 $\pm$ 14.756 & 80.736 $\pm$ 18.757& 0.281 $\pm$ 0.054 & 24.180 $\pm$ 6.972\\
v broad H$\alpha$ &  6563.66$\pm$0.182 &  53.725$\pm$1.602 & 18.22 $\pm$ 1.266  & 1042.0$\pm$ 44.1 \\
broad H$\alpha$ &  6564.723$\pm$0.164 &  27.597$\pm$0.982 & 15.55 $\pm$ 1.312  & 456.7 $\pm$ 54.2 \\
red wing H$\alpha$ &  6597.597$\pm$3.213 & 182.721$\pm$7.209 & 1.894 $\pm$ 0.135 & 368.4 $\pm$ 16.8 \\ \hline

\multicolumn{4}{l}{${\dagger}$ Errors where given are 68\% confidence intervals.}
    \end{tabular}
    \caption{Parameters for Gaussian fits in the vicinity of broad Ly$\alpha$ (upper panel), broad  C~{\sc iv} (middle panel) and broad H$\alpha$, H$\beta$, He~{\sc i} and He~{\sc ii} (lower panel).} 
    \label{mega_tab}
\end{table}

\twocolumn

\begin{minipage}{0.25\textheight}
\vspace{1cm}
\includegraphics[width=1.3\textwidth]{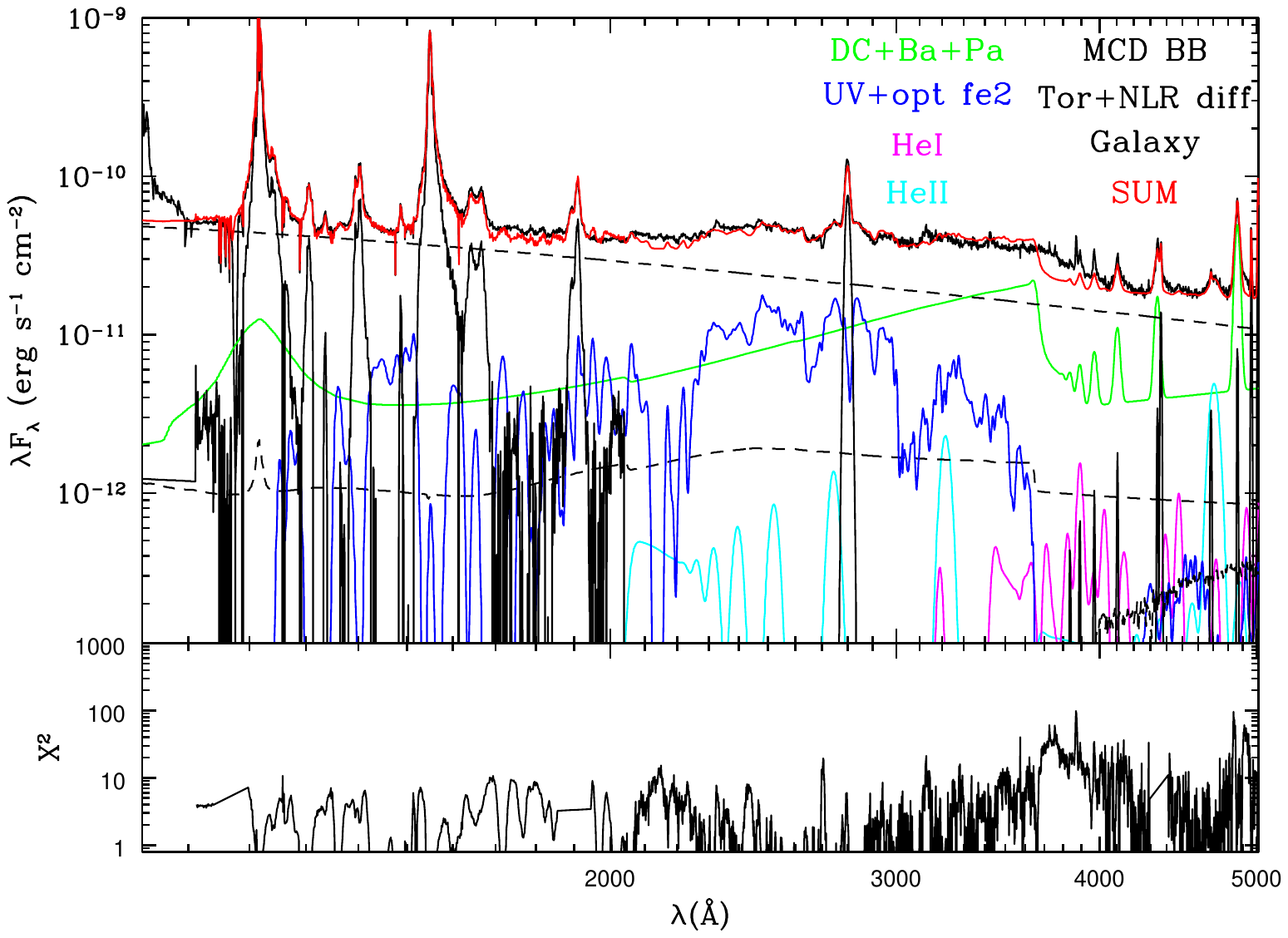}
\includegraphics[width=1.3\textwidth]{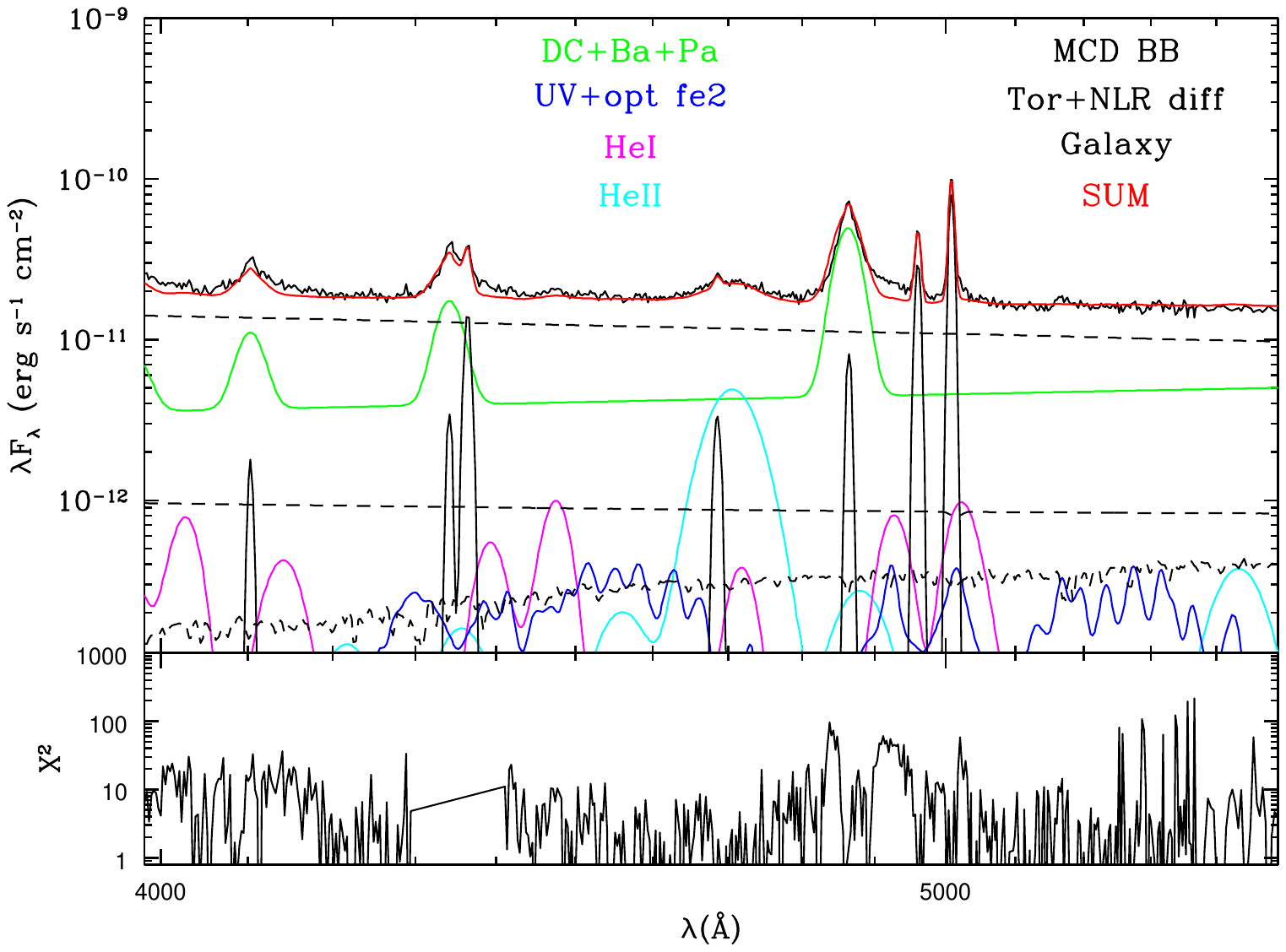}
\includegraphics[width=1.3\textwidth]{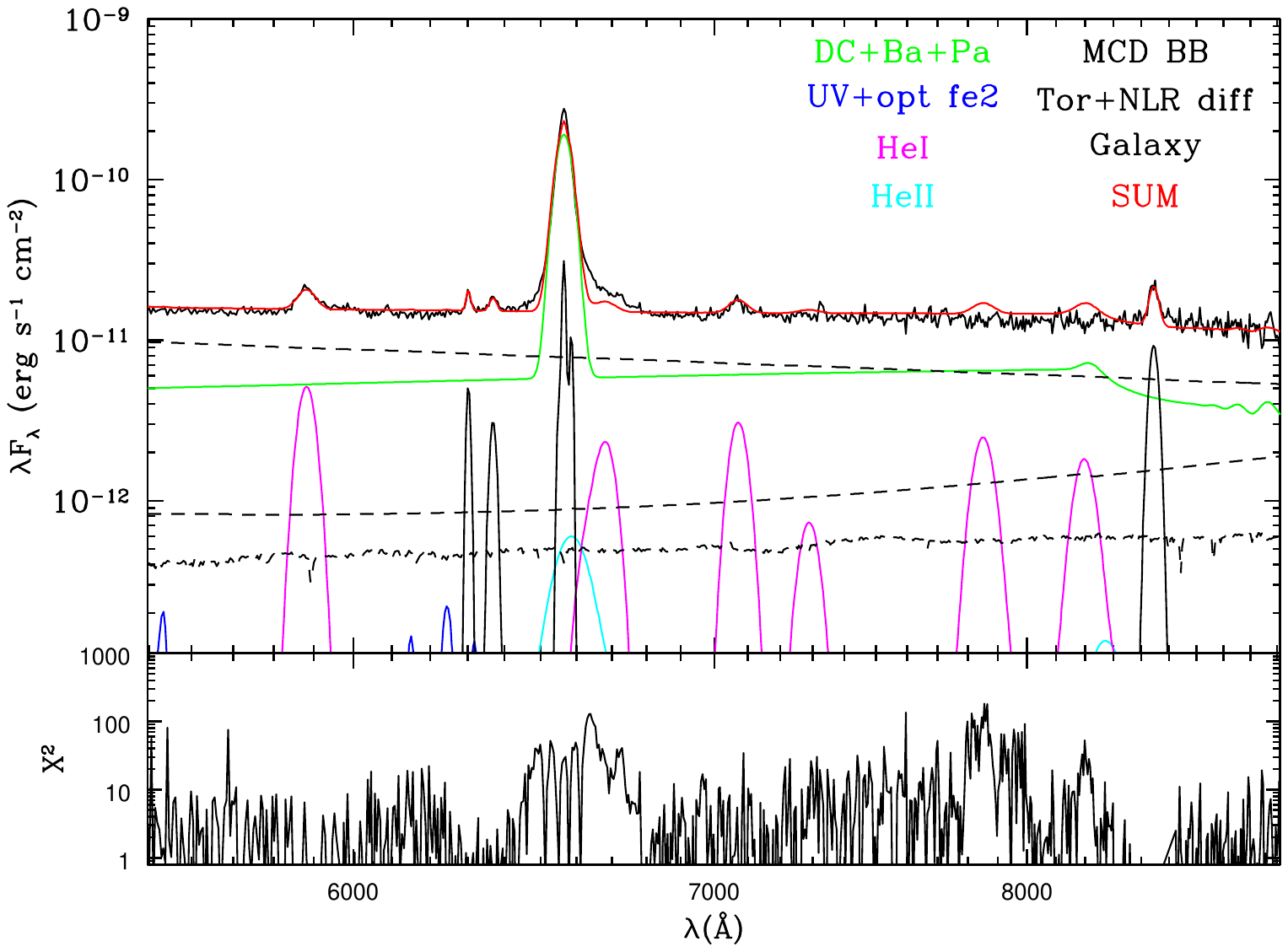}
\includegraphics[width=1.3\textwidth]{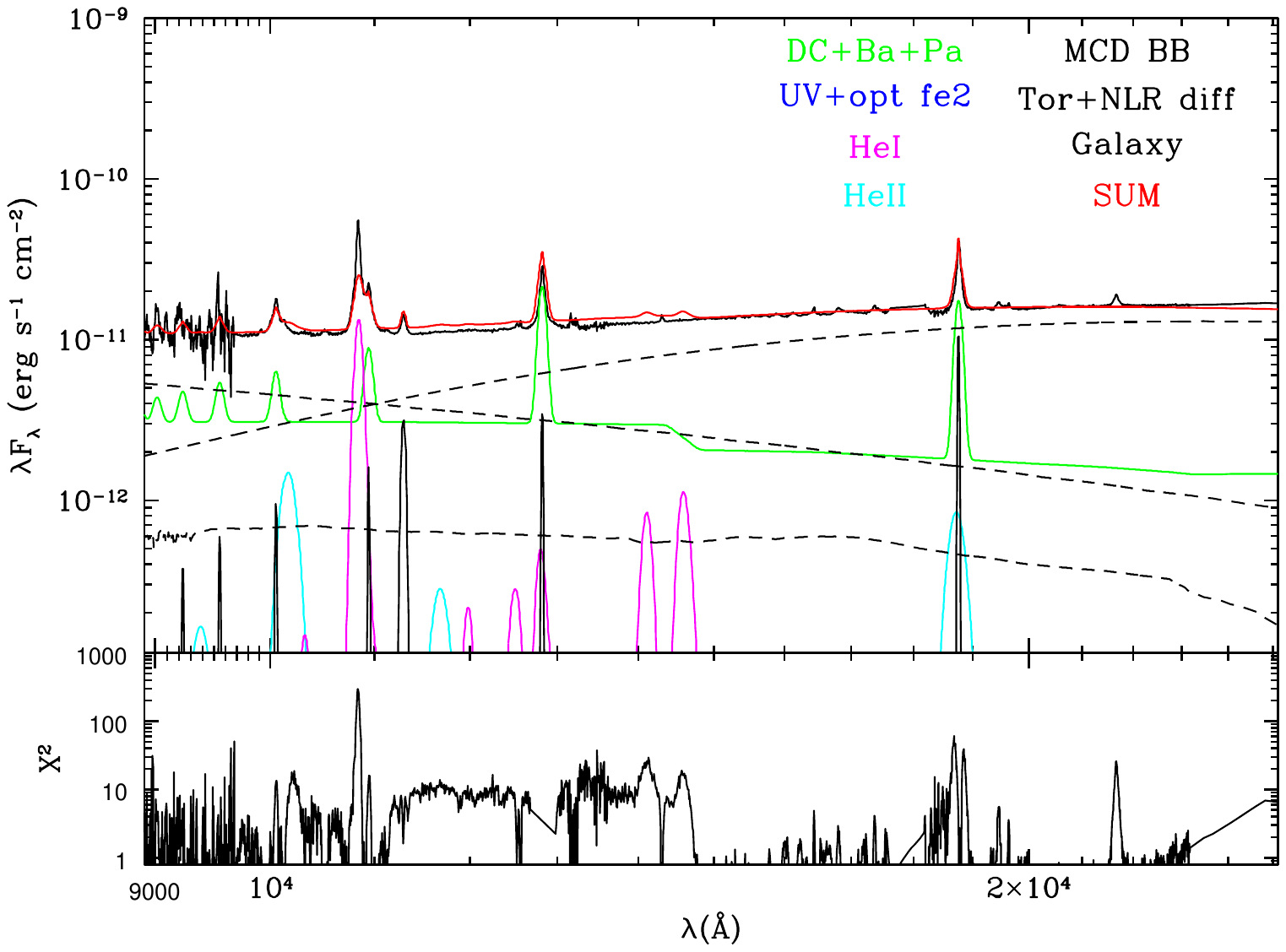}
\label{zero_turb}
\end{minipage}
\hspace{0.05\textwidth}
\begin{minipage}{0.4\textwidth}
\centering
Figure~\ref{zero_turb}1. Close up view of Figure~15 $v_{\rm turb}=0$ km~s$^{-1}$.
\end{minipage}

\begin{minipage}{0.25\textheight}
\vspace{1cm}
\includegraphics[width=1.3\textwidth]{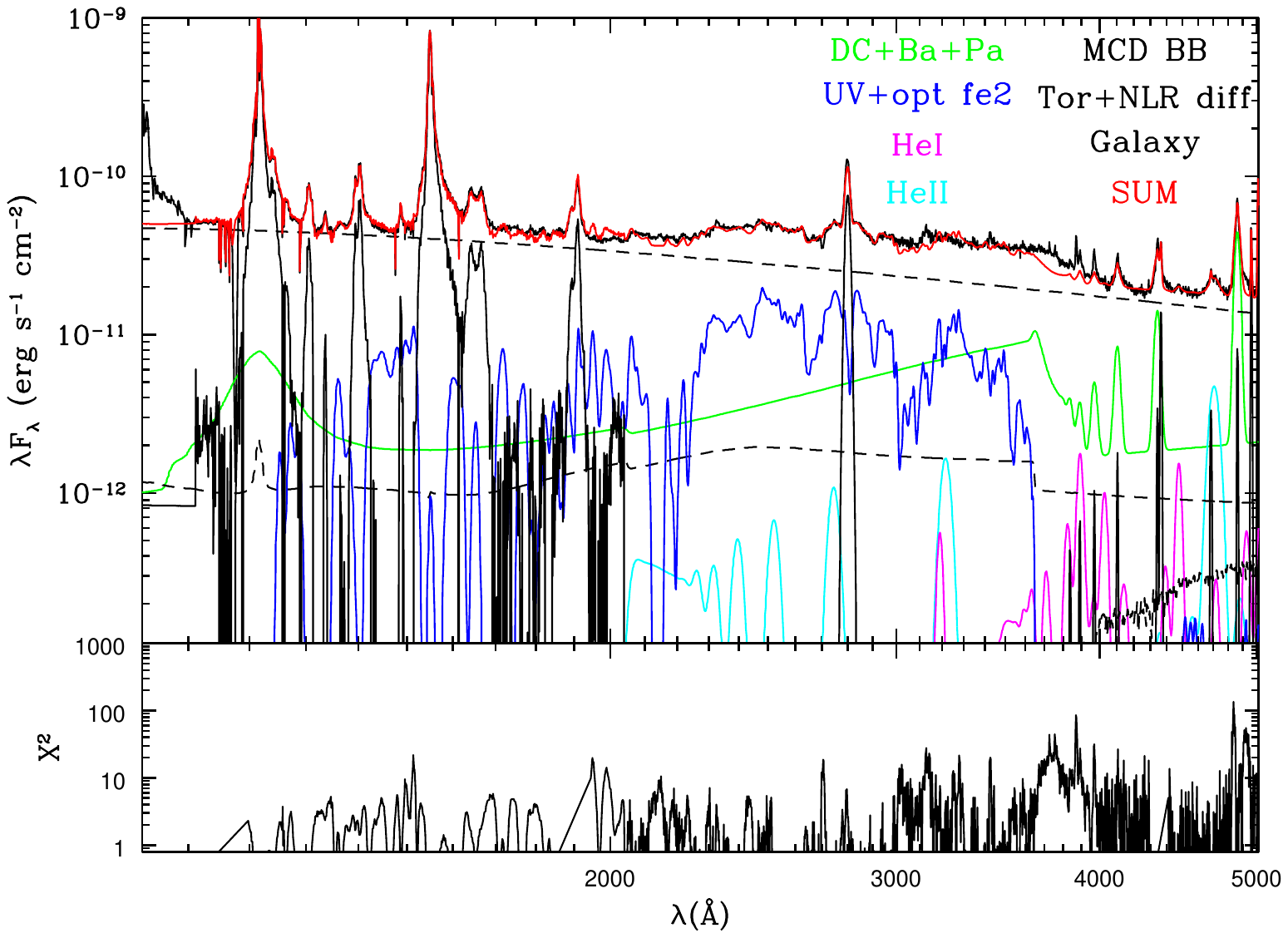}
\includegraphics[width=1.3\textwidth]{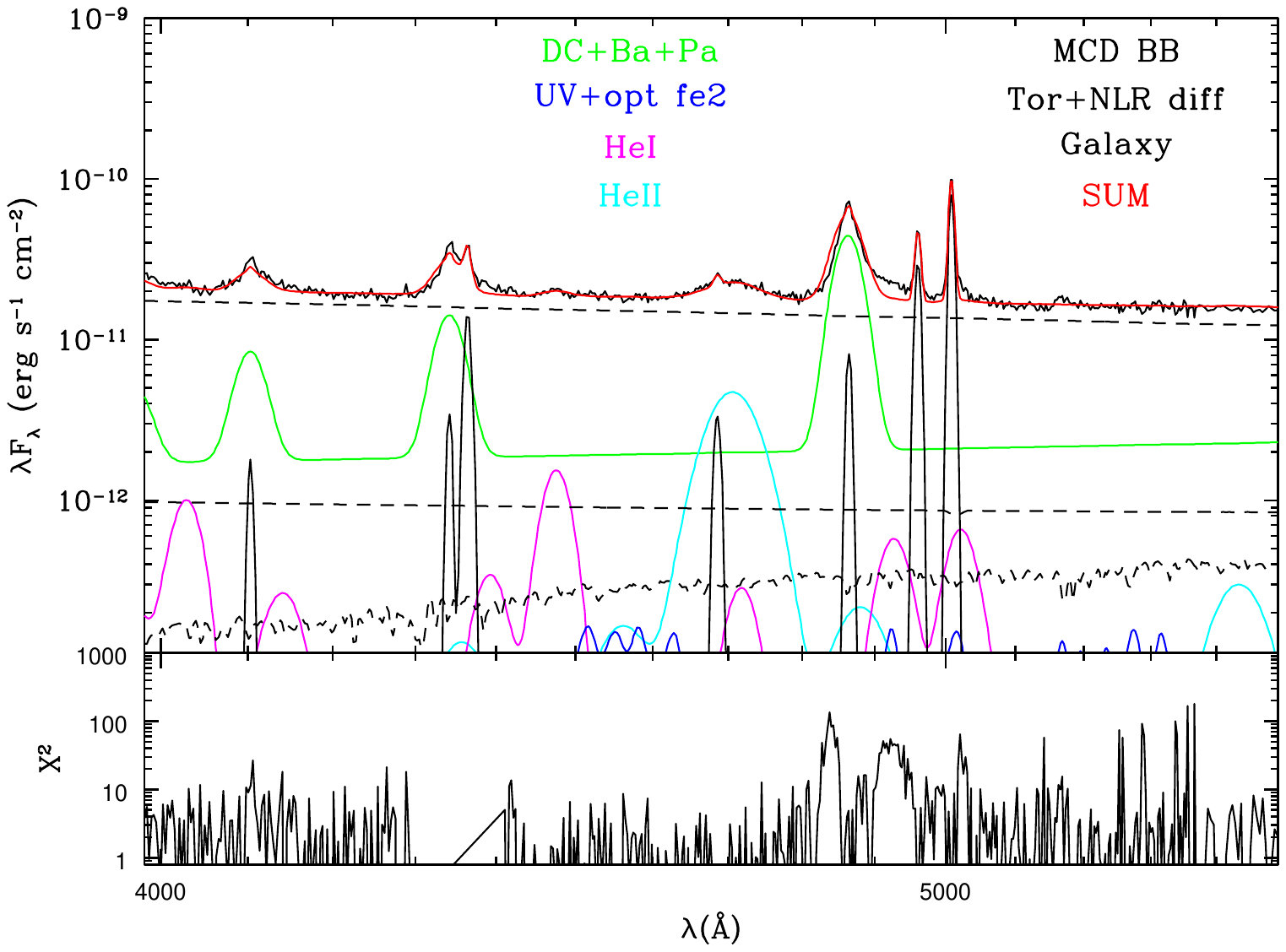}
\includegraphics[width=1.3\textwidth]{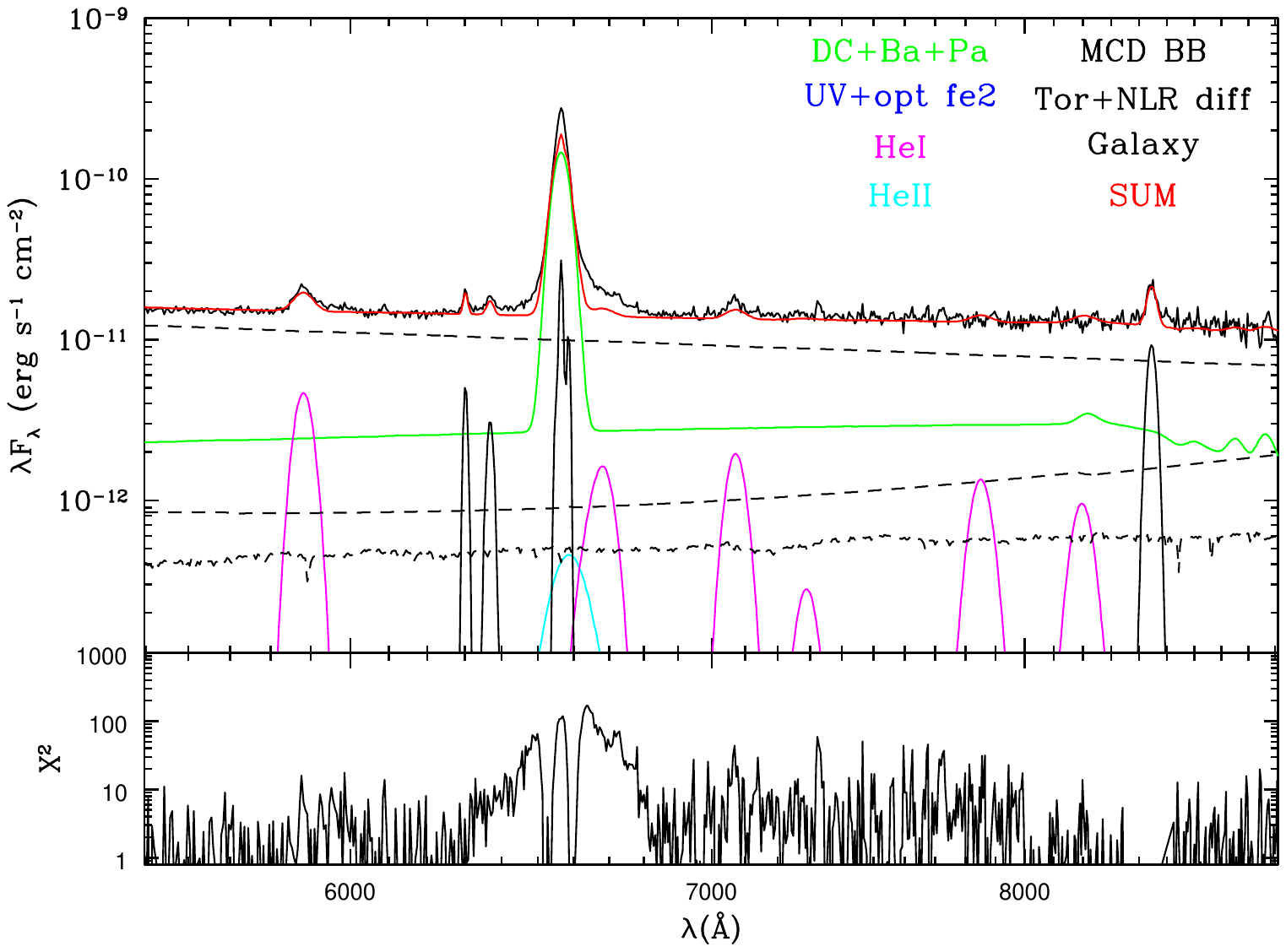}
\includegraphics[width=1.3\textwidth]{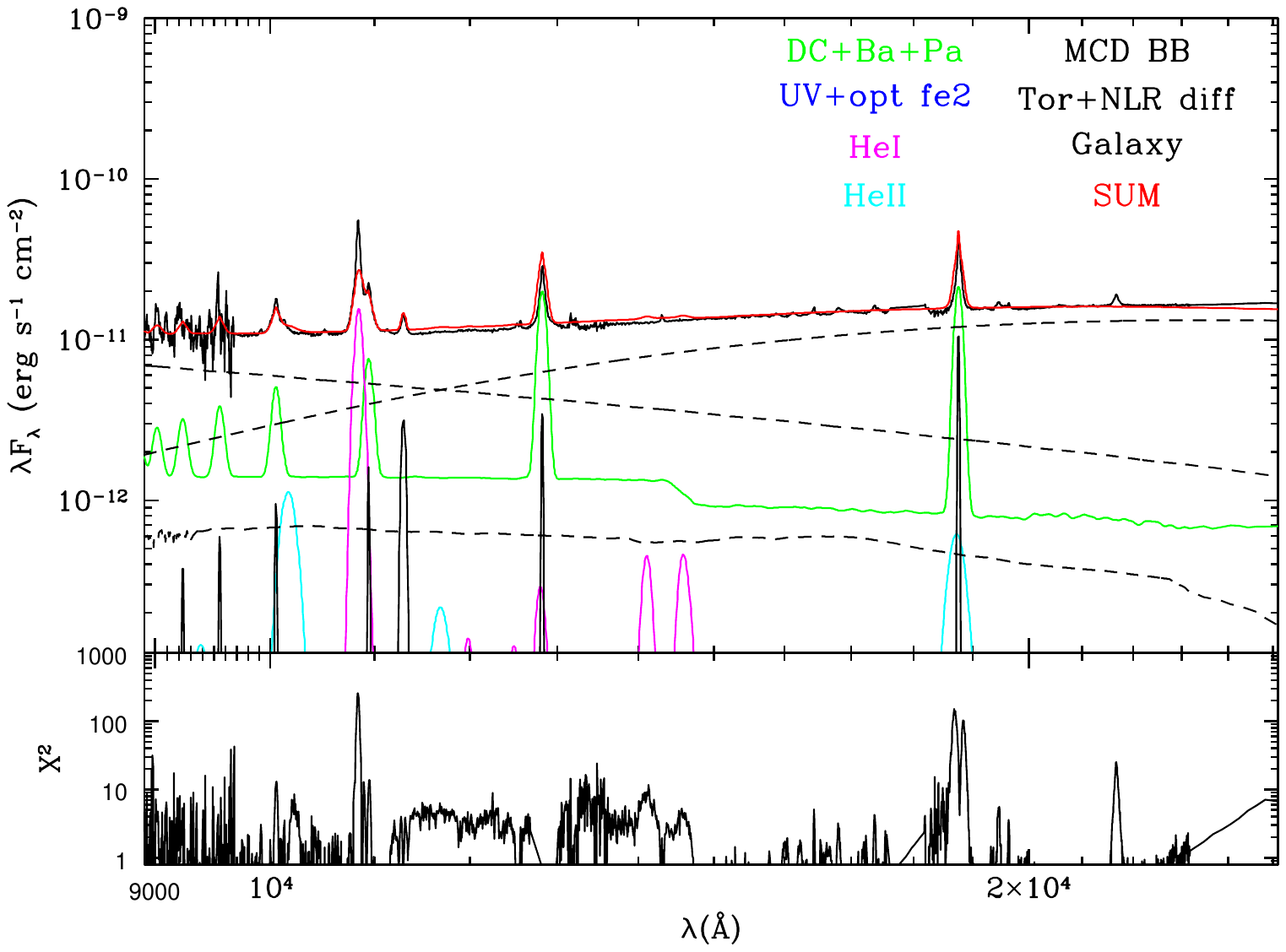}
\label{turb_100}
\end{minipage}
\hspace{0.05\textwidth}
\begin{minipage}{0.4\textwidth}
\centering Figure~\ref{turb_100}2. Close up view of Figure~17, $v_{\rm turb}=100$ km~s$^{-1}$.
\end{minipage}

%%%%%%%%%%%%%%%%%%%%%%%%%%%%%%%%%%%%%%%%%%%%%%%%%%

% Don't change these lines
\bsp	% typesetting comment
\label{lastpage}
\end{document}